\documentclass[prd, aps, nofootinbib, preprintnumbers, showpacs, superscriptaddress, twocolumn, floatfix]{revtex4}

\usepackage{amssymb}
\usepackage[T1]{fontenc}
\usepackage[english]{babel}
\usepackage{graphicx}
\usepackage[latin9]{inputenc}
\usepackage{mathpazo}
\usepackage{mathrsfs}
\usepackage{float}

\newcommand{\be}{\begin{equation}}
\newcommand{\ee}{\end{equation}}

\begin{document}

\title{Non-axisymmetric instability and fragmentation of general relativistic quasi-toroidal stars}

\author{Burkhard~Zink} 
\affiliation{Center for Computation and
Technology, Louisiana State University, Baton Rouge, LA 70803, USA}
\affiliation{Horace Hearne Jr. Institute for Theoretical Physics,
Louisiana State University, Baton Rouge, LA 70803, USA}

\author{Nikolaos~Stergioulas}   \affiliation{Department of Physics,
Aristotle University of Thessaloniki, Thessaloniki 54124, Greece}

\author{Ian~Hawke}  \affiliation{School of Mathematics,  University of
Southampton, Southampton SO17 1BJ, UK}

\author{Christian~D.~Ott}    \affiliation{Max-Planck-Institut f\"ur
Gravitationsphysik, Albert-Einstein-Institut, 14476 Golm, Germany}
\affiliation{Department of Astronomy and Steward Observatory,
The University of Arizona, Tucson, AZ, USA}

\author{Erik~Schnetter}   \affiliation{Center for Computation and
Technology, Louisiana State University, Baton Rouge, LA 70803, USA}
\affiliation{Max-Planck-Institut f\"ur
Gravitationsphysik, Albert-Einstein-Institut, 14476 Golm, Germany}

\author{Ewald~M\"uller}     
\affiliation{Max-Planck-Institut f\"ur Astrophysik,
Karl-Schwarzschild-Str.\ 1, 85741 Garching bei M\"unchen, Germany}

\begin{abstract}

In a recent publication, we have demonstrated that differentially rotating
stars admit new channels of black hole formation via fragmentation instabilities. 
Since a higher order instability of this kind could potentially transform a differentially
rotating supermassive star into a multiple black hole system embedded in a massive accretion
disk, we investigate the dependence of the instability on parameters of the
equilibrium model. We find that many of the models constructed exhibit non-axisymmetric
instabilities with corotation points, even for low values of $T/|W|$,
which lead to a fission of the stars into one, two or three fragments, depending 
on the initial perturbation.

At least in the models selected here, an $m=1$ mode becomes unstable at lower
values of $T/|W|$, which would seem to favor a scenario where one black hole with a
massive accretion disk forms. In this case, we have gained evidence that low values
of compactness of the initial model can lead to a stabilization of the resulting fragment,
thus preventing black hole formation in this scenario.

\end{abstract}


\maketitle

\section{Introduction}
\label{sec:intro}

The study of oscillations and stability of stars has a long history (see
e.g. \cite{Tassoul-1978:theory-of-rotating-stars}). While its classical
results tried to address the limits of possible models for main-sequence stars,
and the question of binary star formation from protostellar clouds, it has been
extended in the last century to examine the properties of relativistic fluid 
equilibria, and their connection to the formation of black holes by axisymmetric
instabilities.

This work is an extension of our publication \cite{Zink2005a}, where we have 
studied the formation of black holes from fragmentation in $N = 3$ polytropes,
which was induced by the growth of a non-axisymmetric instability
in a strongly differentially rotating star. We would like to focus attention
here on the question of whether the instability we have observed is generic for
strongly differentially rotating polytropes, and how changes in parameters
of the initial model affect its development. In addition, we will present
some tentative results related to arguments previously presented
by Watts et al. \cite{Watts:2003nn} on a possible connection between 
low-$T/|W|$\footnote{Here, and subsequently, $T/|W|$ shall denote the ratio
of rotational kinetic to gravitational binding energy in the axisymmetric
initial model. For a definition of these quantities, see \cite{Stergioulas98}.}
and spiral-arm instabilities and the location of the corotation band in a 
sequence of increasing rotational energy.

Due to the recent prospect of detecting gravitational radiation directly, 
the connection between the local dynamics of collapse and gravitational wave 
emission is currently receiving increased attention (e.g. \cite{Thorne97, Baiotti04b}). In
this context, a non-axisymmetric instability in a star is expected to change 
the nature of the signal, and to enhance the chances of detecting it \cite{Houser:1994ds}. 
We shall discuss a number of scenarios for gravitational collapse and black hole
formation to illustrate this point.

\begin{enumerate}

\item \emph{Stars retaining spherical symmetry:} If the initial matter distribution has spherical 
symmetry, no gravitational waves are emitted as a consequence of Birkhoff's theorem.
The exact solution by Oppenheimer and Snyder \cite{Oppenheimer39a} already
exhibits many features of the local dynamics, while the connection to
dynamical stability in general relativity has been made explicit by Chandrasekhar 
\cite{Chandrasekhar64}. The assumption of spherical symmetry, whilst restrictive, already admits 
simple models of phenomena like mass limits for compact stars, some generic properties of black 
hole formation from supermassive stars and neutron stars 
(e.g. \cite{May66, Shapiro79z, Shapiro80}), and the dynamics of apparent and 
event horizons (ibido).

\item \emph{Stars retaining (approximate) axisymmetry:} If the symmetry assumption is relaxed to
axisymmetry, models of gravitational collapse admit a number of additional
features, the
most important in this context being the emission of gravitational
radiation. Stars in axisymmetry
can be rotating, which changes the radial modes of the non-rotating member of a sequence
into a quasi-radial mode. If that is unstable, the star may collapse to a black hole
in a manner which is similar to the spherically symmetric case in its bulk properties, and
it proceeds by (i) contraction due to a quasi-radial instability, (ii) formation of an event 
horizon centered on the axis, and (iii) ring-down to a Kerr black hole with a disk. We
will call this process the \emph{canonical scenario} to represent that it provides the
expected properties of the collapse of slowly rotating stars. It has been studied
extensively in numerical investigations \cite{Nakamura80a, Nakamura80, Nakamura81, Nakamura83,
Stark85, Shibata00a, Shibata99e, Saijo2002, Shibata02, Font01b, Shibata:2003iy, Duez:2002bn,
Duez04, Baiotti04b, Saijo2004, Baiotti04}, and seems even appropriate to describe the 
quality of collapse of most rapidly and differentially rotating neutron stars \cite{Duez04}. It
should not be considered implicit here that the canonical scenario is generic for 
axisymmetric collapse: see e.g. \cite{Hughes94a, Abrahams94a} for systems involving 
toroidal black holes.

\item \emph{Stars not retaining (approximate) axisymmetry:} As already mentioned, even
in many numerical models with three spatial dimensions, the collapse to a black hole proceeds
in an almost axisymmetric manner, although the initial data is represented on discrete
Cartesian grids. It has been found that even when non-axisymmetric perturbations are applied
to the collapsing material, no large deviations from axisymmetry are seen during the collapse
\cite{Saijo2004}. Judging from the perturbative theory of Newtonian polytropes 
\cite{Tassoul-1978:theory-of-rotating-stars}, this indicates that either the amount of rotational 
over gravitational binding energy $T/|W|$ is insufficient, or that the
collapse time is too short to admit growth of initial deviations from the symmetric state
to significant levels.

The situation can be quite different when the system is not unstable to axisymmetric perturbations,
or if the collapse stabilizes around a new equilibrium with higher $T/|W|$. The classical
limit of $T/|W| \approx 0.27$ for Maclaurin spheroids \cite{Chandrasekhar69c} indicates
the onset of a dynamical instability to transition to the $x = +1$ Riemann S-type 
sequence \cite{Chandrasekhar69c, Christodoulou95i}. To which extent this idealized 
behaviour is also realized in general relativistic compressible polytropes, and,
more specifically, how it is connected to the formation of black holes, is the issue 
we would like to address in part here.

\end{enumerate}

If a general relativistic star encounters a non-axisymmetric instability, the nature
of its subsequent evolution may be characterizable by certain properties of the 
equilibrium model, like the rotation law, $T/|W|$, compactness, and equation of 
state. For the limit of uniformly rotating, almost homogeneous models of low
compactness, we expect, for $T/|W| > 0.27$, a dynamical transition to an ellipsoid 
by a principle of correspondence with Newtonian gravity.

By relaxing all but the assumption of low compactness, we can make use of the rich
body of knowledge about the stability and evolution of stars in Newtonian gravity.
Two classical applications of stability theory are the oscillations of disks and 
the fission problem \cite{Tassoul-1978:theory-of-rotating-stars}. These matters have been 
investigated extensively over the 
last decades \cite{Tohline85, Durisen86, Tohline90, Bonnell94, Loeb94, Bonnell95,
Pickett96, Houser96, Rampp98, Brown2000, Imamura2001, Centrella:2001xp, Colpi2002, Davies2002, 
Imamura2003, Banerjee2004}, and a number of possible scenarios have emerged for
the non-linear evolution of non-axisymmetric dynamical instabilities in 
rotating polytropes:

\begin{enumerate}

\item The polytrope develops a bar-mode instability similar to the Maclaurin case,
and possibly retains this shape over many rotational periods (e.g. \cite{Brown2000}).

\item Two spiral arms and an ellipsoidal core region develop, where the latter transports
angular momentum to the spiral arms by gravitational torques, is spun down, and
collapses (e.g. \cite{Durisen86}). This scenario is interesting for black hole formation,
since the rotational support of the initial model can be partially removed by such
a mechanism. If this transport is efficient enough, the core ellipsoid might collapse,
resulting in a Kerr black hole with a disk of material around it. One might conjecture
that, if the equation of state is soft enough, the disk itself may be subject to 
fragmentation, and form several smaller black holes which are subsequently 
accreted.

\item One spiral arm develops, and the mode saturates at some amplitude 
\cite{Pickett96, Ott:2005gj}, leaving a central condensation. This might also lead to central
black hole formation. Note that the onset for this dynamical instability
in terms of $T/|W|$ can be significantly lower than the Maclaurin limit. 

\item For polytropes with strong differential rotation, the initial model may
be \emph{quasi-toroidal}, i.e. it has at least one isodensity
surface which is homeomorphic to a torus. If models of this kind, or purely
toroidal ones, are subject to the development of a non-axisymmetric instability,
they may exhibit fragmentation \cite{Rampp98, Centrella:2001xp}. This is
clearly the most interesting setting for the fission problem, but has been 
discussed also in the context of core collapse (``collapse, pursuit and
plunge'' scenario, Fig. 24.3 in \cite{Misner73}, also \cite{Bonnell95, Davies2002}).

\end{enumerate}
It is this last kind of instability we will investigate here in the context 
of general relativity, and its relation to the formation of black holes.

Concerning the \emph{nature} of the spiral-arm and low-$T/|W|$ instabilities, Watts et
al. (\cite{Watts:2003nn}, see also \cite{Watts:2002ik, Watts:2003uj}) have suggested a possible relation of these instabilities to their location
with respect to the corotation band\footnote{The
\emph{corotation band} in a differentially rotating star is the set of frequencies
associated with modes having at least one \emph{corotation point}, i.e. a 
point where the local pattern speed of the instability matches the local angular
velocity.}.
That corotation has a bearing on instabilities in differentially rotating disks has been suggested for some time;
the perturbation operator is singular at corotation points, which gives rise
to a continuous spectrum of ``modes.'' While the initial-value problem of 
perturbations associated with the continuous spectrum of stars is not well 
understood even in Newtonian gravity, there is evidence \cite{Papaloizou84,
Papaloizou85, Balbinski85b, Watts:2002ik} that a mode entering corotation
may be subject to a shear-type instability, or that it merges with another mode
inside the corotation band, which appears to admit a certain class of solutions
showing similar properties as the solutions in the discrete spectrum 
\cite{Watts:2003uj}. While three-dimensional Newtonian simulations would appear,
at least as a first step, most appropriate to gain more intuition
in these matters \cite{Saijo:2005gb},
we will collect some evidence on corotation points in the evolutions presented here as well.

Since
the parameter space of possible initial models is large, and given that three-dimensional
simulations of this kind are still quite expensive in terms of computational resources,
we restrict attention to several isolated sequences, where just one
initial model parameter is varied to gain evidence on its systematic effects, and to a plane
in parameter space defined by a constant central rest-mass density and a fixed parameter
$\Gamma = 4/3$ in the $\Gamma$-law equation of state $P = (\Gamma-1) \rho \epsilon$. 
We will find that, at least as long as we 
are concerned with the question when certain 
modes become dynamically unstable on a sequence, the consideration of models of constant central 
density is not overly restrictive as far as the  development of the instability is concerned, 
while the nature of the final remnant might be rather sensitive to it. This latter issue,
namely whether a black hole forms or not, will not be answered in full here, 
since we will only determine whether the fragment stabilizes during collapse,
and re-expands, or if it does not. We leave the location of the apparent horizon
with adaptive mesh-refinement techniques and the subsequent evolution with excision
to future work, and concentrate here on the general structure of the parameter
space and its relation to the non-axisymmetric instability.

The choice $\Gamma = 4/3$ is well known to approximately correspond to the adiabatic 
coefficient of a degenerate, relativistic 
Fermi gas or a radiation pressure dominated gas \cite{Shapiro83}, and is thus closely
connected to iron cores and supermassive stars.
We would like to point out that a collapsing iron core, even if its initial 
state is assumed to be determined mostly by electron pressure, is subject to a complex set 
of nuclear reactions, which involve the generation of neutrinos and transition 
to nuclear matter at high densities. It is because of this complexity, which we do not
take into account here in order to reduce the number of free parameters, that we do not
suggest the use of the fragmentation and black hole formation investigated in 
\cite{Zink2005a} as a highly idealized model of core collapse.
For supermassive stars, the situation is different, since an event horizon can
form before thermonuclear reactions become important, depending on the metallicity
and mass of the progenitor \cite{Fuller86}. Because of this, it is conceivable
that the type of evolution in \cite{Zink2005a} can be used as an approximate
model of supermassive black hole formation. Finally, we would like to mention the
possibility that gravitational wave detection may uncover so-far unexpected processes
involving black hole formation, and in that case it is useful to have a
general understanding of possible dynamical scenarios.

\section{Previous work}
\label{sec:previous}

The background for this study comes from three areas: the study of (i) fragmentation
in Newtonian polytropes, (ii) non-axisymmetric instabilities in general relativistic
polytropes, and (iii) black hole formation by gravitational collapse. The first
area is represented by a large number of publications (some of them cited in the introduction),
but we would like to mention specifically the work by Centrella,
New et al. \cite{New:2001nq, Centrella:2001xp}, since the kind of initial model and
subsequent evolution studied in these publications are similar to the ones presented 
by us in \cite{Zink2005a}
and here, apart from the fact that Newtonian gravity and a softer equation
of state ($\Gamma = 1.3$) was used. New and Shapiro \cite{New:2001qs} investigated
equilibrium sequences of differentially rotating Newtonian polytropes with 
$\Gamma = 4/3$ to present an evolutionary scenario where supermassive stars 
develop a bar-mode instability instead of collapsing axisymmetrically. This kind
of scenario (see also \cite{Bodenheimer73, Houser:1994ds}) is also important when connecting
the general relativistic fragmentation presented here to the evolution of
supermassive stars.

Non-axisymmetric dynamical instabilities in general relativistic, self-gravitating 
fluid stars have been studied by several authors \cite{Shibata:2000jt, Saijo:2000qt, 
Shibata:2004kb, Duez04, Saijo:2005yn, Baiotti06b}. Some evidence of fragmentation has been found
in \cite{Duez04} in a ring resulting from a ``supra-Kerr'' collapse with
$J/M^2 > 1$
(here $J$ denotes the total angular momentum, and $M$ the ADM mass), 
but no black hole was identified, although the pressure in the initial
data was artificially reduced by a large factor in order to induce collapse.
Finally, black hole formation by gravitational collapse has been studied extensively,
(see references in the introduction), and in recent years also in three spatial 
dimensions \cite{Shibata99e, Shibata02, Duez:2002bn, Baumgarte02b, Duez04, Baiotti04b,
Baiotti04, Duez:2005cj}. The collapse of differentially rotating supermassive stars
in the approximation of spatial conformal flatness has been investigated by Saijo 
\cite{Saijo2004}.

In addition, the work on low-$T/|W|$ instabilities by Watts et al. 
\cite{Watts:2002ik, Watts:2003uj, Watts:2003nn} has already been described in the 
introduction, and recent numerical studies of related interest can be found
in \cite{Shibata:2002mr, Shibata:2003yj, Saijo:2005gb}.

\section{Numerical methods}
\label{sec:methods}

All model calculations presented here are performed in 
full general relativity. We use Cartesian meshes with the 
only symmetry assumption being reflection invariance with respect
to the equatorial plane of our models. We employ the 
{\tt Cactus} computational framework \cite{Goodale02a, url:cactus} in our 
simulations. We use finite-differencing methods of second
order for the spacetime variables, and finite-volume, high-resolution
shock-capturing techniques for the hydrodynamical variables. Hydrodynamics
and metric evolution are coupled by means of the \emph{method 
of lines}, using the time-explicit iterative Crank-Nicholson 
integrator with three intermediate steps 
\cite{Alcubierre99d,Gustafsson95,New98}.

In the following we give a brief overview on the
3-metric evolution, hydrodynamics and mesh refinement methods
used. We set $c=G=K=1$ to fix our system of units, where $c$ denotes
the speed of light, $G$ the gravitational constant and $K$ the
polytropic constant in the relation $P = K \rho^\Gamma$ between
pressure $P$ and rest-mass density $\rho$
used to compute the initial models. Latin indices run from 1 to 3,
Greek from 0 to 3. We use a spacelike signature $(-,+,+,+)$.
Unless explicitly mentioned otherwise, we assume  
Einstein's summation convention.

\subsection{Metric evolution}

We evolve the 3-metric with the 3+1 Cauchy
free-evolution code developed at the
Albert Einstein Institute \cite{Alcubierre99d, cactuseinsteinweb}. The code
provides at each time step a solution to the Einstein 
equations in the ADM 3+1 formulation \cite{Arnowitt62},
while internally evolving the spacetime in the 
NOK-BSSN conformal-traceless recast of the ADM 
equations \cite{Nakamura87,Shibata95,Baumgarte99} which
has been proven to lead to much more stable numerical
evolutions of Einstein's equation than the original ADM 
system \cite{Alcubierre99d}. The details of our NOK-BSSN 
implementation can be found in \cite{Alcubierre99d,Alcubierre02a}. 
Here we only briefly summarize the ADM system through which we 
couple spacetime and hydrodynamics~\cite{Baiotti04}. 

In ADM, the four-dimensional spacetime is foliated into 
a set of three-dimensional and non-intersecting spatial 
hypersurfaces. Individual surfaces are related through
the lapse function $\alpha$, which describes the
rate of advance of time along a timelike unit
normal vector n$^\alpha$, and through the shift vector $\beta^i$
which indicates how the coordinates change from one
slice to the next. The gauge-freedom in ADM allows a free
choice of both lapse and shift, but care must be taken for
the choice of gauge may have an effect on numerical 
stability~\cite{Alcubierre02a}.

The ADM line element reads
\begin{equation}
ds^2 = -(\alpha^{2} -\beta_{i}\beta ^{i}) dt^2 + 
        2 \beta_{i} dx^{i} dt +\gamma_{ij} dx^{i} dx^{j}
\end{equation}
and the first-order form of the evolution equations for the spatial 3-metric
$\gamma_{ij}$ and the extrinsic curvature $K_{ij}$
read
\begin{eqnarray}
\partial_t \gamma_{ij} &=& - 2 \alpha K_{ij}+\nabla_i
        \beta_j + \nabla_j \beta_i, 
\label{dtgij} \\
        \partial_t K_{ij} &=& -\nabla_i \nabla_j \alpha + \alpha \Biggl[
        R_{ij}+K\ K_{ij} -2 K_{im} {K^m}_j  \nonumber \\
        &\ & - 8 \pi \left( S_{ij} - \frac{1}{2}\gamma_{ij}S \right)
        - 4 \pi {\rho}_{_{\rm ADM}} \gamma_{ij}
        \Biggr] \nonumber \\ 
        &\ & + \beta^m \nabla_m K_{ij}+K_{im} 
        \nabla_j \beta^m+K_{mj} \nabla_i \beta^m ,
        \nonumber \\
\label{dtkij}
\end{eqnarray}
where $\nabla_i$ denotes the covariant derivative with respect
to the coordinate direction $\partial_i$ and the 3-metric $\gamma_{ij}$.
$R_{ij}$ is the coordinate representation of the 3-Ricci tensor
and $K=\gamma^{ij} K_{ij}$ is the trace of the extrinsic curvature.
$\rho_{\rm ADM}$ is the energy density as measured by an Eulerian
observer \cite{York73}. $S_{ij}$ is the projection of the
stress-energy tensor on the spacelike hypersurfaces 
and $S = \gamma^{ij}S_{ij}$.

\subsubsection*{Gauge choices}

We evolve the lapse function with the 
hyperbolic K-driver condition \cite{Bona95b,Alcubierre02a} 
of the form
\begin{eqnarray}
\partial_t \alpha &=& - \alpha^2 f(\alpha) (K - K_0), 
\label{eq:hypKdriver}
\end{eqnarray}
where $f(\alpha)$ is an arbitrary positive function of $\alpha$ and
$K_0 = K(t=0)$.  We choose $f(\alpha) = 2 / \alpha$ which 
is referred to as \emph{1+log slicing} and has excellent 
singularity-avoiding properties in the sense that the lapse
tends to zero near a physical singularity, \emph{freezing}
the evolution in that region.

For the model simulations presented in this paper we find
that a dynamical evolution of the 
spatial gauge $\beta^i$ is not necessary, and we keep it
fixed to its initial direction and magnitude.

\subsection{General-relativistic hydrodynamics}

The equations of general-relativistic hydrodynamics
are derived from the conservation equations
of the stress-energy tensor $T^{ab}$ and 
the matter current density $J^a$:
\begin{equation}
\nabla_a T^{ab} = 0\;,\;\;\;\;\;\;
\nabla_a J^a = 0\,\,,
\end{equation}
where $J^a = \rho u^a$, $\rho$ is the rest-mass
density and $u^a$ the 4-velocity of the fluid. We
use the perfect-fluid stress energy tensor
\begin{equation}
T^{ab} = \rho h u^{a} u^{b} + P g^{ab},
\end{equation}
with $P$ being the fluid pressure, 
$h = 1 + \epsilon + P/\rho$ the relativistic
specific enthalpy and $\epsilon$ the specific internal energy
of the fluid. 

For evolving the hydrodynamical fields we employ
the {\tt Whisky} code \cite{Baiotti04, url:whisky} which implements
the general relativistic hydrodynamics equations in the hyperbolic first-order 
flux-conservative form proposed and tested in \cite{Banyuls97,Font98b}.
This code requires us to add an artificial atmosphere to the
computational domain in regions of very low density. We typically
choose an atmospheric density of $10^{-5}$ of the maximal density of
the initial model.
The evolved state vector $\vec{\cal{U}} = (D,S_i,\tau)^T$ is 
defined in terms of the {\it primitive} hydrodynamical 
variables $\rho$, $P$, and $v^i$, the 3-velocity, measured
by an Eulerian observer:
\begin{equation}
\vec{\cal{U}} = 
\left[ 
\begin{array}{c}
{D} \\
{S_j} \\
{\tau} \\
\end{array}  
\right]
= \left[ 
\begin{array}{c}
\sqrt{\gamma} W \rho \\
\sqrt{\gamma} \rho h W^2 v_j \\
\sqrt{\gamma} (\rho h W^2 - P - W \rho) \\
\end{array}  
\right],
\end{equation}
where $\gamma = det \gamma_{ij}$ and 
$W = \alpha u^0 = 1/\sqrt{1 - \gamma_{ij} v^i v^j}$ is the
Lorentz factor.

The set of equations then reads
\begin{equation}
{\partial}_t \vec{\cal{U}} + {\partial}_i \vec{F^i} = \vec{S},
\end{equation}
with the three flux vectors given by
\begin{equation}
\vec{F^i} = \left[ \begin{array}{c}
                    \alpha (v^i - \frac {1}{\alpha} {\beta}^i) {D} \\
                    \alpha ((v^i - \frac {1}{\alpha} {\beta}^i) 
                         {S_j} + \sqrt{\gamma} P {\delta}^i_j) \\
                    \alpha ( (v^i - \frac {1}{\alpha} {\beta}^i) 
                         {\tau} + \sqrt{\gamma} v^i P)
                 \end{array}  \right].
\end{equation}
The source vector $\vec{S}$, which contains the 
curvature-related force and work terms, but no derivatives of the
primitive variables, is given by
\begin{equation}
\vec{S} = \left[ \begin{array}{c}
                  0 \\
                  \alpha \sqrt{\gamma} T^{\mu \nu} g_{\nu \sigma}
                  { {\Gamma}^{\sigma} }_{\mu j} \\
                  \alpha \sqrt{\gamma} (T^{\mu 0} {\partial}_{\mu} \alpha -
                  \alpha T^{\mu \nu} { {\Gamma}^0}_{\mu \nu})
                 \end{array}  \right],
\end{equation}
where ${ {\Gamma}^{\alpha} }_{\mu \nu}$ are the standard 4-Christoffel 
symbols. 

We choose the ideal fluid $\Gamma$-law equation of state,
\begin{equation}
P(\rho,\epsilon) = (\Gamma - 1) \rho \epsilon
\end{equation}
to close the system of hydrodynamic equations. 

\subsection{Mesh refinement}

In order to ensure adequate spatial resolution whilst
keeping the computational resource requirements of our 
three-dimensional simulations to a minimum, we use Berger-Oliger
style \cite{Berger84} mesh-refinement with subcycling in time
as implemented by the open-source {\tt Carpet} \cite{Schnetter-etal-03b, url:carpetcode} 
driver for the {\tt Cactus} code. {\tt Carpet} provides
fixed, progressive \cite{Baiotti04b} and adaptive 
mesh refinement. In this study we
use a predefined refinement hierarchy with five levels
of refinement, arranged in a box-in-box manner centered
on the origin. The resolution factor between levels is two. 
We point out that adaptive
(or at least progressive) mesh refinement will be necessary
to track black-hole formation in detail, and has been performed
on one model in \cite{Zink2005a}.

\subsection{Mode extraction}
To evaluate and quantify the stability or instability of a given model
to non-axisymmetric perturbations, we extract
azimuthal modes $e^{im\varphi}$ by means of a Fourier analysis of the
rest-mass density on a ring of fixed coordinate radius in the equatorial
plane\footnote{These quantities are not gauge-invariant, but they provide
a useful way of characterizing the representation of the instability within 
our choice of coordinates.}.
Following Tohline et al.~\cite{Tohline85}, we compute complex weighted averages
\begin{equation}
C_{m} = \frac{1}{2\pi} \int^{2\pi}_0 \rho(\varpi,\varphi,z=0)\, e^{im\varphi} d\varphi
\end{equation}
and define normalized real mode amplitudes 
\begin{equation}
A_{m} = \frac{|C_m|}{C_0}\,\,.
\end{equation}
Here $\varpi = \sqrt{x^2 + y^2} = const.$ and is chosen to correspond
to the initial equatorial radius of maximum density in our quasi-toroidal
models, if not mentioned otherwise. The index $m$ corresponds to the number 
of azimuthal density nodes and is used to characterize the modes.

\section{Initial data}
\label{sec:id}

\subsection{Quasi-toroidal polytropes}
\label{sec:quasi_toroidal_polytropes}

We focus on general relativistic, differentially rotating 
polytropes which are \emph{quasi-toroidal}: Such a polytrope has at least
one isodensity surface which is homeomorphic to a torus. 
To construct equilibrium polytropes
of this kind, an extended version of the Stergioulas-Friedman (RNS) 
code is used \cite{Stergioulas95, Stergioulas96, Stergioulas98}, which uses
the numerical methods developed in Komatsu, Hachisu and Eriguchi
\cite{Cook94b, Komatsu89, Komatsu89b}.
The code assumes a certain gauge in a stationary, axisymmetric spacetime, such
that we can write the line element in terms of potentials $\nu, \psi, \omega$ and
$\mu$, and the Killing fields $\partial_t$ and $\partial_\phi$ as \cite{Stergioulas98}
\be
ds^2 = -e^{2\nu} dt^2 + e^{2\psi} (d\phi - \omega dt)^2 + e^{2\mu} (dr^2 + r^2 d\theta^2).
\ee

To compute an equilibrium polytrope, the central rest-mass density $\rho_c$, 
the coordinate axis ratio $r_p/r_e$ and a barotropic relation $P(\rho)$ need to be 
specified\footnote{Purely toroidal models have $r_p/r_e = 0$.}. In
addition, the equations of structure \cite{Stergioulas98} contain an additional freely
specifiable function $F(\Omega)$. We will use the common choice
\be
F(\Omega) = \tilde{A}^2 (\Omega_c - \Omega)
\ee
where $\Omega_c$ denotes the angular velocity at the star's center. This rotation law
reduces to uniform rotation in the limit $\tilde{A} \rightarrow \infty$, and to 
the constant specific angular momentum case in the limit $\tilde{A} \rightarrow 0$.
We will, however, often use the normalized parameter $A = \tilde{A} / r_e$, where
$r_e$ is the coordinate radius of the intersection of the stellar surface with the
equatorial plane $\theta = \pi/2$. For construction of a polytrope, 
the equation of state is constrained to the polytropic relation
\be
P = K \rho^{\Gamma}
\ee
with the polytropic constant $K$ and the coefficient $\Gamma$, which can also be expressed
by the polytropic index $N = (\Gamma - 1)^{-1}$. Without loss of
generality we set $K = 1$ in all cases. The equation of state of the initial model is 
completed by the energy relation $\epsilon = P/[(\Gamma-1) \rho]$.

The resulting set of equations is solved iteratively \cite{Stergioulas98}, where the initial
trial fields are a suitable solution of the TOV system. To converge to the desired model, it
may be necessary to select a number of intermediate attractors as trial fields. Some 
models are thus constructed by first obtaining a specific quasi-toroidal model, and then
moving in parameter space in the quasi-toroidal branch to the target model. In
this work, a \emph{hook model} with parameters $A_{hook} = 0.3$ and $(r_p/r_e)_{hook} =
0.15$ is generated, which then is used as initial guess to construct the target model.

If we include the polytropic coefficient $\Gamma$, we have to consider a four-dimensional parameter space 
$(\Gamma, \rho_c, A, r_p/r_e)$. We will not study the whole parameter space
here: Rather, we first use a reference model and explore sequences in $\rho_c$, $\Gamma$ and $r_p/r_e$ 
containing this model, and will subsequently concentrate on the important case $\Gamma=4/3$, 
since it approximately
represents a radiation-pressure dominated star. 

Most polytropes have been constructed with a meridional grid resolution of $n_r = 601$ radial zones
and $n_{\cos \theta} = 301$ angular zones, a maximal harmonic index $\ell_{max} = 10$ for the angular
expansion of the Green function and a solution accuracy of $10^{-7}$. Selected models have been tested 
for convergence with
resolutions up to $n_r = 2401$, $n_{\cos \theta} = 1201$, and $\ell_{max} = 20$.

To investigate the stability of the polytropes constructed with the RNS code, two kinds of perturbations
are applied: the pressure is reduced by $0.1 \%$, and a cylindrical density perturbation of the 
form 
\be
\rho(x) \rightarrow \rho(x) \Big[1 +  \frac{1}{r_e}
	\sum_{m=1}^{4} \lambda_m B f(\varpi) \sin(m \phi) \Big]
\label{eq:perturbation}
\ee
is added to the equilibrium polytrope. Here, $m \in \{1,2,3,4\}$,
$\lambda_m$ is either $0$ or $1$, $\varpi$ is the cylindrical radius and $f(\varpi)$ is a radial trial function.
Experiments have been made with $f(\varpi) = \varpi$ and $f(\varpi) = \varpi^m$, but the exact choice was found not to 
affect the results significantly. This is true quite generally, since we only require the trial function
to have some reasonable overlap with a set of quasi-normal modes. It is beyond our scope to
investigate the full spectrum of quasi-normal modes of general relativistic polytropes; therefore, we determine
stability only with respect to specific trial functions. The choice made is not completely arbitrary,  
however: a quasi-toroidal polytrope has an off-center toroidal region of maximal density, and it is 
this region which will dominate the dynamics if a fragmentation instability sets in. 
A linear perturbation without nodes in this region can be expected to be compatible (have non-zero scalar
product) with most low-frequency quasi-normal modes. The function $f(\varpi) = \varpi^m$ has the additional 
property of smoothness at the center, but, as already noted, numerical experiments have shown the
difference to be negligible in practice. An additional note on the use of language: If we find that
a perturbation with $\lambda_i = \delta_{ij}$, $j \in \{1,2,3,4\}$ leads to an instability 
with the associated number of node lines in the equatorial plane, we will denote this instability 
with the term \emph{$m=j$ mode} (and the corresponding perturbation
\emph{$m=j$ perturbation}). 
This is a simplification insofar as each $m$ is expected to represent a (discrete) infinite spectrum of
modes \cite{Schutz83b}, from which we will observe only the fastest-growing unstable member.
While we will attempt to discuss the nature of the global evolution to some extent, we will, 
for the reasons stated above, concentrate on the high-density torus, and mostly neglect the dynamics 
of its halo.

\begin{figure}
\begin{center}
\includegraphics[width=\columnwidth]{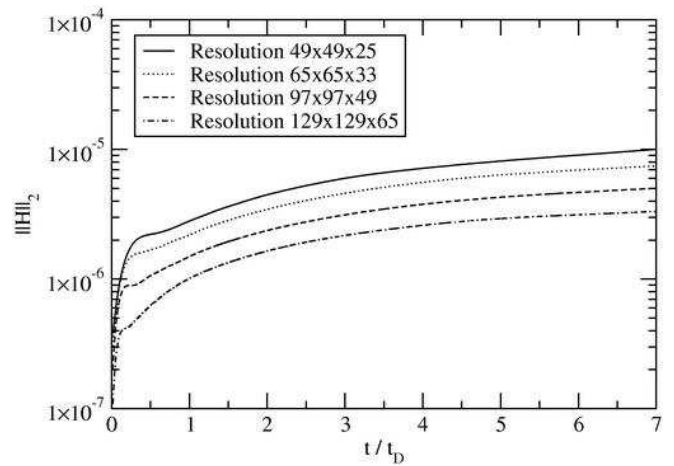}
\end{center}
\caption{Time  evolution   of  the  $L^2$  norm   of  the  Hamiltonian
constraint for  different resolutions (the numbers refer to grid point on a single
patch of our mesh refinement grid). The  time is normalized  to the
dynamical timescale $t_{dyn} = R_e \sqrt{R_e/M}$.}
\label{fig:ham_convergence} 
\end{figure}

The perturbations applied are both constraint-violating. This is no significant issue, 
since the discrete evolution of the NOK-BSSN system will introduce constraint violations even
if some minimization technique has been applied to the initial data. Fig.~\ref{fig:ham_convergence}
shows the evolution of the $L^2$ norm of the Hamiltonian constraint for
different grid resolutions, for a model perturbed with $\lambda_i = \delta_{i1}$ and 
$B = 10^{-3}$. (The time in this figure, and in the subsequent
ones, is normalized to $t_{dyn} = R_e \sqrt{R_e/M}$, where $R_e$ is the circumferential equatorial
radius, and $M$ is the ADM mass.)
Note that, for a typical perturbation amplitude of $\delta \rho/\rho \approx 10^{-3}$, the 
Hamiltonian constraint will be violated by $\delta H \approx 16 \pi \rho \cdot 10^{-3} \approx 
2.5 \cdot 10^{-7}$, which is sufficiently smaller than the violations during evolution in 
Fig.~\ref{fig:ham_convergence}. 
Also, tests have been performed where the perturbation amplitude $B$ (cf. eqn.~\ref{eq:perturbation})
is reduced by a factor of $10$, and 
found that this does not affect the growth rate of the perturbation, as expected from small
perturbations. To conveniently compare different resolutions, the amplitude is kept constant; however,
it is possible to reduce the perturbation amplitudes for resolutions significantly higher than the ones
used here (specifically, in the regime where $\delta H$ would become comparable to the constraint
violations during evolution) to obtain a system where convergence, now to the equilibrium system,
does only depend on the well-posedness of the continuum initial-boundary value problem and the stability 
of the discrete 
system\footnote{In addition, since our technique of solving the equations of hydrodynamics require us
to add an artificial atmosphere in the vacuum region, one would also need to reduce its density 
with resolution.}.

\subsection{The reference polytrope and associated sequences}

We start with a polytrope with the same central rest-mass density 
($\rho_c = 3.38 \cdot 10^{-6}$) as Saijo's series of
differentially rotating supermassive star models \cite{Saijo2004}. To obtain experience
with the influence of certain parameters on the stability properties of the relativistic
quasi-toroidal polytropes, some sequences containing this reference model have
been constructed.

\subsubsection{The reference model}
\label{sec:id_reference}

\begin{table}
\begin{center}
\begin{tabular}{|l|l|l|}
\hline
Polytropic index $\Gamma$ & $\Gamma$ & $4/3$ \\
Central rest-mass density & $\rho_c$ & $3.38 \cdot 10^{-6}$  \\  
Degree of differential rotation & $A$ & $1/3$ \\
Coordinate axis ratio & $r_p/r_e$ & $0.24$ \\ 
Density ratio & $\rho_{max}/\rho_c$ & $16.71$ \\
ADM mass & $M$ & $7.003$ \\ 
Rest mass & $M_0$ & $7.052$ \\ 
Equatorial inverse compactness & $R_e/M$ & $11.71$ \\  
Angular momentum & $J$  & $52.20$  \\ 
Normalized angular momentum & $J/M^2$  & $1.064$  \\  
Kinetic over binding energy & $T/|W|$ &  $0.227$  \\ 
(See caption) & $\Omega_e/\Omega_K$ & $0.467$ \\
\hline
\end{tabular}
\end{center}
\caption{Parameters and integral quantities of the reference quasi-toroidal polytrope
\cite{Zink2005a}. The quantities $\Gamma$, $\rho_c$, $A$ and $r_p/r_e$ are parameters.
The quantity $\Omega_e$ is the angular velocity on the equator, while $\Omega_K$ is
the associated Keplerian velocity of the same model. Therefore, the mass-shedding sequence is located
at $\Omega_e/\Omega_K = 1$.}
\label{tab:reference_polytrope}
\end{table}

\begin{figure}
\begin{center}
\includegraphics[width=\columnwidth]{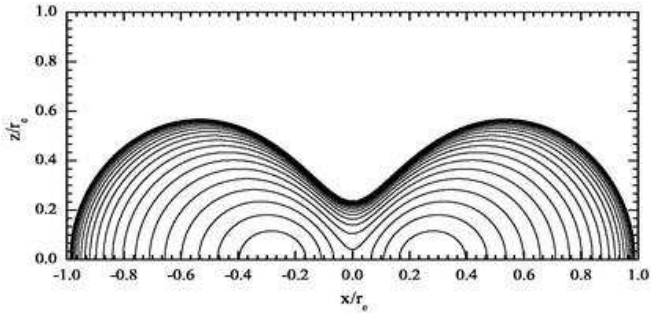}
\end{center}
\caption{Decadic logarithm of the density in a meridional plane of the model constructed with
the parameters in Table~\ref{tab:reference_polytrope}. The model is of quasi-toroidal
nature. Note that this contour plot is included to 
demonstrate the qualitative properties of the initial model. Therefore, no detailed scale
is provided.}
\label{fig:contours_reference} 
\end{figure}

The reference model is identical to the model used in \cite{Zink2005a}; its parameters and
integral quantities are shown in Table~\ref{tab:reference_polytrope}. 
Fig.~\ref{fig:contours_reference} is a graph of the density in the 
meridional plane for this model. This model has been found to be unstable to perturbations
with $m = 1$ and $m = 2$ (see below, and \cite{Zink2005a}), which lead to 
fragmentation. In the case of
$m = 1$, black hole formation has been demonstrated by locating an apparent
horizon centered on the fragment \cite{Zink2005a}.

\subsubsection{The sequence of axis ratios}
\label{sec:id_ratio}

\begin{table*}
\begin{center}
\begin{tabular}{|l|c|c|c|c|c|c|c|c|c|c|c|c|}
\hline
Model              & $\rho_c$             & $r_p/r_e$ & $\rho_{max}/\rho_c$ & $M$     & $M_0$   & $R_e/M$ & $J$     & $J/M^2$ & $T/|W|$ & $\Omega_e/\Omega_K$ \\
\hline
\emph{R0.20} & $3.38 \cdot 10^{-6}$  & $0.20$    & $38.12$ & $6.181$ & $6.200$ & $9.660$ & $38.59$ & $1.010$ & $0.235$ & $0.487$            \\
\emph{R0.22} & $3.38 \cdot 10^{-6}$ & $0.22$    & $25.69$ & $6.662$ & $6.710$ & $10.41$ & $45.46$ & $1.024$ & $0.228$ & $0.475$            \\
\emph{R0.24} & $3.38 \cdot 10^{-6}$ & $0.24$    & $16.76$ & $6.989$ & $7.037$ & $11.71$ & $52.00$ & $1.065$ & $0.227$ & $0.467$            \\
\emph{R0.26} & $3.38 \cdot 10^{-6}$ & $0.26$    & $11.07$ & $7.334$ & $7.391$ & $13.10$ & $58.99$ & $1.097$ & $0.223$ & $0.460$            \\
\emph{R0.28}  & $3.38 \cdot 10^{-6}$ & $0.28$    & $7.312$ & $7.585$ & $7.646$ & $14.83$ & $65.19$ & $1.133$ & $0.219$ & $0.455$            \\
\emph{R0.30} & $3.38 \cdot 10^{-6}$ & $0.30$    & $4.733$ & $7.764$ & $7.825$ & $17.13$ & $70.82$ & $1.175$ & $0.213$ & $0.452$            \\
\emph{R0.32} & $3.38 \cdot 10^{-6}$ & $0.32$    & $2.934$ & $7.847$ & $7.905$ & $20.46$ & $75.48$ & $1.226$ & $0.207$ & $0.452$            \\
\emph{R0.34} & $3.38 \cdot 10^{-6}$ & $0.34$    & $1.539$ & $7.755$ & $7.803$ & $27.42$ & $78.72$ & $1.309$ & $0.196$ & $0.463$            \\
\hline
\end{tabular}
\end{center}
\caption{Parameters and integral quantities of the \emph{R} sequence of axis ratios,
which contains the reference model for $r_p/r_e = 0.24$. Each member of the sequence is denoted
by the term \emph{R<$r_p/r_e$>}. All models have $\Gamma = 4/3$ and $A = 1/3$.}
\label{tab:sequence_ratio}
\end{table*}

\begin{figure}
\begin{center}
\includegraphics[width=\columnwidth]{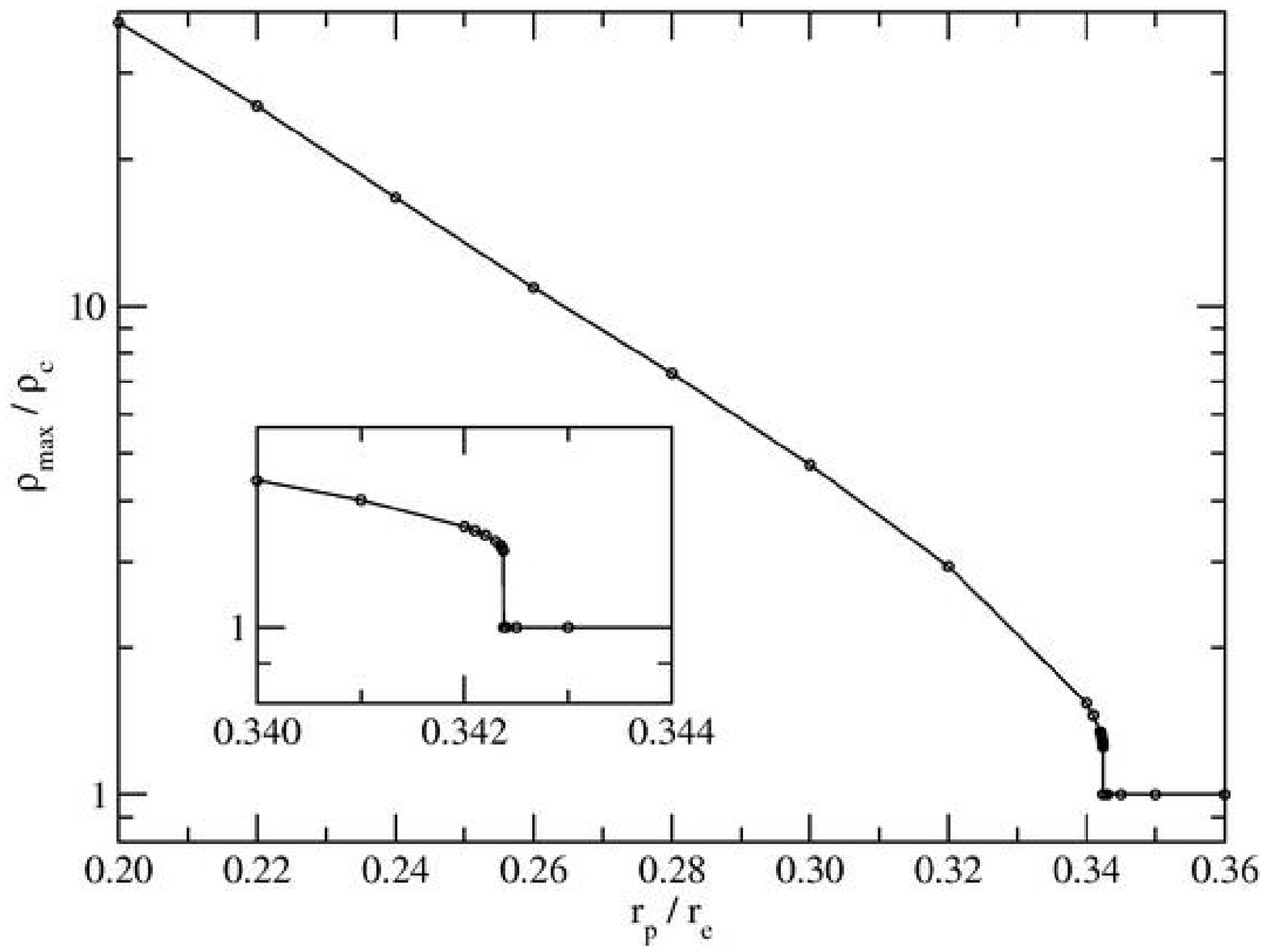}
\includegraphics[width=\columnwidth]{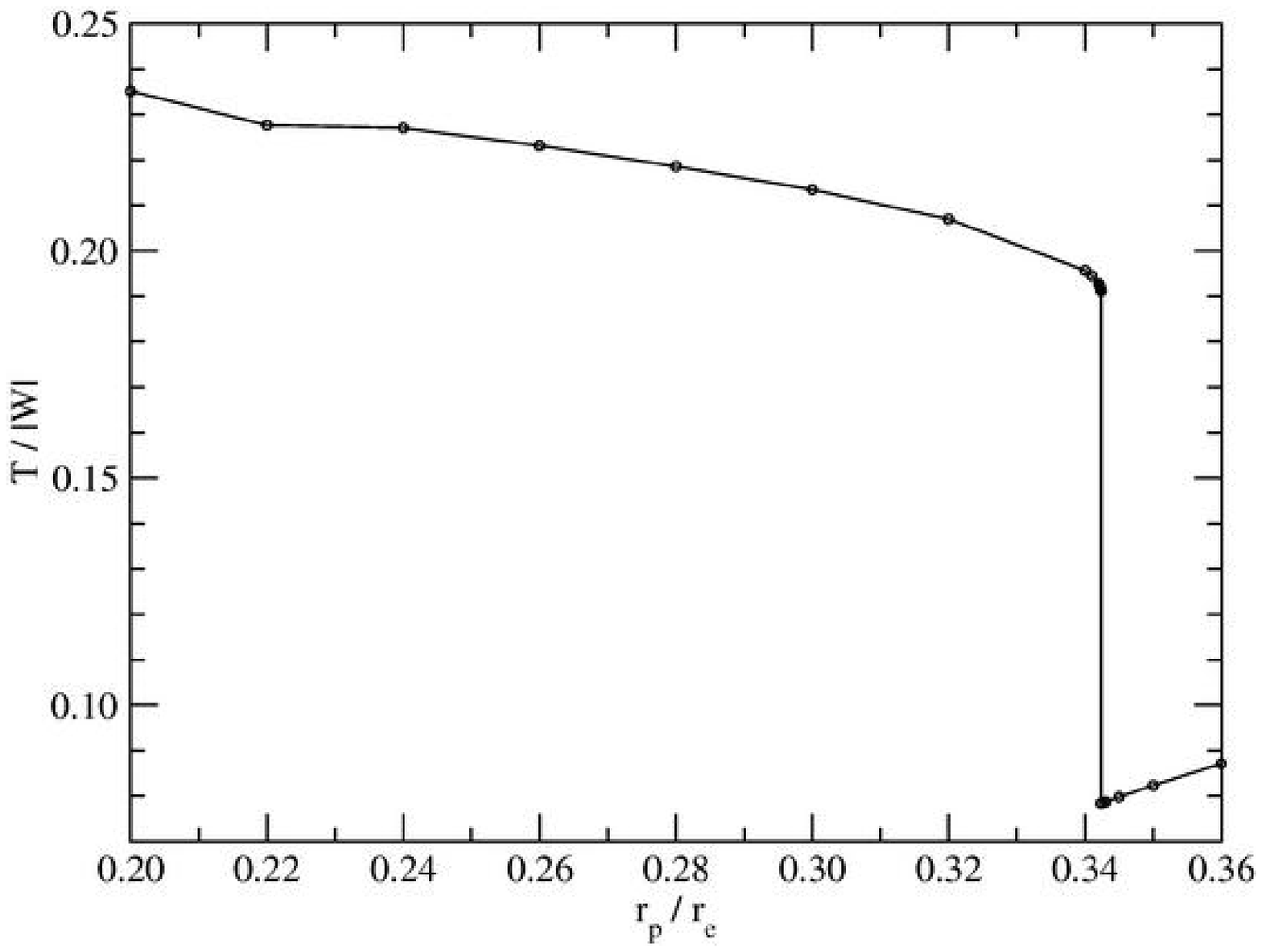}
\end{center}
\caption{$\rho_{max}/\rho_c$ (top) and $T/|W|$ (bottom) for the \emph{R}
  sequence. Beyond an axis ratio of $r_p/r_e \approx 0.3423$, the initial data solver converges to spheroidal models.}
\label{fig:sequence_ratio_figures} 
\end{figure}

A number of sequences containing this model have been constructed to study the
stability properties when varying typical parameters. The first of these
sequences is parameterized by the coordinate axis ratio $r_p/r_e$. Its members
are denoted by \emph{R<$r_p/r_e$>}, and their properties can be
found in Table~\ref{tab:sequence_ratio}. It is apparent that, with
increasing axis ratios, the quantity $T/|W|$ and the ratio of maximal to 
central rest-mass density $\rho_{max}/\rho_c$ both decrease monotonically. None of the sequence
members is close to the mass-shedding limit. Below $r_p/r_e = 0.20$, no models could
be constructed due to failure of convergence. We will also not consider models with
larger $r_p/r_e$. The reason for this is indicated in Fig.~\ref{fig:sequence_ratio_figures}:
Beyond an axis ratio of $r_p/r_e \approx 0.3423$, 
quasi-toroidal models could not be constructed as the numerical code fails to
converge.

\subsubsection{The sequence of stiffnesses}
\label{sec:id_gamma}

\begin{table*}
\begin{tabular}{|l|c|c|c|c|c|c|c|c|c|c|c|c|}
\hline
Model               & $\Gamma$ & $\rho_c$             & $\rho_{max}/\rho_c$ & $M$     & $M_0$   & $R_e/M$ & $J$     & $J/M^2$ & $T/|W|$ & $\Omega_e/\Omega_K$ \\
\hline
\emph{G1.333} & $4/3$    & $3.38 \cdot 10^{-6}$ & $16.76$             & $6.989$ & $7.037$ & $11.71$ & $52.00$ & $1.065$ & $0.227$ & $0.467$            \\
\emph{G1.4}   & $1.4$    & $1.32 \cdot 10^{-5}$ & $16.44$             & $2.805$ & $2.856$ & $11.69$ & $8.337$ & $1.059$ & $0.211$ & $0.424$            \\
\emph{G1.45}  & $1.45$   & $3.7 \cdot 10^{-5}$  & $13.83$             & $1.624$ & $1.662$ & $11.70$ & $2.767$ & $1.049$ & $0.202$ & $0.404$            \\
\emph{G1.5}   & $1.5$    & $9.2 \cdot 10^{-5}$  & $11.53$             & $1.038$ & $1.066$ & $11.65$ & $1.115$ & $1.035$ & $0.194$ & $0.390$            \\
\emph{G1.6}   & $1.6$    & $3.75 \cdot 10^{-4}$ & $8.349$             & $0.5197$ & $0.5363$ & $11.80$ & $0.2758$ & $1.021$ & $0.183$ & $0.370$         \\
\emph{G2.0}   & $2.0$    & $8.2 \cdot 10^{-3}$   & $3.880$             & $0.1270$ & $0.1323$ & $11.69$ & $0.0155$ & $0.9633$ & $0.159$ & $0.331$ \\
\hline
\end{tabular}
\caption{Parameters and integral quantities of the \emph{G} sequence of polytropic coefficient $\Gamma$,
which contains the reference model for $\Gamma = 4/3$. Each member of the sequence is denoted
by the term \emph{G<$\Gamma$>}. The central density is adjusted to yield approximately the
same inverse equatorial compactness $R_e/M$ as in the reference model. The models \emph{G1.7} to
\emph{G1.9} have not been constructed, since the models \emph{G1.6}
and \emph{G2.0} were found to be stable. All models have $A = 1/3$ and $r_p/r_e = 0.24$.}
\label{tab:sequence_gamma}
\end{table*}

This sequence is a variation of the parameter $\Gamma$ in the polytropic relation $P=K\rho^\Gamma$,
which also determines the stiffness of the ideal fluid equation of state $P=(\Gamma-1)\rho\epsilon$ used
for evolution. To obtain a sequence of comparable compactness, we adjust the central density $\rho_c$ to
yield approximately the same $R_e/M$. The parameters and integral quantities are shown in 
Table~\ref{tab:sequence_gamma}. 
Along the sequence of increasing $\Gamma$, the value $T/|W|$ decreases from $0.227$ to $0.159$. Therefore,
it would also be interesting to consider a sequence of models with varying $\Gamma$, but constant
$T/|W|$, by adjusting the axis ratio $r_p/r_e$ accordingly. Unfortunately, 
such a sequence could not be obtained, since the initial data solver did not converge to models with the required (low) 
axis ratios.

\subsubsection{The sequence of compactnesses}
\label{sec:id_comp}

\begin{table*}
\begin{center}
\begin{tabular}{|l|c|c|c|c|c|c|c|c|c|c|c|c|}
\hline
Model              & $\rho_c$             & $r_p/r_e$ & $\rho_{max}/\rho_c$ & $M$     & $M_0$   & $R_e/M$ & $J$     & $J/M^2$ & $T/|W|$ & $\Omega_e/\Omega_K$ \\
\hline
\emph{C1}     & $3.38 \cdot 10^{-6}$ & $0.24$    & $16.76$             & $6.989$ & $7.037$ & $11.71$ & $52.00$ & $1.065$ & $0.227$ & $0.467$            \\
\emph{C2}     & $1 \cdot 10^{-7}$    & $0.24$    & $31.06$             & $10.65$ & $10.74$ & $22.72$ & $167.0$ & $1.474$ & $0.225$ & $0.434$            \\
\emph{C4}     & $7.5 \cdot 10^{-9}$  & $0.24$    & $37.01$             & $12.54$ & $12.60$ & $45.65$ & $326.0$ & $2.073$ & $0.225$ & $0.423$            \\
\emph{C8}     & $8 \cdot 10^{-10}$   & $0.24$    & $39.64$             & $13.47$ & $13.51$ & $89.91$ & $525.6$ & $2.897$ & $0.225$ & $0.419$            \\
\hline
\end{tabular}
\end{center}
\caption{Parameters and integral quantities of the \emph{C} sequence of compactnesses,
which contains the reference model for $\rho_c = 3.38 \cdot 10^{-6}$. Each member 
of the sequence is denoted
by the term \emph{C<$a$>}, where $a$ denotes the approximate ratio of $R_e/M$ to 
$(R_e/M)_{ref}$ of the
reference model \emph{C1}. The sequence is obtained by varying the central rest-mass density.
All models have $\Gamma = 4/3$ and $A = 1/3$.}
\label{tab:sequence_comp}
\end{table*}

The next sequence is a variation of the central rest-mass density $\rho_c$ while leaving
all other parameters fixed. The resulting sequence, as is apparent from Table~\ref{tab:sequence_comp},
is also a sequence of \emph{inverse equatorial compactnesses} $R_e/M$. The members have been selected to 
represent models which are half to one eighth as compact as the reference polytrope. 
The sequence shows that $T/|W|$ is only slightly
affected by the choice of $\rho_c$, but $R_e/M$ and $J/M^2$ change significantly. 

\subsection{Quasi-toroidal and spheroidal models of constant central rest-mass density}
\label{sec:id_plane}

In addition to sequences containing the reference model, we explore a more extended part of the parameter 
space of models. We use $\Gamma = 4/3$ and $\rho_c = 10^{-7}$ to define a surface in the
parameter space spanned by the axis ratio $r_p/r_e$ and the degree of differential
rotation $A$. While an ideal fluid with $\Gamma=4/3$ is an approximation for the material
properties of radiation-pressure dominated stars, the restriction to $\rho_c = 10^{-7}$ 
is arbitrary. However, as discussed in Section~\ref{sec:qts_results}, the non-axisymmetric 
stability of the quasi-toroidal models is probably less sensitive to $\rho_c$ than to 
$r_p/r_e$ or $A$. The restriction to a plane is necessary since three-dimensional simulations
of quasi-toroidal relativistic stars are still expensive; however, selected models will 
also be studied with different 
$\rho_c$ in Section~\ref{sec:qts_results}.

\begin{figure*}
\begin{tabular}{cc}
\includegraphics[width=\columnwidth]{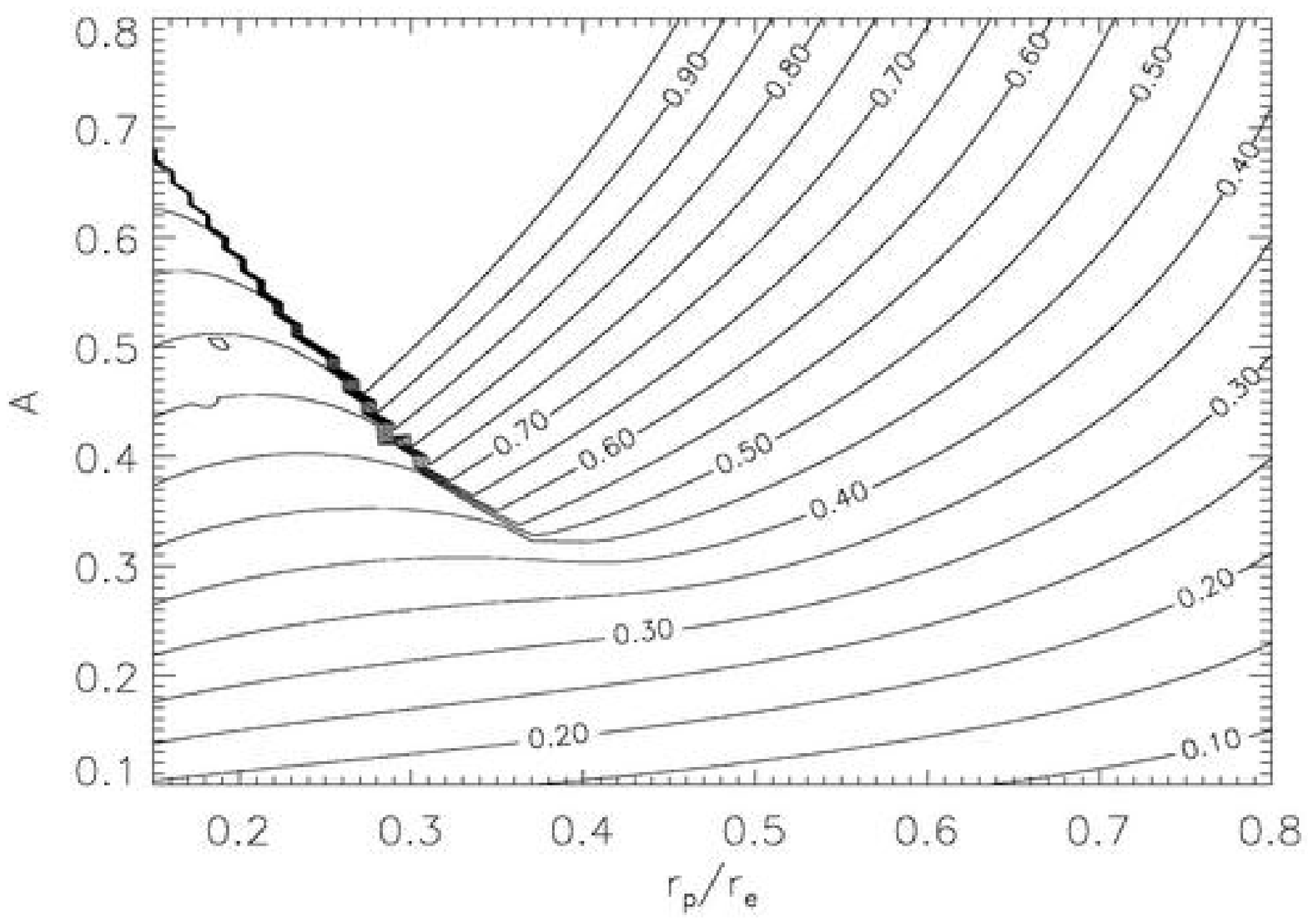} &
\includegraphics[width=\columnwidth]{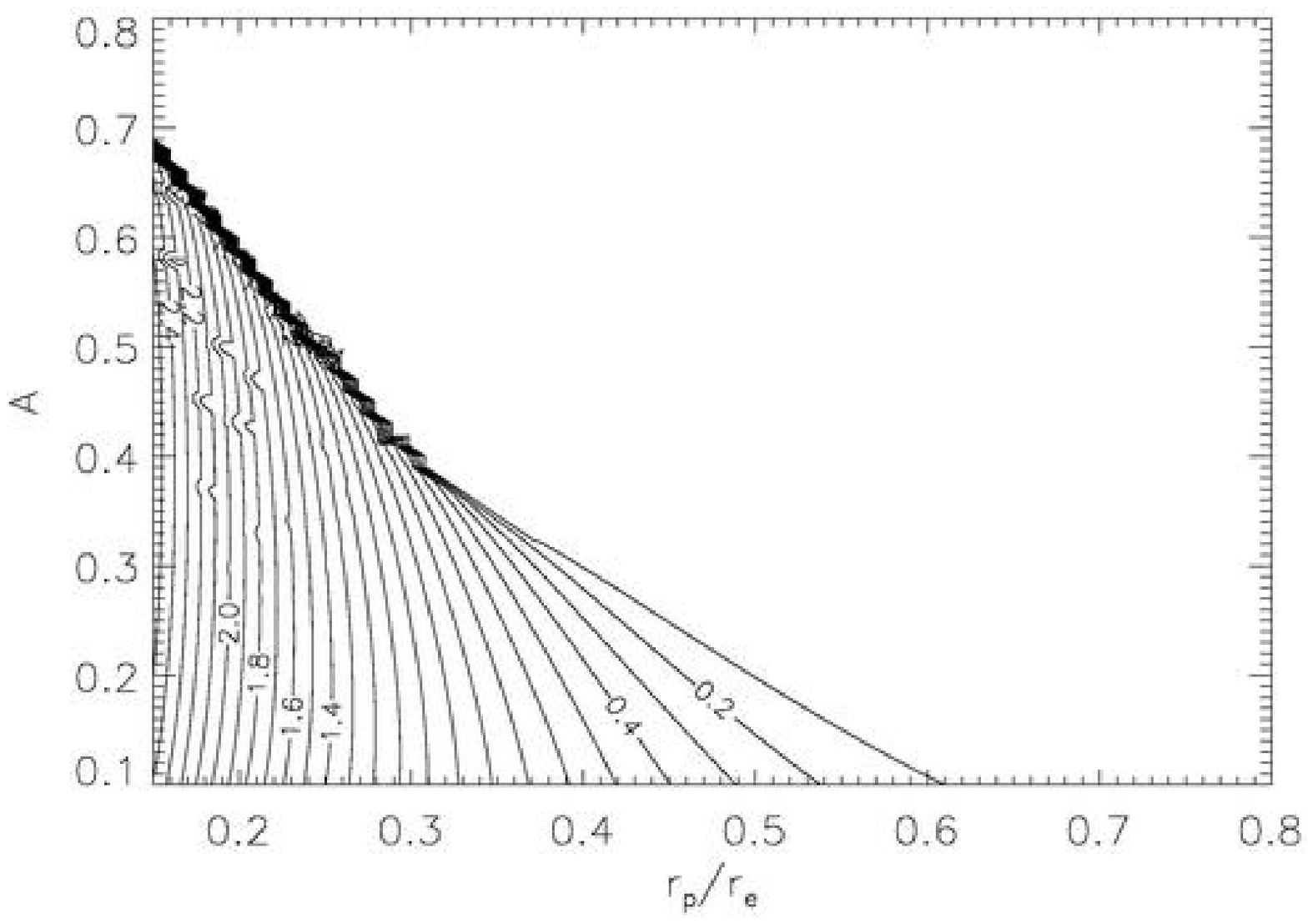} \\
\includegraphics[width=\columnwidth]{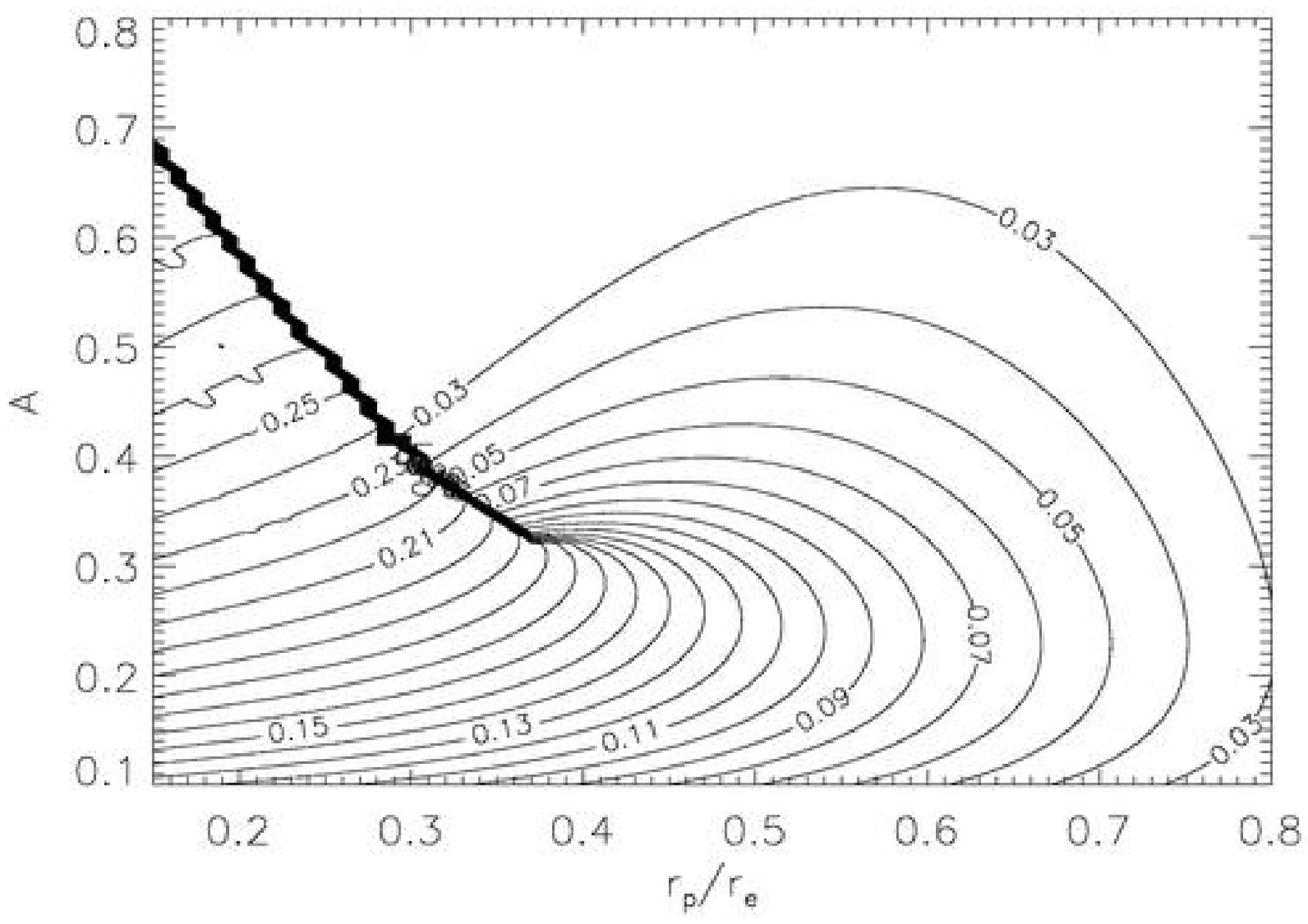} & 
\includegraphics[width=\columnwidth]{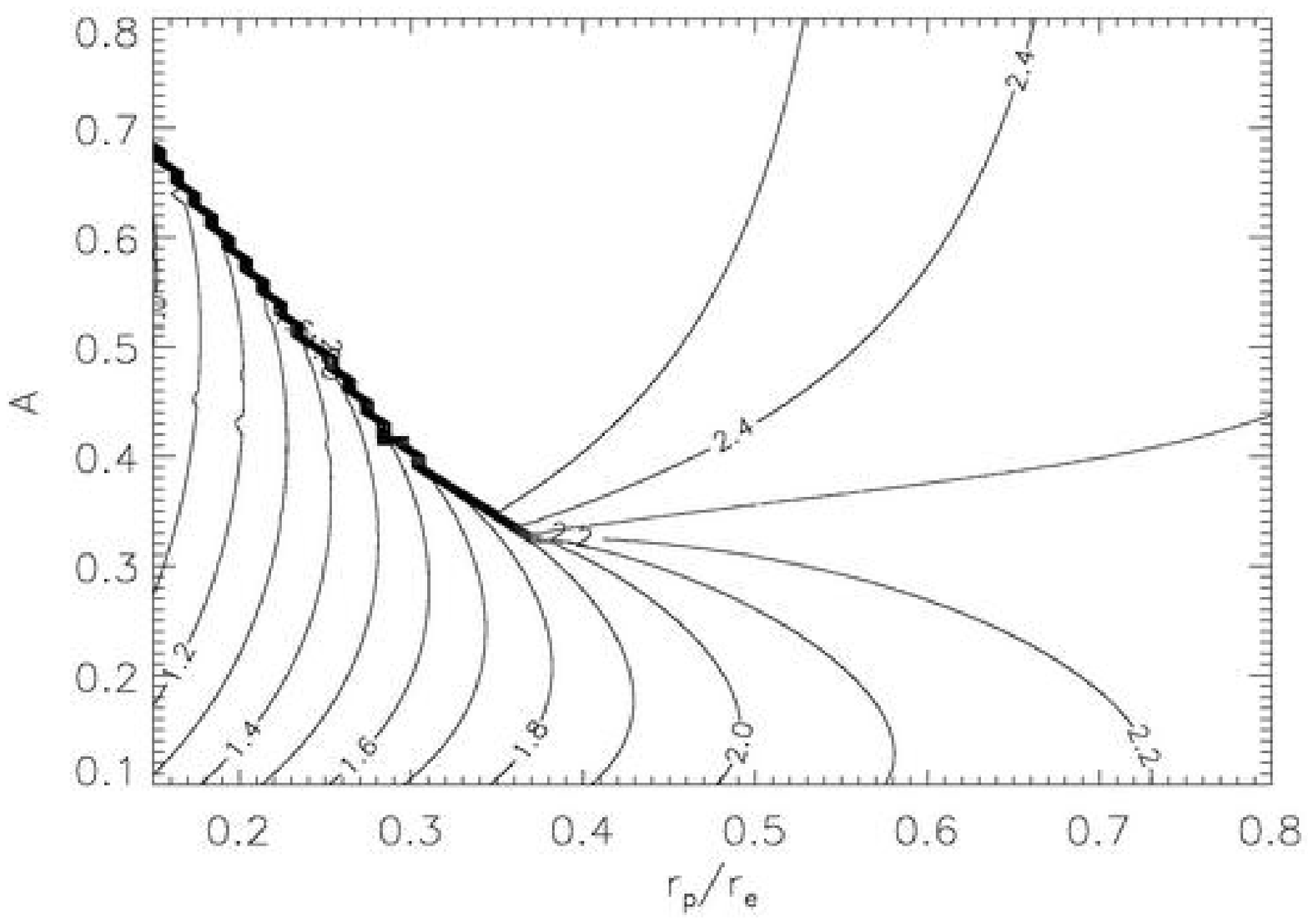} \\
\includegraphics[width=\columnwidth]{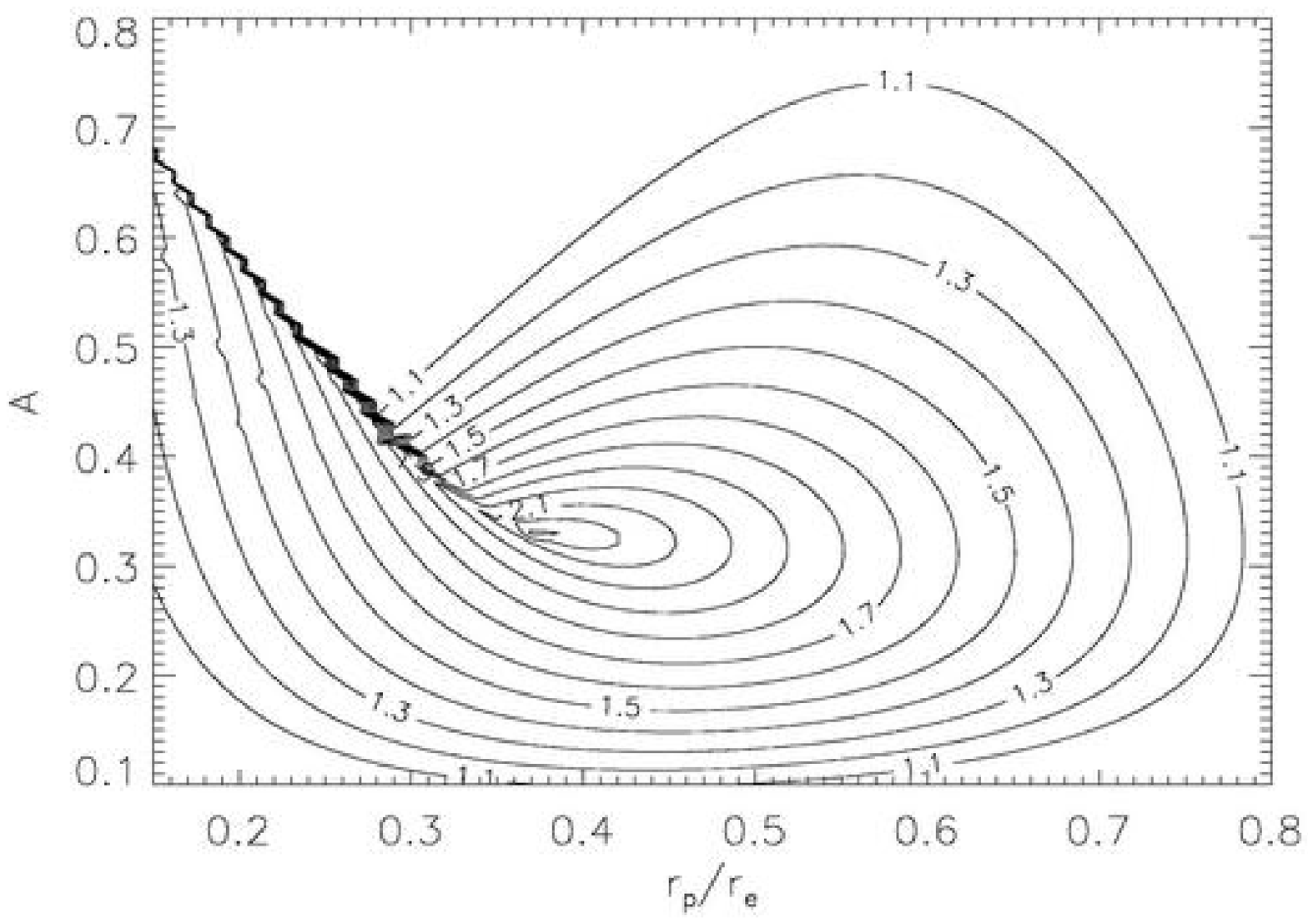} &
\includegraphics[width=\columnwidth]{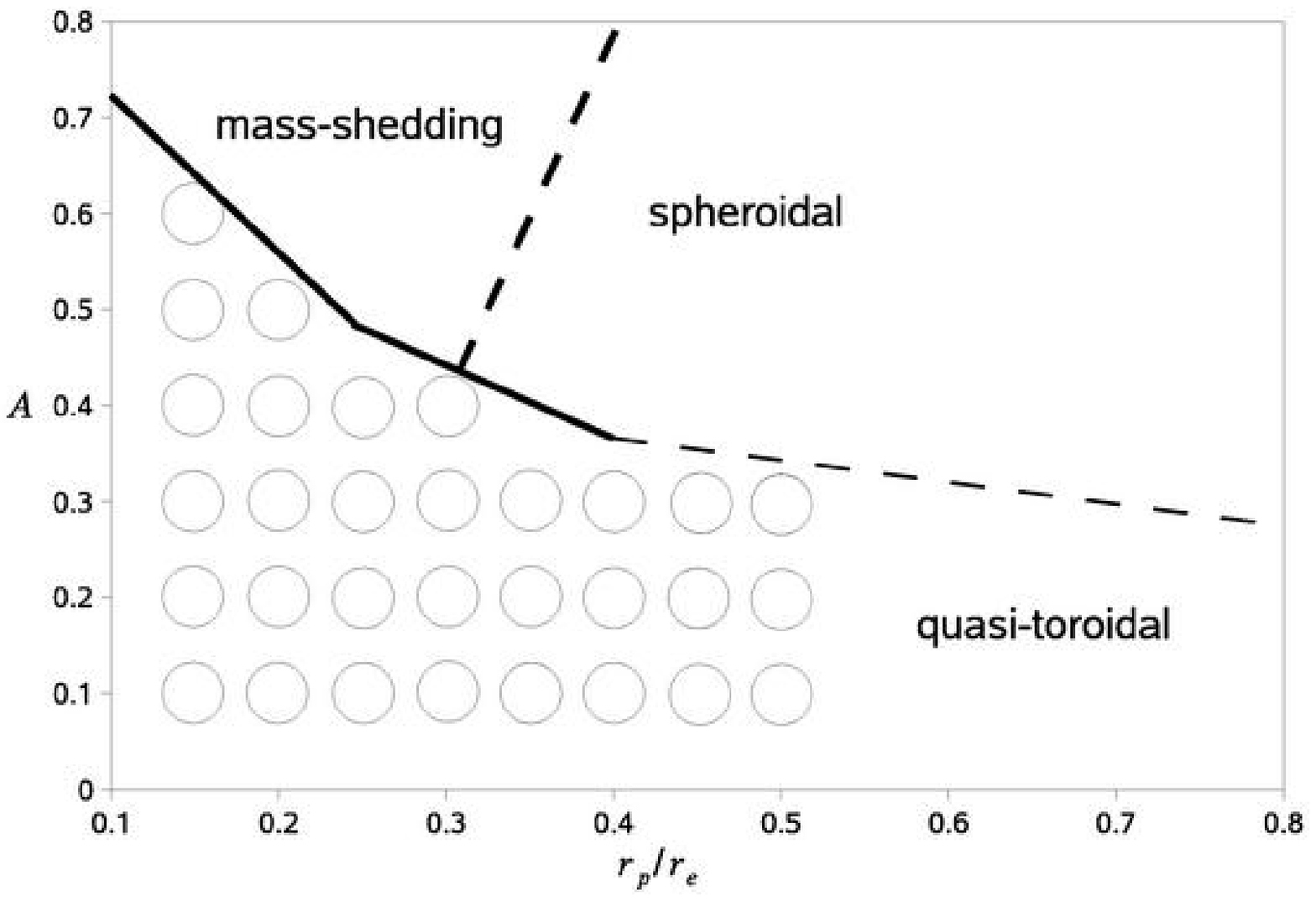}
\end{tabular}
\caption{Polytropic models constructed with the RNS code. The plots show the
parameter plane spanned by the axis ratio $r_p/r_e$ and the differential
rotation parameter $A$, constrained by $\rho_c = 10^{-7}$ and $\Gamma = 4/3$, and resolved by 
construction of $6400$ ($80 \times 80$) models. 
\emph{Top left:} contours of the equatorial stellar 
over Keplerian angular velocity $\Omega_e/\Omega_K$ (the mass shedding
limit is the isosurface $\Omega_e/\Omega_K = 1$). \emph{Top right:} 
decadic logarithm of the ratio of maximal over central density
$log_{10}(\rho_{max}/\rho_c)$. \emph{Middle left:} rotation parameter
$T/|W|$. \emph{Middle right:} decadic logarithm of the
inverse equatorial compactness $R_e/M$. \emph{Bottom left:} normalized
angular momentum $J/M^2$. \emph{Bottom right:} Selection of evolved initial
models. The thick continuous line marks the jump between quasi-toroidal and spheroidal models 
apparent in 
the top left to bottom left plots, the thick dashed line is the mass-shedding limit,
and the thin dashed line indicates an approximate division between spheroidal
and quasi-toroidal models. The models selected for numerical evolution are marked
by circles.}
\label{fig:models_plane} 
\end{figure*}

\begin{table*}
\begin{center}
\begin{tabular}{|l|c|c|c|c|c|c|c|c|c|c|c|c|}
\hline
Model            & $\rho_c$  & $A$    & $r_p/r_e$ & $\rho_{max}/\rho_c$ & $M$     & $M_0$   & $R_e/M$ & $J$     & $J/M^2$ & $T/|W|$  & $\Omega_e/\Omega_K$ \\
\hline
\emph{A0.1R0.15} & $10^{-7}$ & $0.1$  & $0.15$    & $246.8$             & $4.896$ & $4.893$ & $20.82$ & $18.62$ & $0.777$ & $0.124$  & $0.195$             \\
\emph{A0.1R0.50} & $10^{-7}$ & $0.1$  & $0.5$     & $1.881$             & $5.387$ & $5.390$ & $106.2$ & $31.69$ & $1.092$ & $0.0706$ & $0.126$             \\ 
\emph{A0.3R0.15} & $10^{-7}$ & $0.3$  & $0.15$    & $356.9$             & $7.964$ & $8.034$ & $12.04$ & $70.66$ & $1.114$ & $0.228$  & $0.434$             \\
\emph{A0.3R0.50} & $10^{-7}$ & $0.3$  & $0.5$     & $1.00005$           & $6.291$ & $6.300$ & $151.4$ & $77.10$ & $1.948$ & $0.108$  & $0.360$             \\
\emph{A0.6R0.15} & $10^{-7}$ & $0.6$  & $0.15$    & $541.2$             & $21.29$ & $21.39$ & $61.62$ & $1360$  & $3.000$ & $0.276$  & $0.650$             \\
\hline
\end{tabular}
\end{center}
\caption{Parameters and integral quantities of selected quasi-toroidal models in the parameter space plane
defined by $\rho_c = 10^{-7}$ and $\Gamma = 4/3$. The models are labelled by \emph{A<$A$>R<$r_p/r_e$>}.
All models have $\Gamma = 4/3$.}
\label{tab:selected_models}
\end{table*}

The general properties of the polytropes obtained with the RNS code is shown for
a central rest-mass density $\rho_c = 10^{-7}$ in Fig.~\ref{fig:models_plane}.\footnote{We have
also generated the same set of plots for the plane $\rho_c = 10^{-10}$, and find the
dimensionless quantities $\Omega_e/\Omega_K$, $\rho_{max}/\rho_c$, and $T/|W|$ to be 
quite similar.}
The top left plot shows the  function $\Omega_e/\Omega_K$, where $\Omega_e$ is 
the equatorial stellar angular velocity,
and $\Omega_K$ is the corresponding Keplerian angular velocity.
The jump indicated in the equilibrium model surface has already been found in 
the \emph{R} sequence, see Fig.~\ref{fig:sequence_ratio_figures}.

The topological nature of the polytropes is shown in the top right panel of Fig.~\ref{fig:models_plane}, 
which plots the ratio of maximal to central rest-mass density $\rho_{max}/\rho_c$. This value measures the 
degree of toroidal deformation of the model, with the limiting cases $\rho_{max}/\rho_c = 1$ 
(purely spheroidal polytrope) and $\rho_{max}/\rho_c = \infty$ (purely toroidal polytrope).
Since we are interested in the properties of quasi-toroidal models, we will concentrate
our study on the part of this plot covered by contour lines. 

Judging from the study of Newtonian polytropes, one would expect that the function $T/|W|$
is related to non-axisymmetric stability. For the sequence of Maclaurin spheroids, the dynamically
unstable subset can be described by the simple inequality $T/|W| \ge (T/|W|)_{dyn}$ 
\cite{Chandrasekhar69c}, suggesting to use $T/|W|$ to parameterize the sequence. While the situation is 
clearly more complicated with relativistic, differentially rotating polytropic models, 
the middle left plot in Fig.~\ref{fig:models_plane} suggests that
the quasi-toroidal models with small axis ratio $r_p/r_e = 0.15$ are more likely to be unstable to 
non-axisymmetric perturbations. We will study this in Section~\ref{sec:qts_results}. 

The bottom right panel in Fig.~\ref{fig:models_plane} is an illustration of
the initial model parameter space. The polytropes which have been evolved numerically 
are marked by circles. The equilibrium parameters and 
associated quantities of a selected set of these polytropes are listed in Table~\ref{tab:selected_models}.

\subsection{A sequence of central rest-mass densities containing the model \emph{A0.2R0.40}}
\label{sec:id_marginal}

\begin{table*}
\begin{center}
\begin{tabular}{|l|c|c|c|c|c|c|c|c|c|c|c|c|}
\hline
Model          & $\Gamma$ & $\rho_c$  & $A$   & $r_p/r_e$ & $\rho_{max}/\rho_c$ & $M$     & $M_0$   & $R_e/M$ & $J$     & $J/M^2$ & $T/|W|$ & $\Omega_e/\Omega_K$ \\
\hline
\emph{L1}      & $4/3$    & $10^{-4}$ & $0.2$ & $0.354$   & $1.874$             & $3.621$ & $3.578$ & $14.85$ & $9.952$ & $0.760$ & $0.144$ & $0.301$             \\
\emph{L2}      & $4/3$    & $10^{-5}$ & $0.2$ & $0.378$   & $2.434$             & $5.173$ & $5.177$ & $20.67$ & $24.06$ & $0.900$ & $0.144$ & $0.280$             \\
\emph{L3}      & $4/3$    & $10^{-6}$ & $0.2$ & $0.392$   & $2.644$             & $6.505$ & $6.524$ & $35.69$ & $49.51$ & $1.170$ & $0.144$ & $0.269$             \\
\emph{L4}      & $4/3$    & $10^{-7}$ & $0.2$ & $0.4$     & $2.689$             & $7.348$ & $7.362$ & $69.41$ & $87.33$ & $1.617$ & $0.144$ & $0.264$             \\
\hline
\end{tabular}
\end{center}
\caption{Parameters and integral quantities of the \emph{L} sequence. This sequence is constructed
by starting from the model \emph{A0.2R0.40} (cf. Section~\ref{sec:id_plane}), which is identical
to the model \emph{L4}, and keeping $T/|W|$ fixed while 
reducing the central density.}
\label{tab:sequence_marginal}
\end{table*}

The \emph{L} sequence (see Table~\ref{tab:sequence_marginal}) is a variation of the \emph{C}
sequence in Section~\ref{sec:id_comp}. It starts from the model \emph{A0.2R0.40} instead of
the reference model. In contrast to the \emph{C} sequence, we do not keep the axis ratio $r_p/r_e$
fixed while varying the central rest-mass density, but rather the quantity $T/|W|$. This sequence
is constructed to study the influence of the compactness on a model near the boundary to the region
denoted by ``I'' in Fig.~\ref{fig:model_stability} (see also Section~\ref{sec:limit_sequence}).

\section{Results}
\label{sec:qts_results}

The models constructed in Section~\ref{sec:id} have been evolved numerically to study their stability
properties. We will start with discussing the reference model, and show that it is unstable
to a non-axisymmetric perturbation which leads to black hole formation. Then, the sequences of axis ratios,
compactness and stiffness from Section~\ref{sec:id} are studied. Finally, the parameter plane
constrained by $\Gamma = 4/3$ and $\rho_c = 10^{-7}$ is sampled, and the coordinate location of the 
corotation point on a sequence in this plane is investigated.

\subsection{Evolution of the reference model}

The main results from evolving the reference polytrope defined in Section~\ref{sec:id_ratio}
have already been discussed in \cite{Zink2005a}. When subject to a perturbation of the form
given in eqn.~\ref{eq:perturbation}, the torus transforms into one ($m = 1$) or two ($m = 2$)
fragments. In the case of the $m = 1$ perturbation, it has been shown that the fragment
is partially covered by an apparent horizon, indicating black hole formation.

In this section, we will take a more detailed look at this model, and discuss some technical
issues relevant to the parameter space study below.

\subsubsection{Development of the instability}
\label{sec:reference_general}

\begin{figure*}
\begin{tabular}{cc}
\includegraphics[width=\columnwidth]{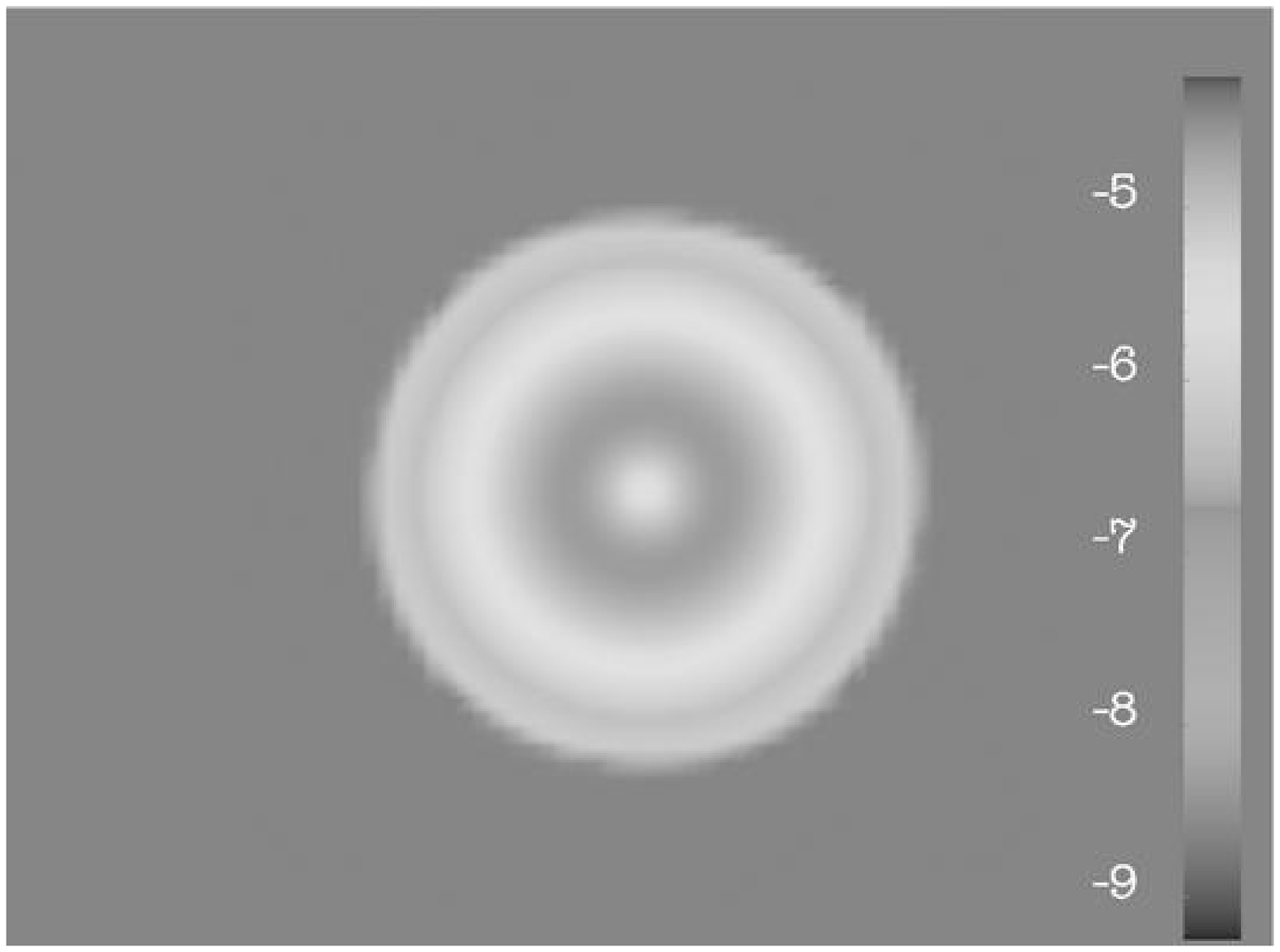} &
\includegraphics[width=\columnwidth]{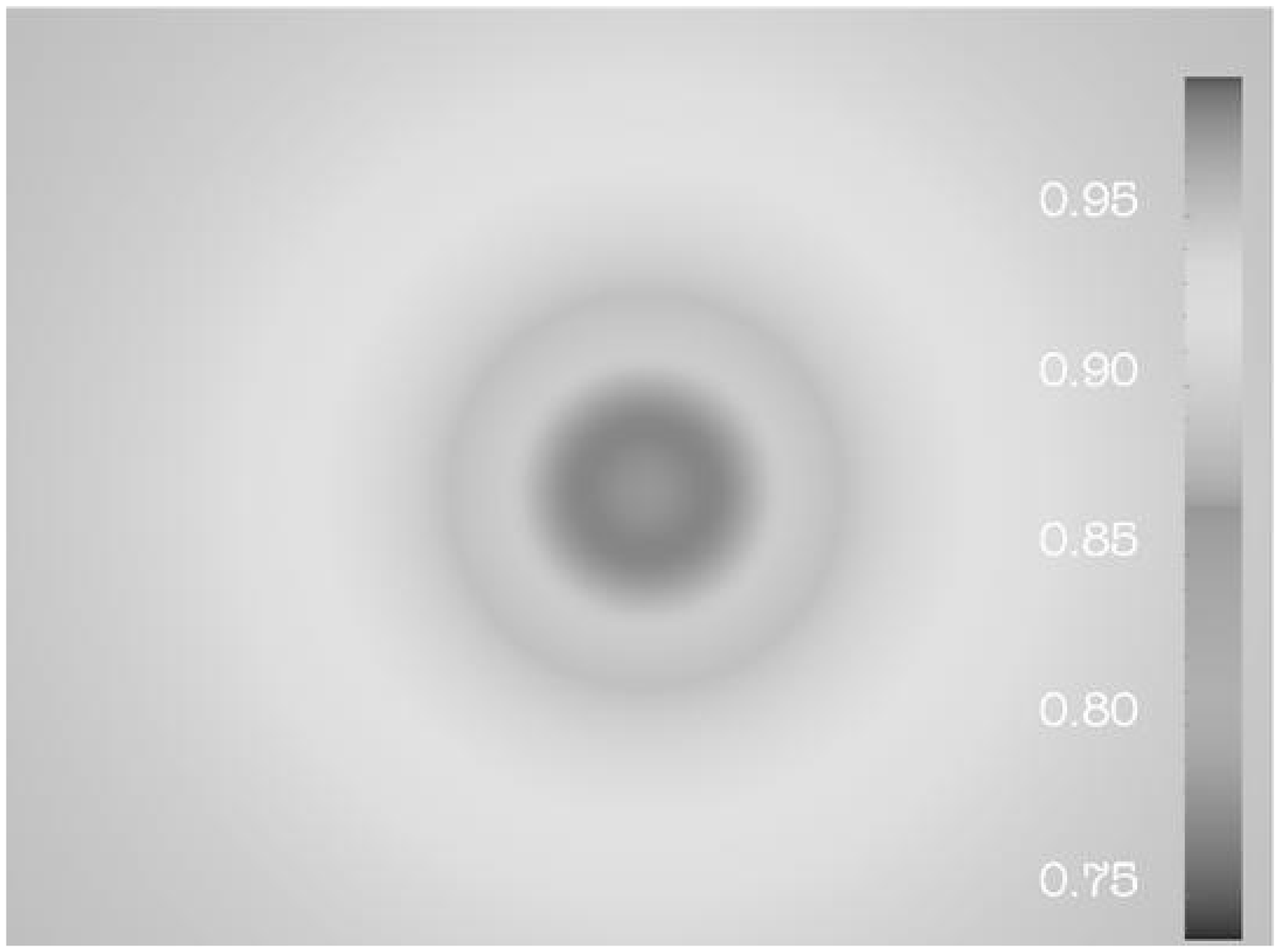} \\
\includegraphics[width=\columnwidth]{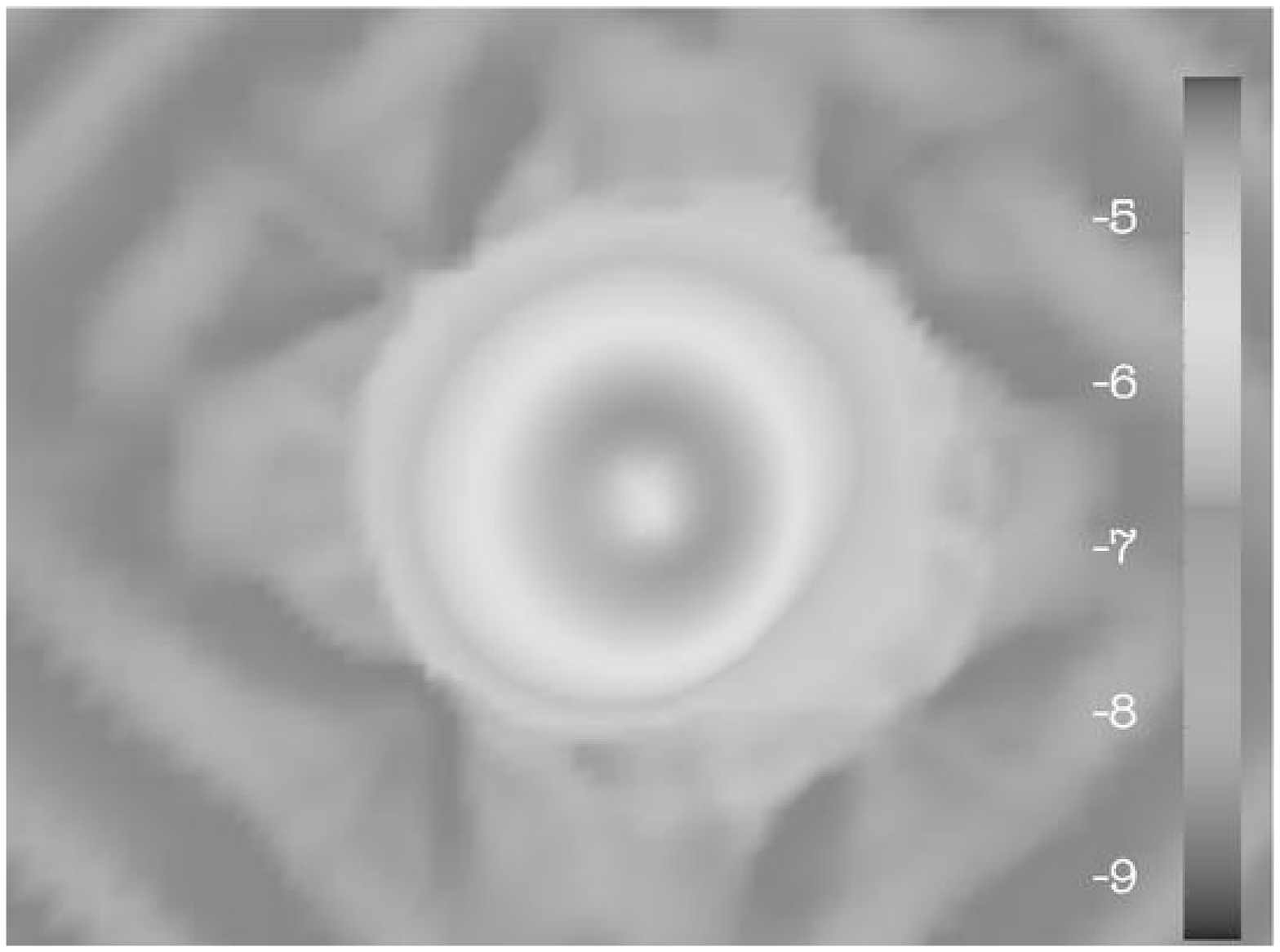} &
\includegraphics[width=\columnwidth]{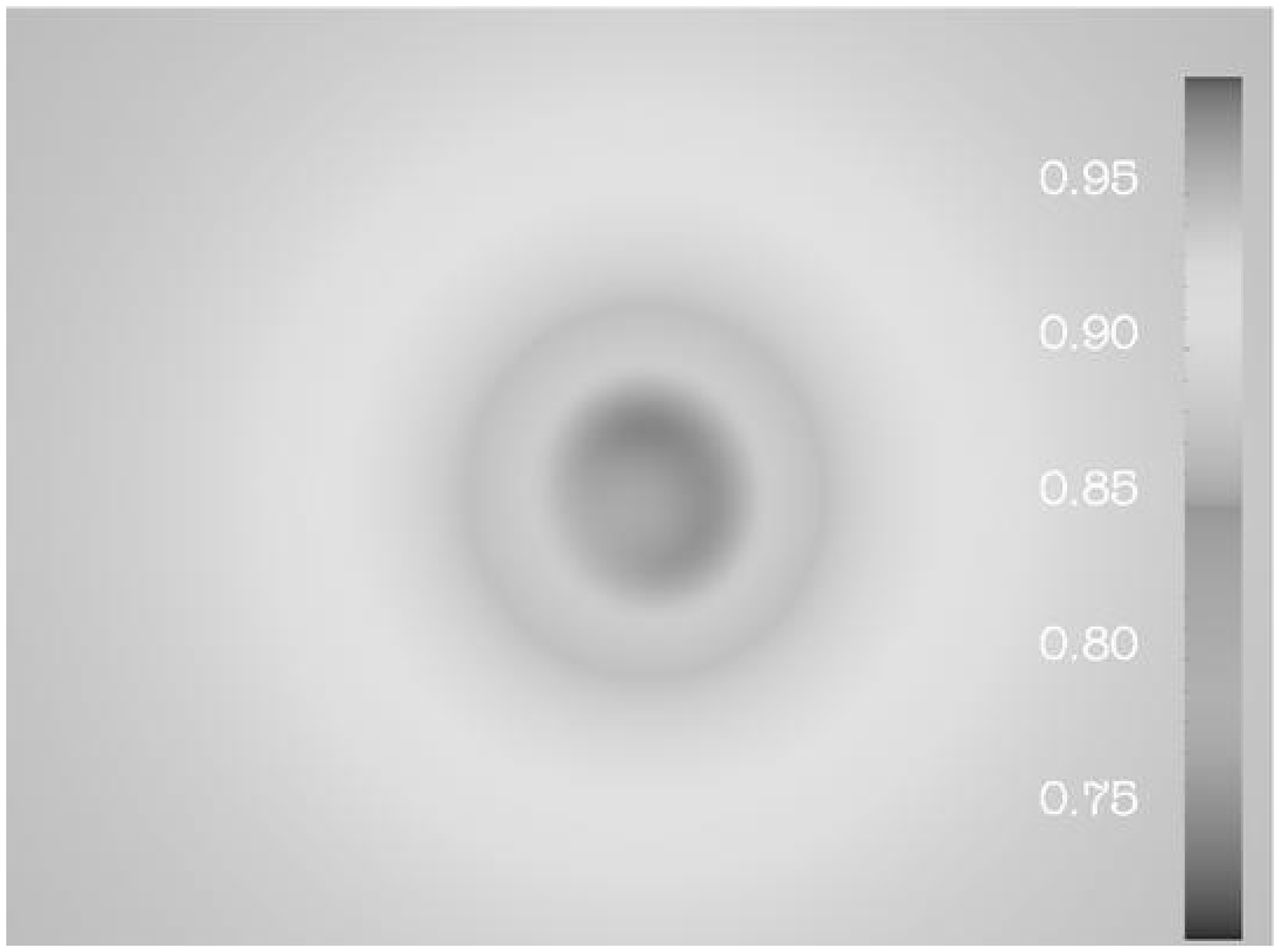} \\
\includegraphics[width=\columnwidth]{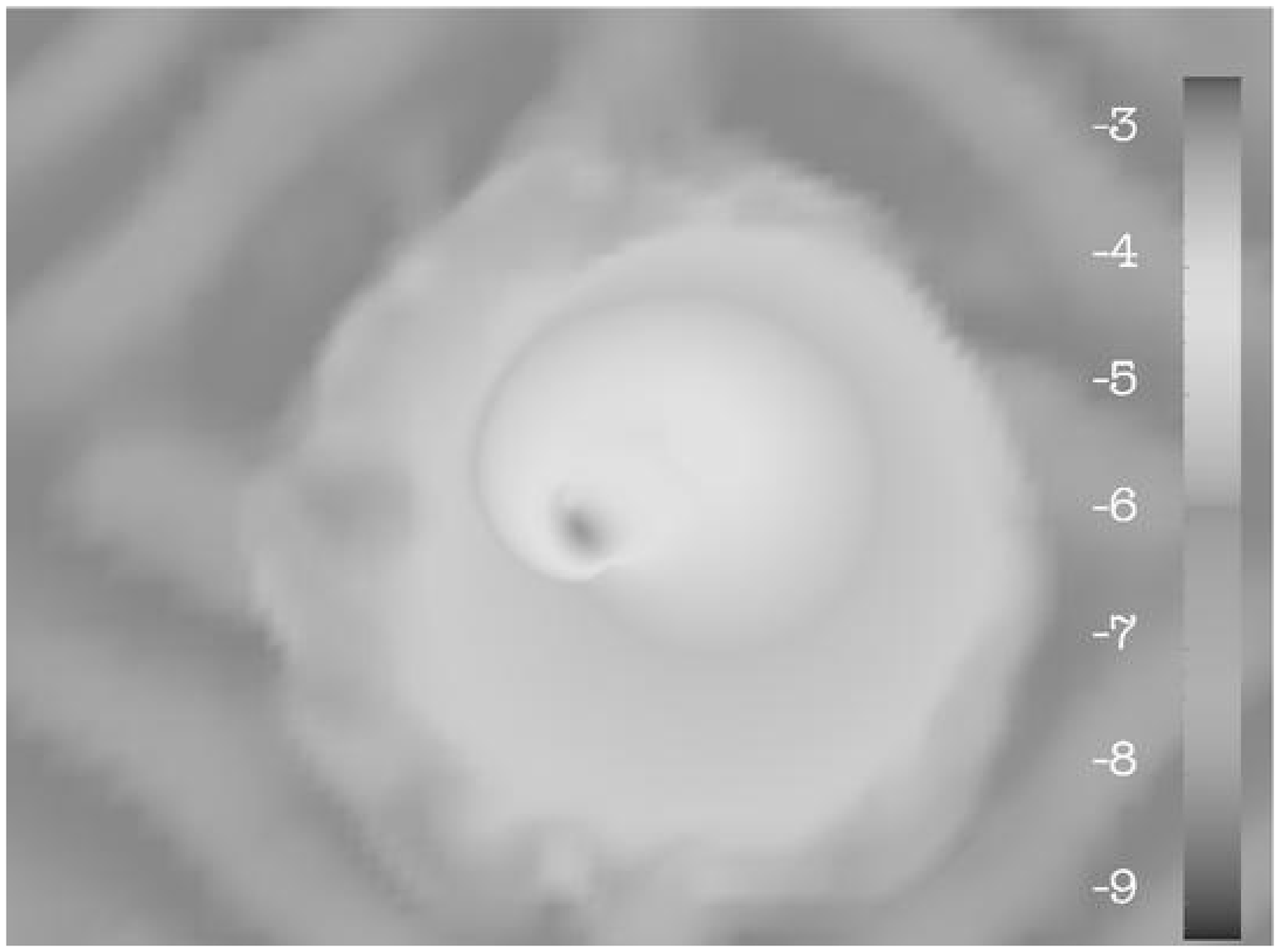} &
\includegraphics[width=\columnwidth]{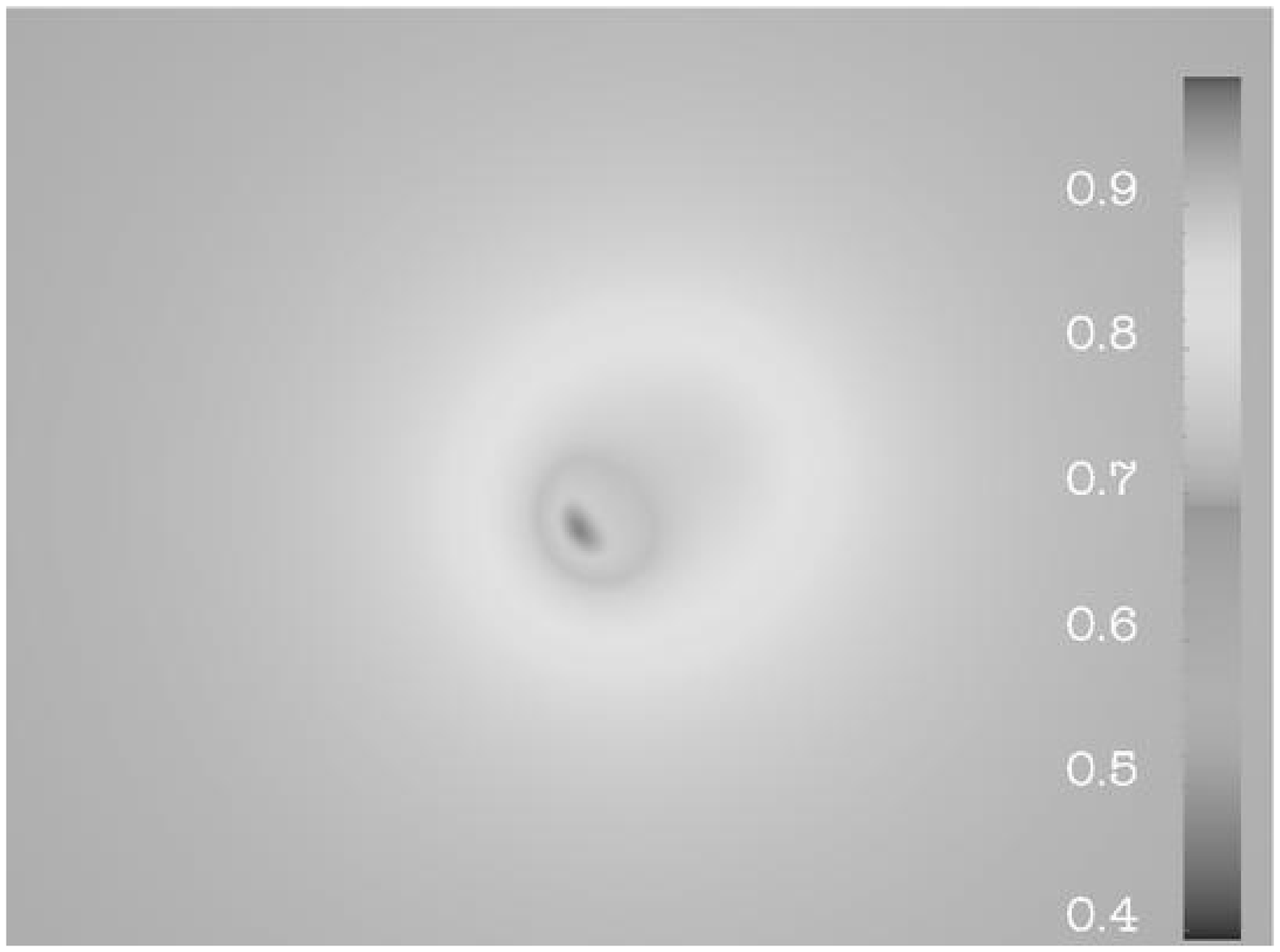}
\end{tabular}
\caption{Development of the fragmentation instability in the reference polytrope.
The left sequence of plots show the decadic logarithm of rest-mass density in the equatorial
plane, the right sequence of plots shows the lapse function. The snapshots correspond to 
times $t/t_{dyn} = 0$ (top), $6.28$ (middle) and $7.48$ (bottom).
The initial model, perturbed by eqn.~\ref{eq:perturbation} with
$\lambda_m = 1$ for $m = 1,\ldots,4$, develops a spiral arm instability and
a collapsing fragment. The domain outside the star initially has the density of the artificial
atmosphere ($\rho_{atm} \approx 5 \cdot 10^{-10}$), and the artifacts outside the star
are caused by interactions of the atmosphere with the stellar surface and the outer boundaries
of the computational domain.
Here, and in all evolution sequences which follow,
the extent of the spatial coordinate domain plotted is the same in all snapshots.
Also, note that the color map is adapted to the range of function values in each plot.}
\label{fig:ref_64_rho_alp}
\end{figure*}

\begin{figure*}
\begin{tabular}{cc}
\includegraphics*[width=\columnwidth]{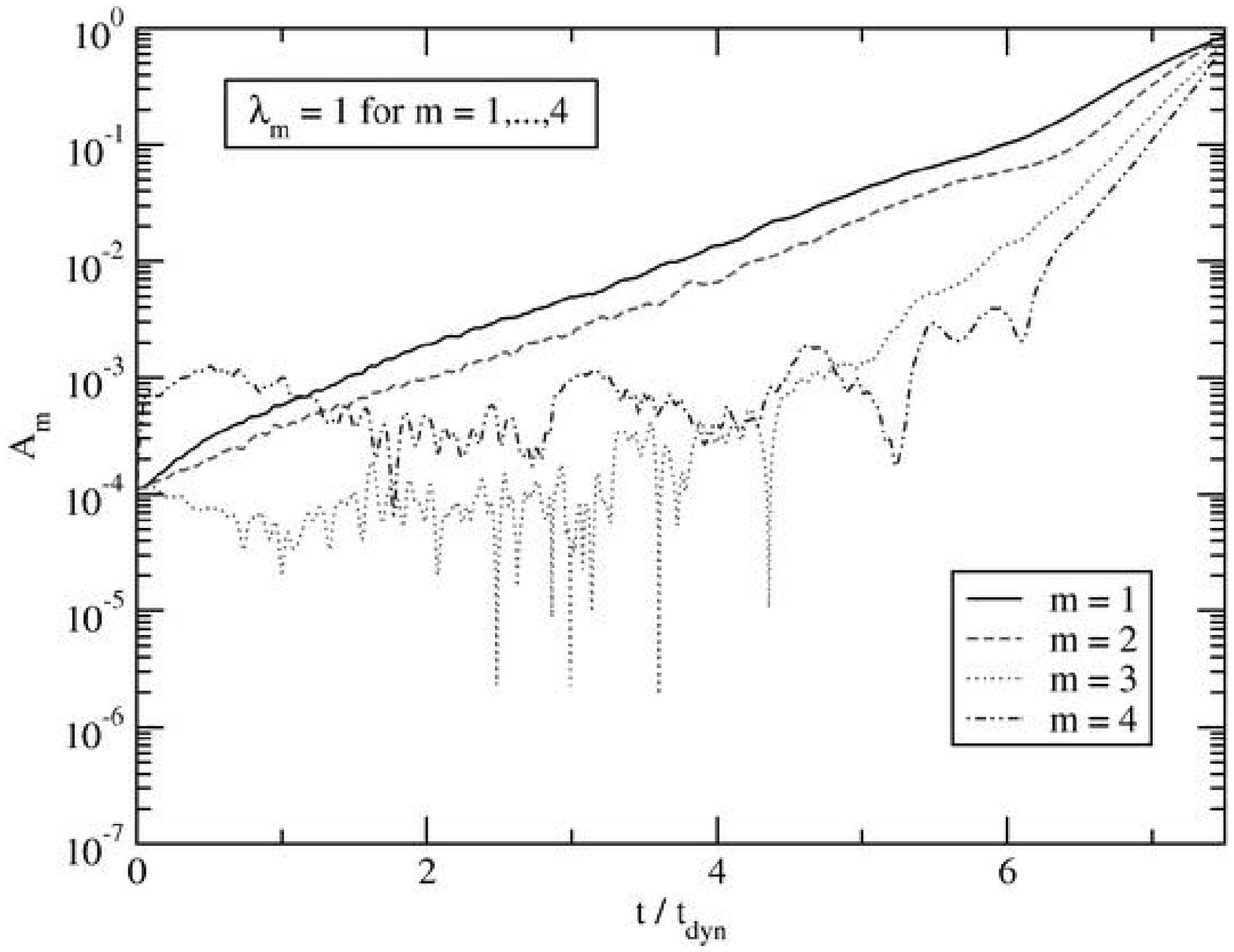} & 
\includegraphics*[width=\columnwidth]{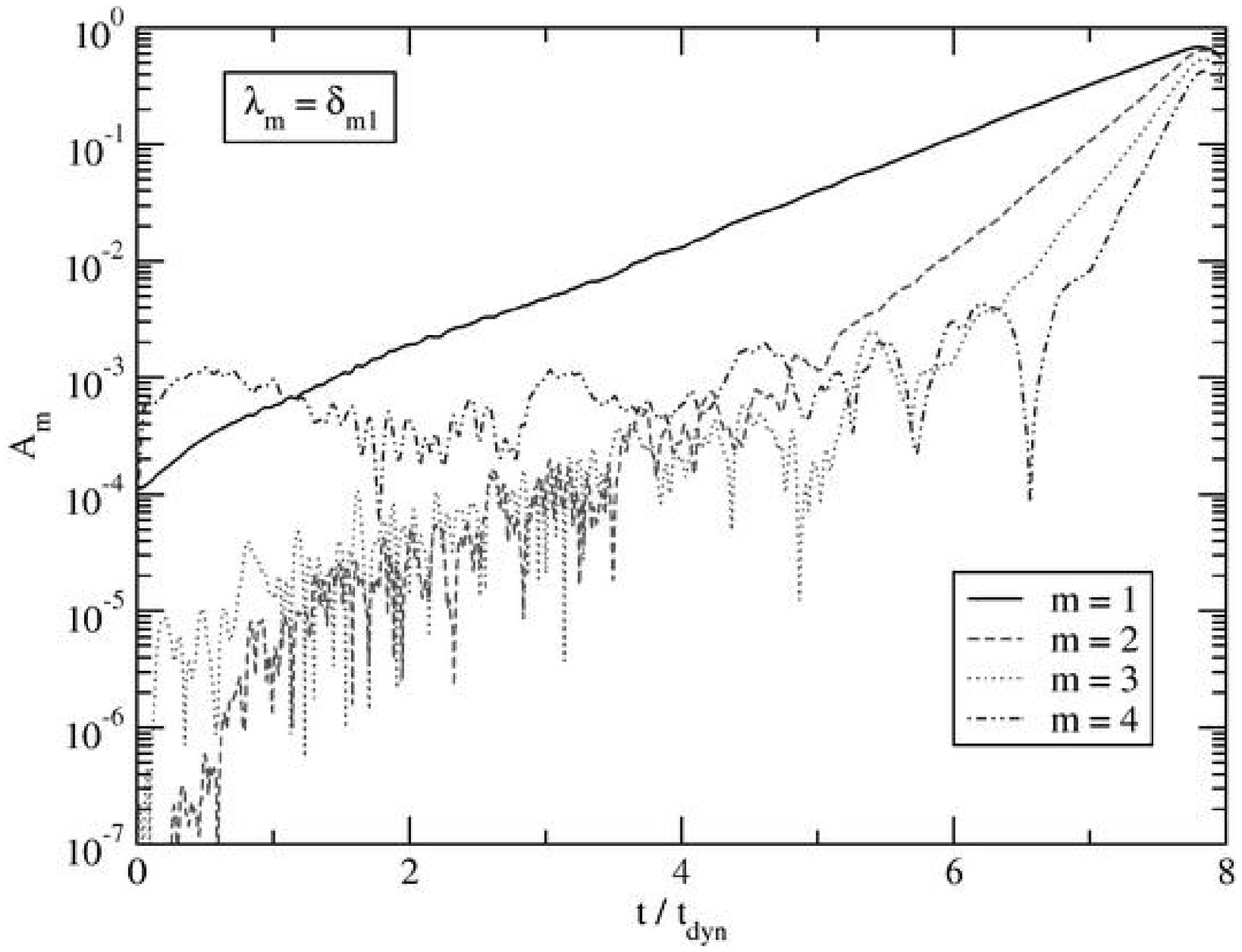} \\
\includegraphics*[width=\columnwidth]{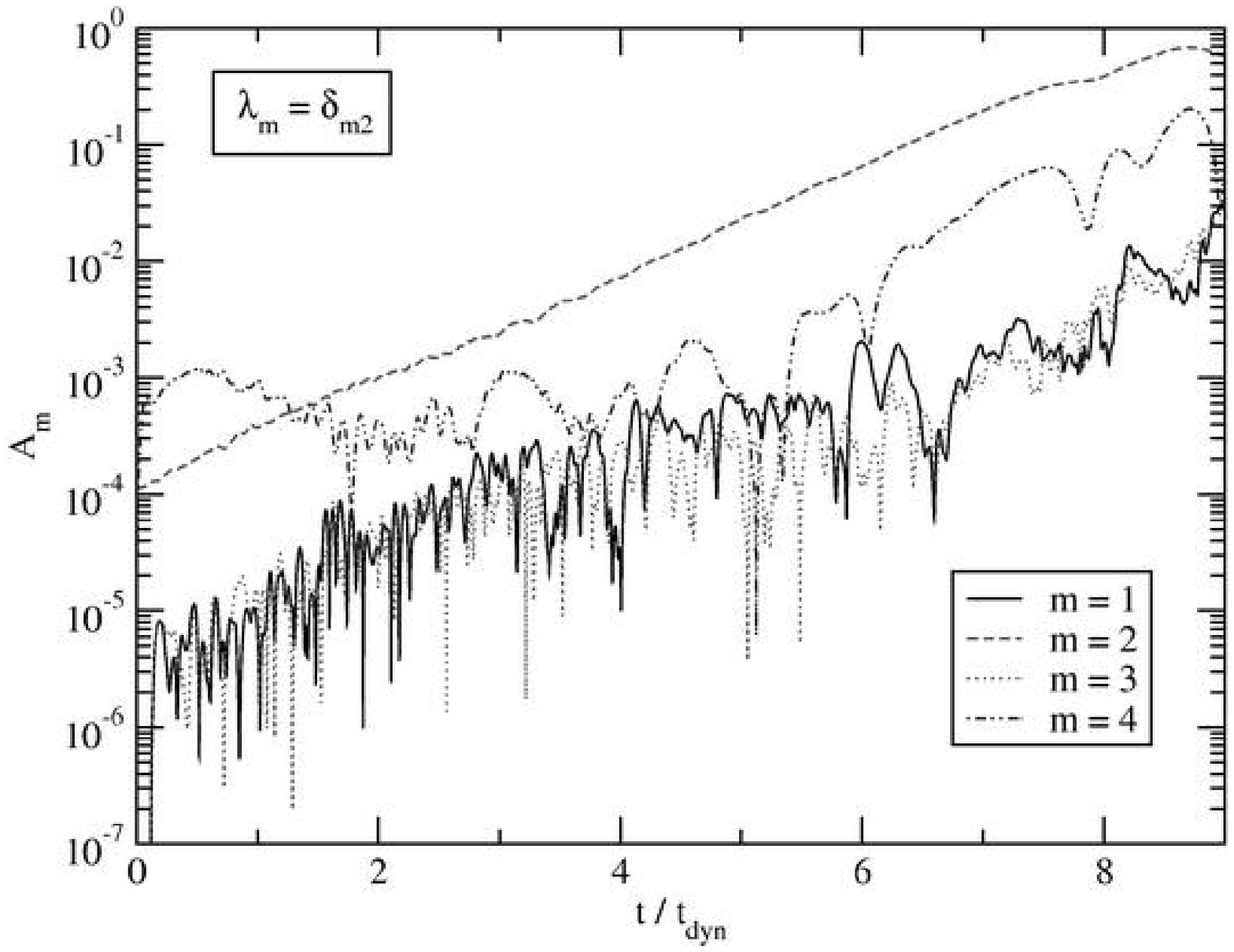} &
\includegraphics*[width=\columnwidth]{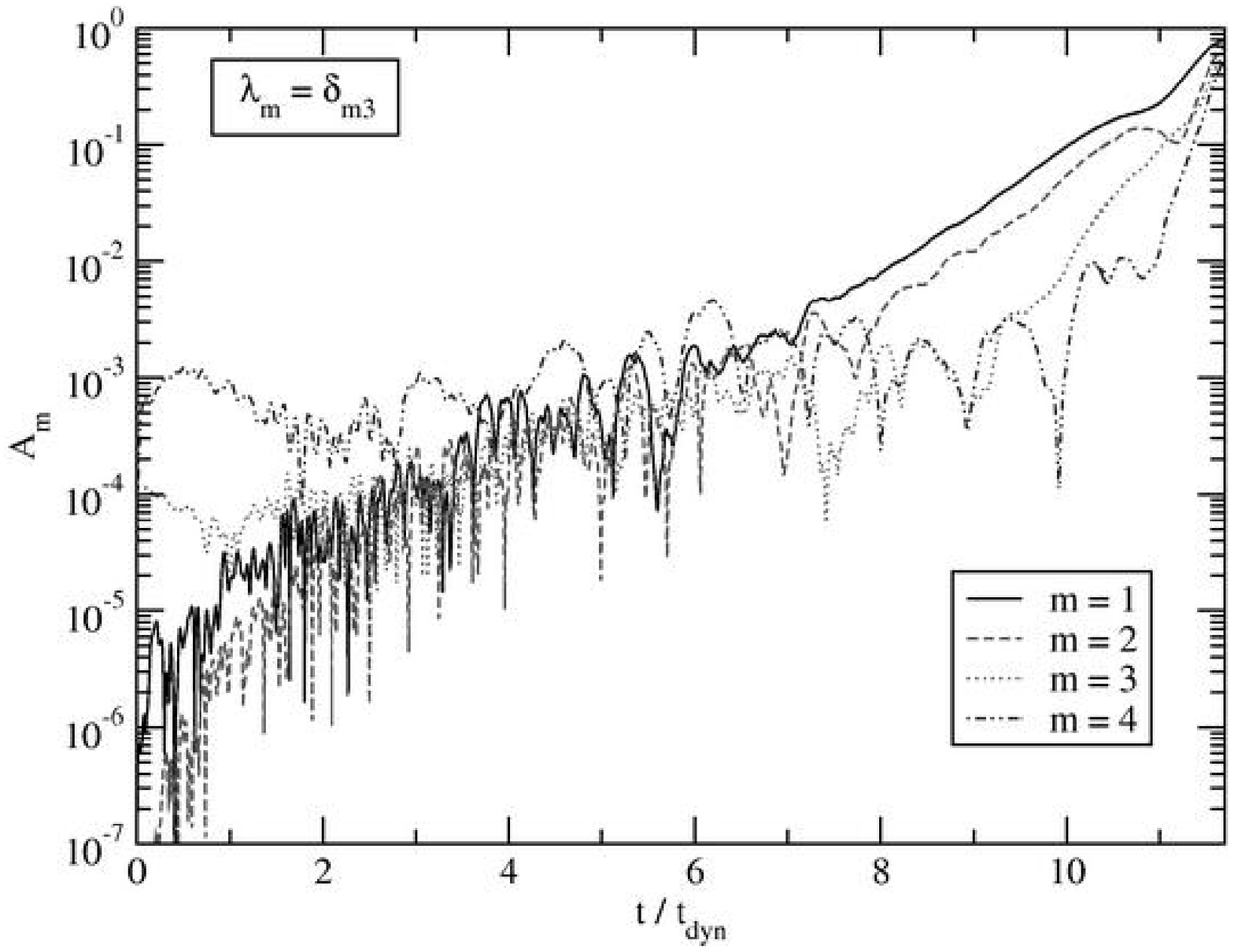} \\
\includegraphics*[width=\columnwidth]{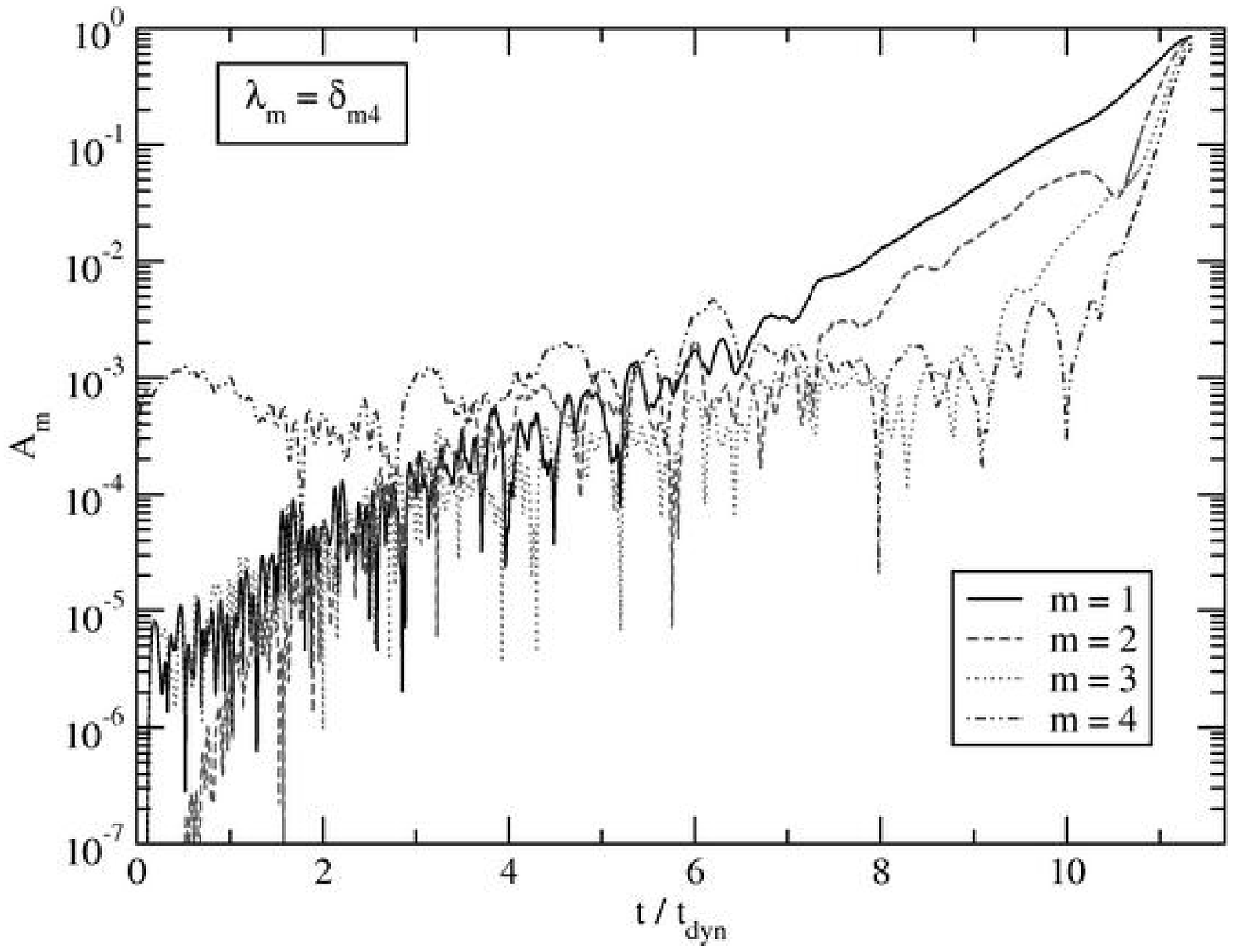} &
\includegraphics*[width=\columnwidth]{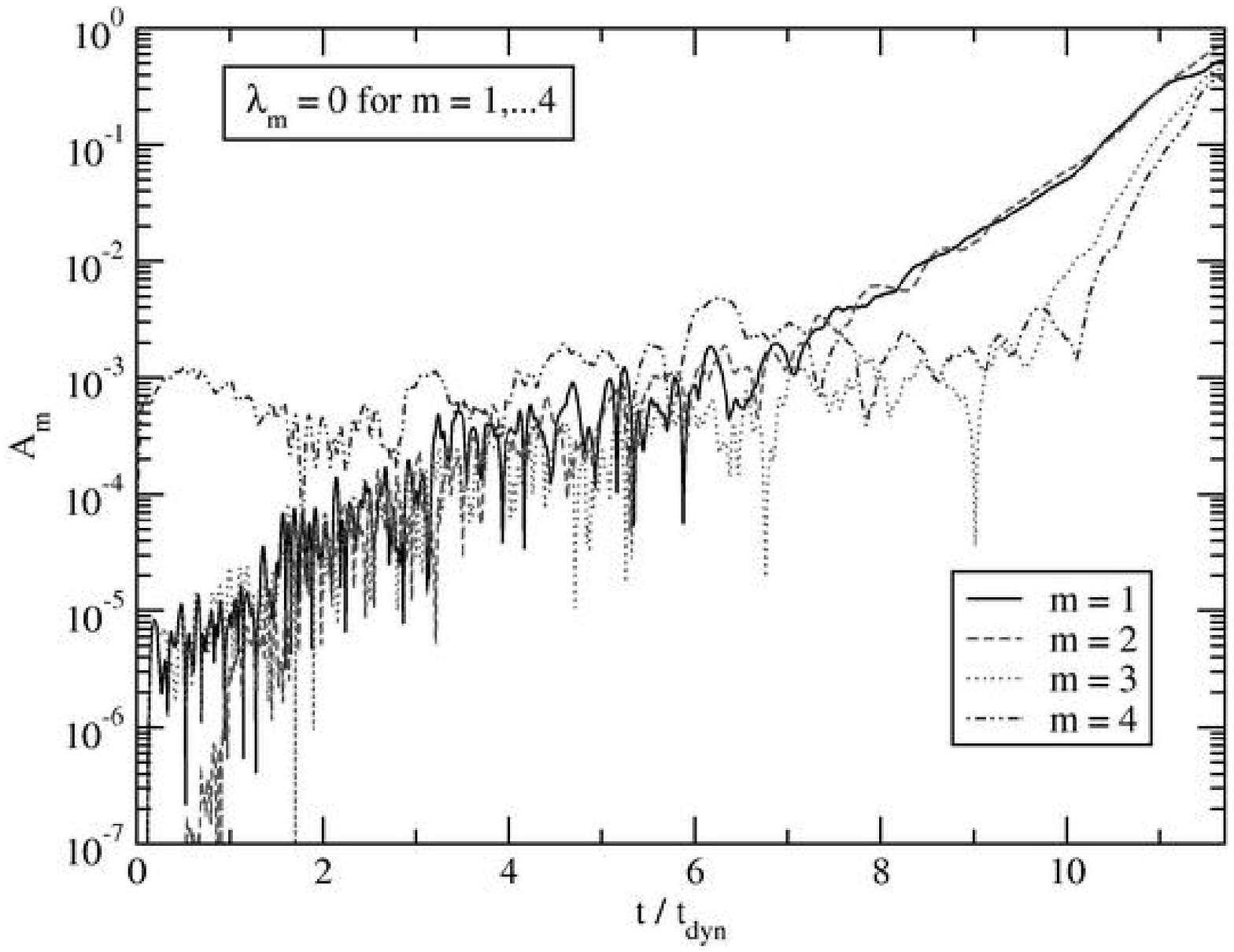} 
\end{tabular}
\caption{Mode amplitudes versus time, extracted at $r = 0.25 R_e$, the radius of 
highest rest-mass density in the initial model, for different initial perturbations.
The amplitude $A_m$ is the $m$-th harmonic Fourier projection of the density, 
normalized to the average value. The perturbations corresponding to each plot are
(cf. also eqn.~\ref{eq:perturbation}): $\lambda_m = 1$ for $m = 1,\ldots,4$ (top left),
$\lambda_m = \delta_{m1}$ (top right), $\lambda_m = \delta_{m2}$ (middle left),
$\lambda_m = \delta_{m3}$ (middle right), $\lambda_m = \delta_{m4}$ (bottom left),
and $\lambda_m = 0$ for $m = 1,\ldots,4$. For details see text.}
\label{fig:ref_64_modes}
\end{figure*}

\begin{figure}
\begin{center}
\begin{tabular}{c}
\includegraphics[width=\columnwidth]{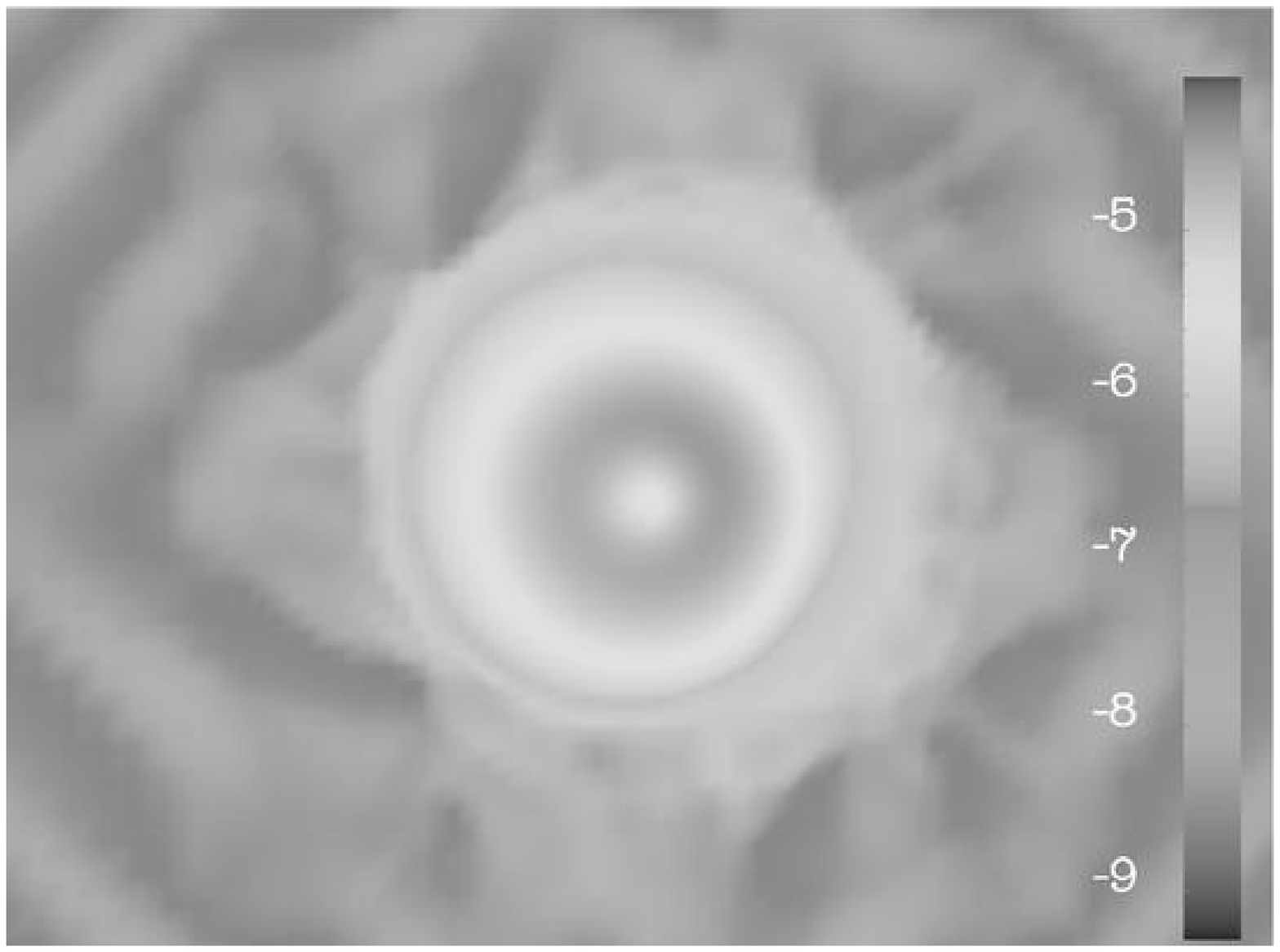} \\
\includegraphics[width=\columnwidth]{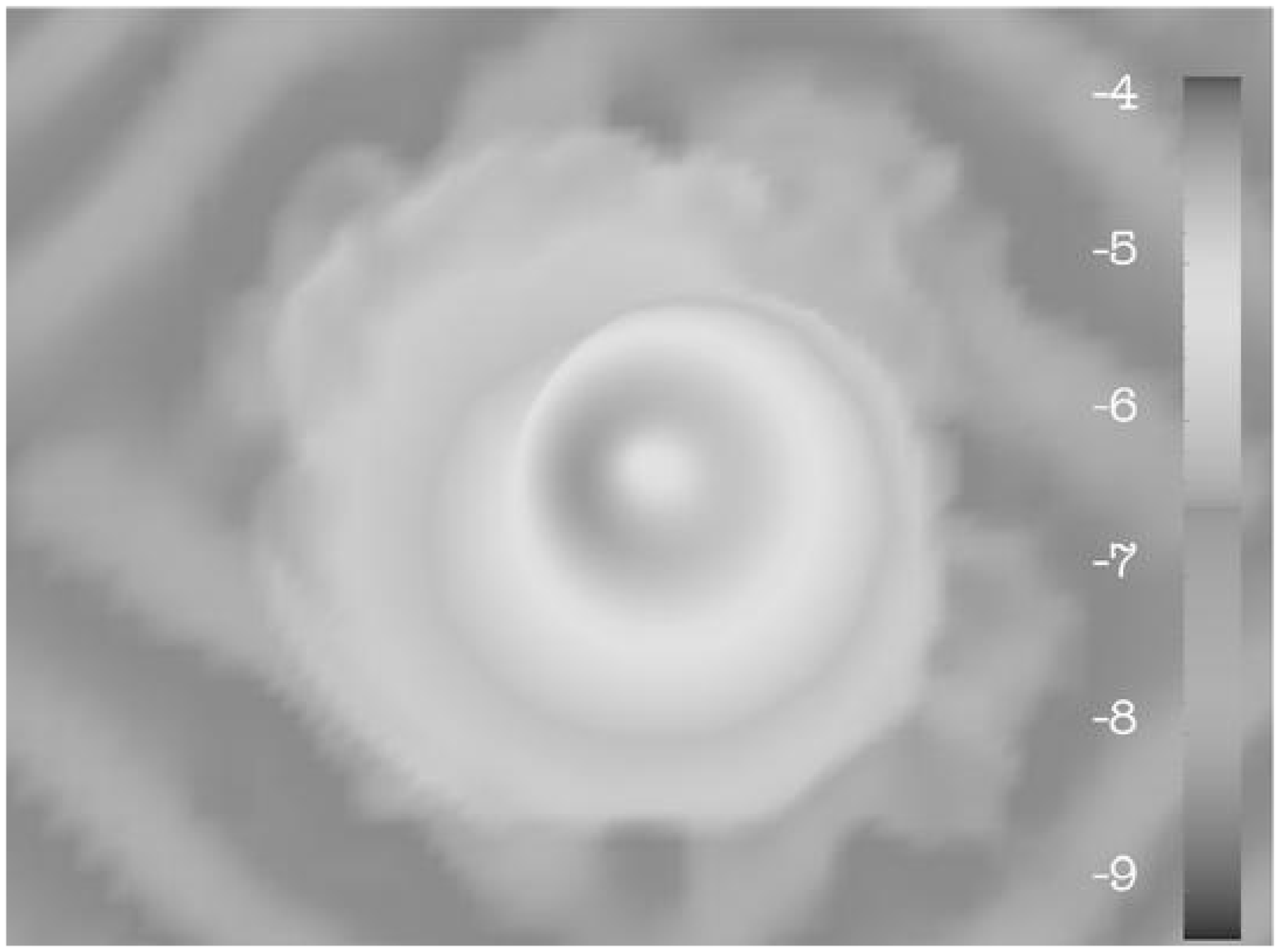} \\
\includegraphics[width=\columnwidth]{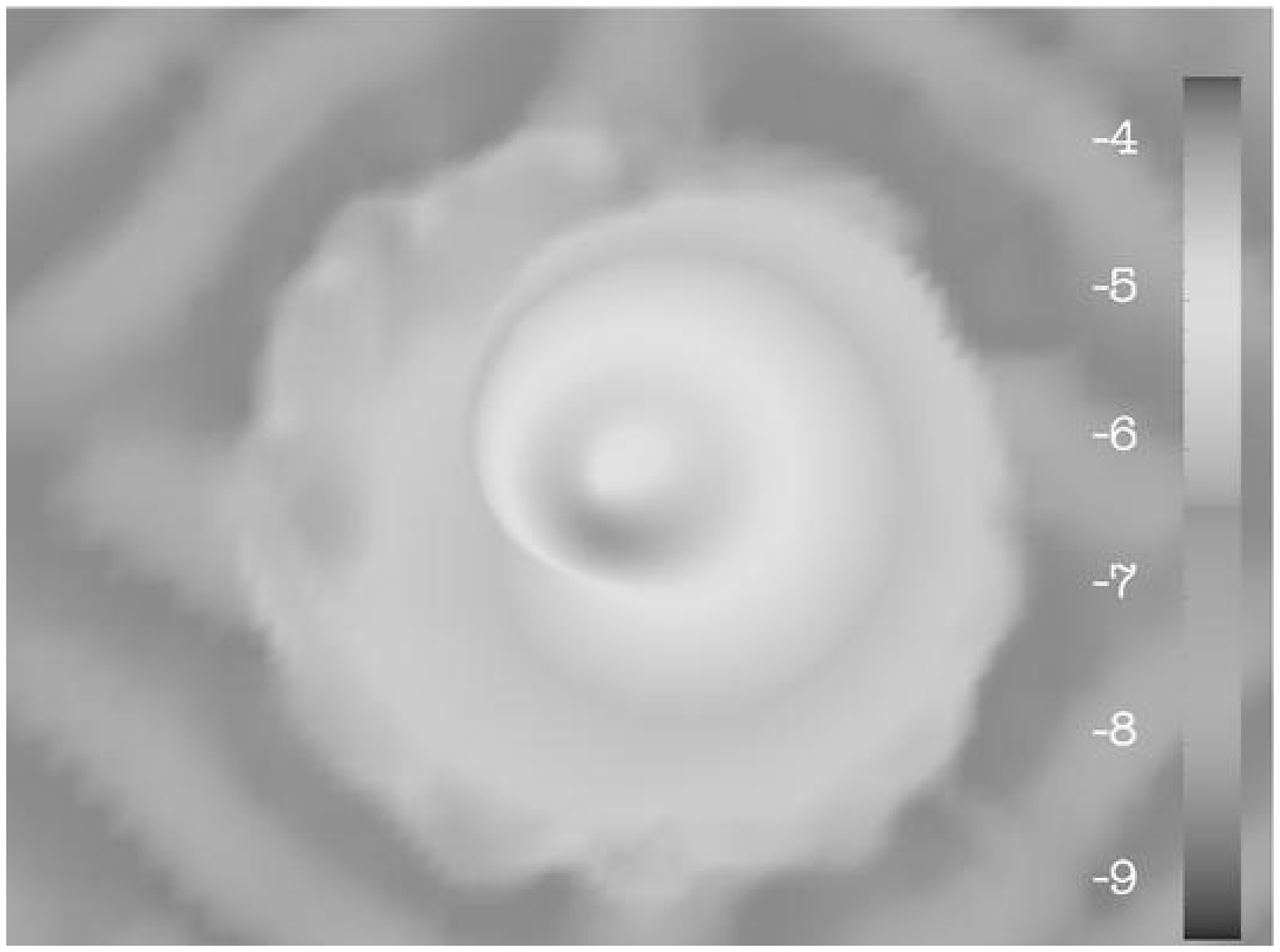} \\
\end{tabular}
\end{center}
\caption{Similar to Fig.~\ref{fig:ref_64_rho_alp}, but now for a perturbation
$\lambda_m = \delta_{m1}$. Shown is the decadic logarithm of the density in the equatorial plane.
The snapshots correspond to times $t/t_{dyn} = 6.28$ (top), $7.11$ (middle) and
$7.48$ (bottom). While the $m = 2$ mode is now suppressed, the 
qualitative evolution is similar to that displayed in Fig.~\ref{fig:ref_64_rho_alp}.}
\label{fig:ref_64_m1_rho}
\end{figure}

\begin{figure}
\begin{center}
\begin{tabular}{c}
\includegraphics[width=\columnwidth]{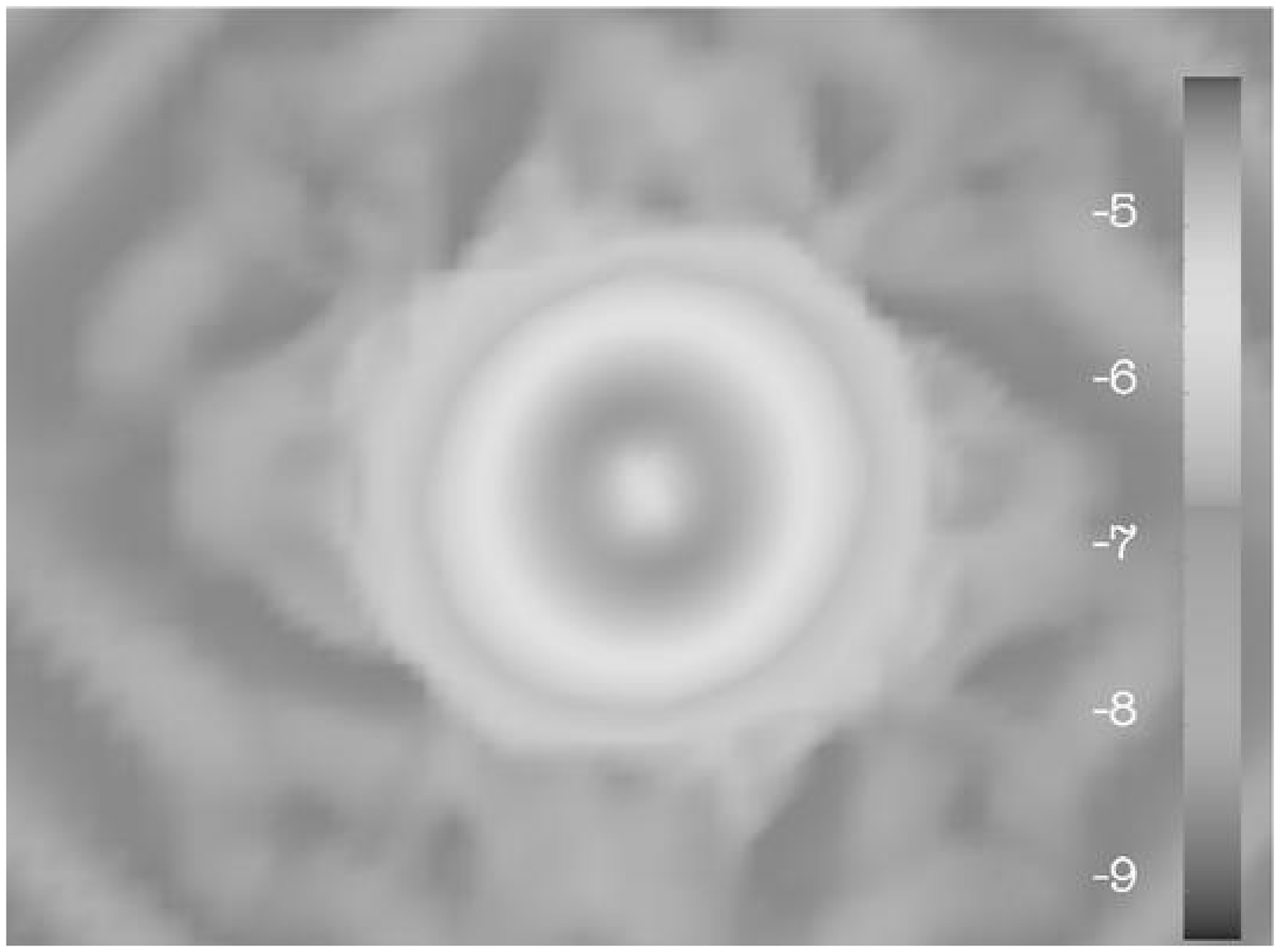} \\
\includegraphics[width=\columnwidth]{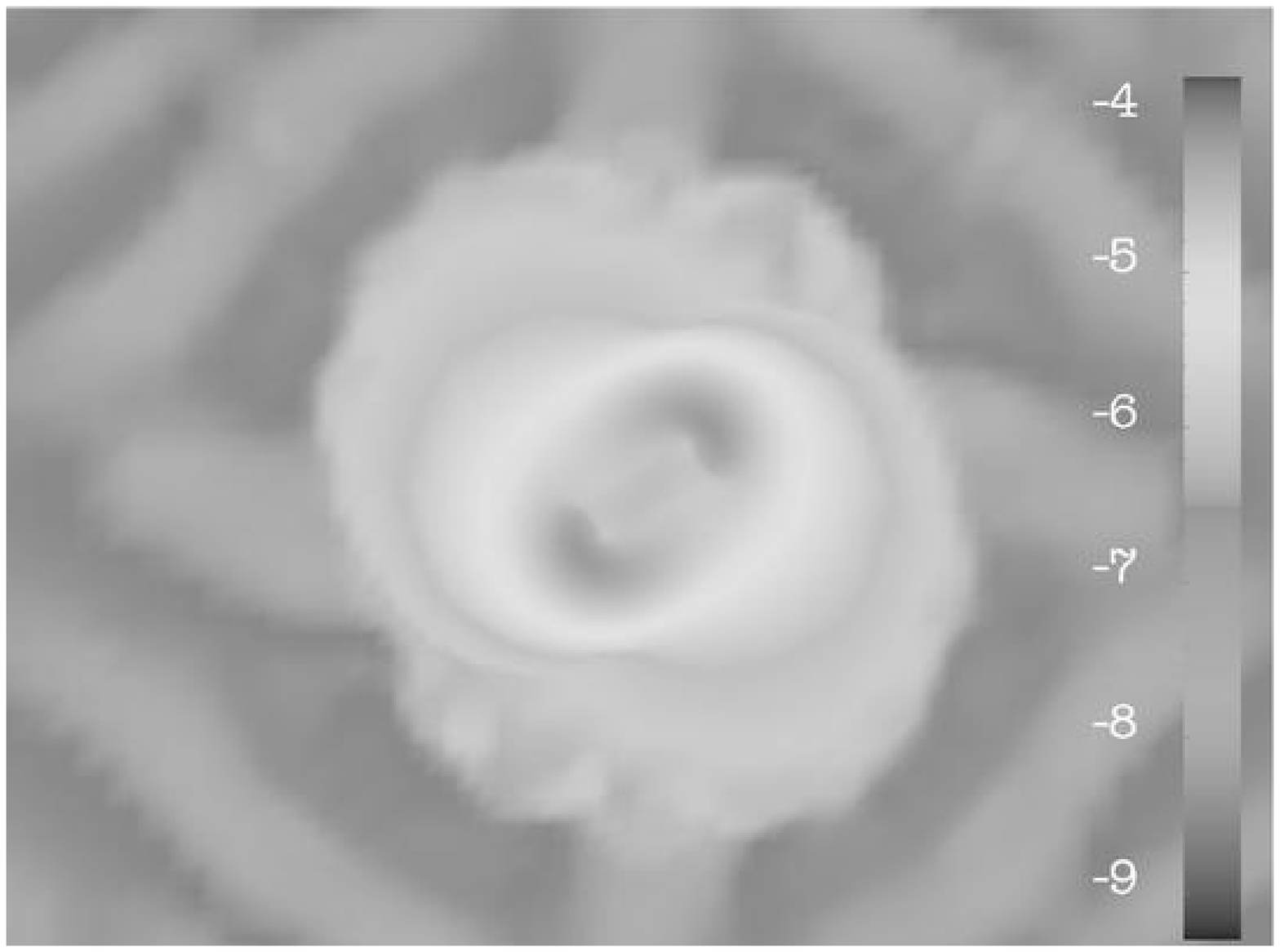} \\
\includegraphics[width=\columnwidth]{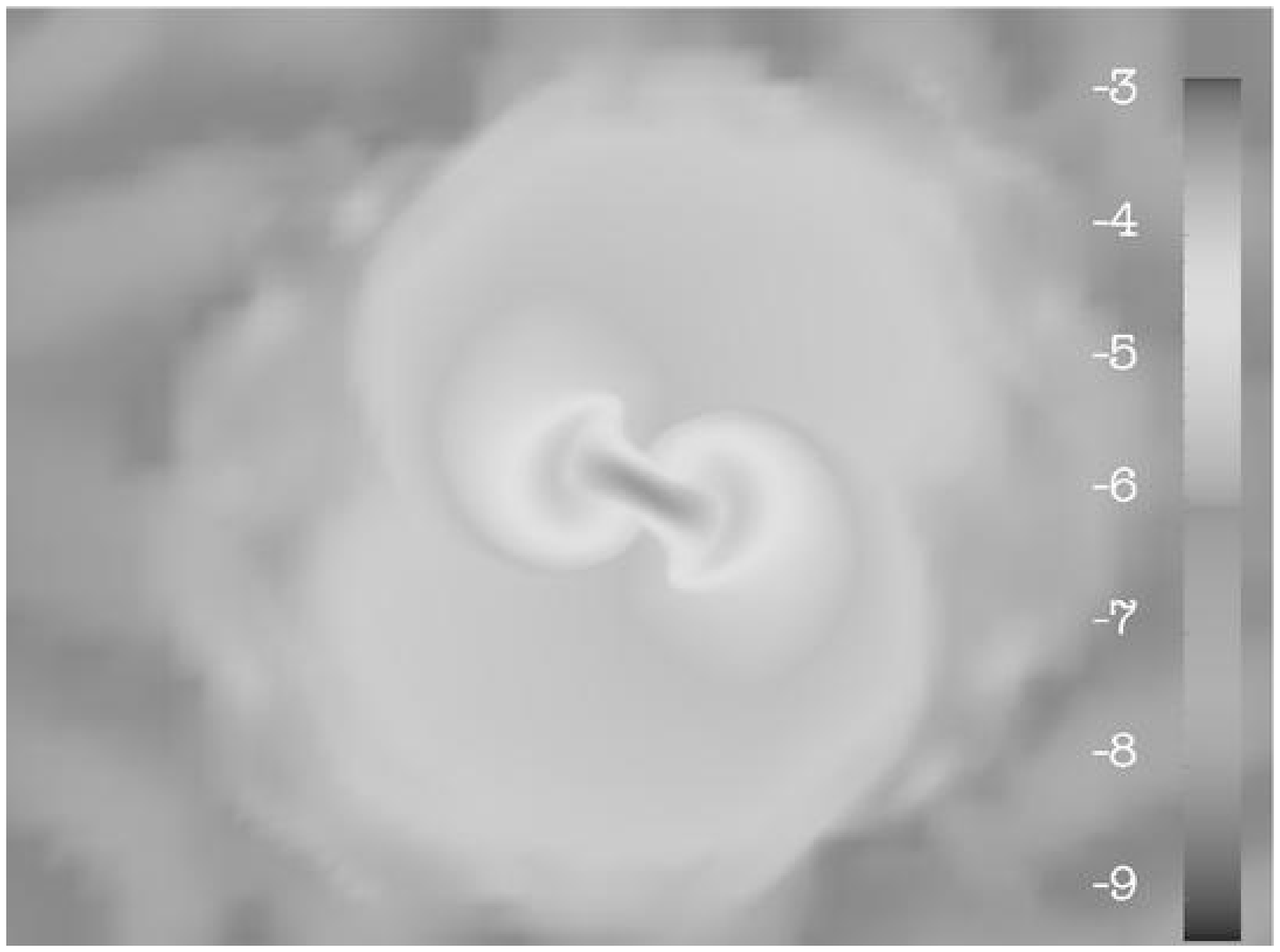} \\
\end{tabular}{c}
\end{center}
\caption{Similar to Fig.~\ref{fig:ref_64_rho_alp}, but for a perturbation
$\lambda_m = \delta_{m2}$. Shown is the decadic logarithm of the density in the equatorial plane.
The snapshots correspond to times $t/t_{dyn} = 6.28$ (top), $7.66$ (middle) and
$8.85$ (bottom). Two fragments develop and encounter a runaway instability while orbiting
each other.}
\label{fig:ref_64_m2_rho}
\end{figure}

\begin{figure}
\begin{center}
\includegraphics[width=\columnwidth]{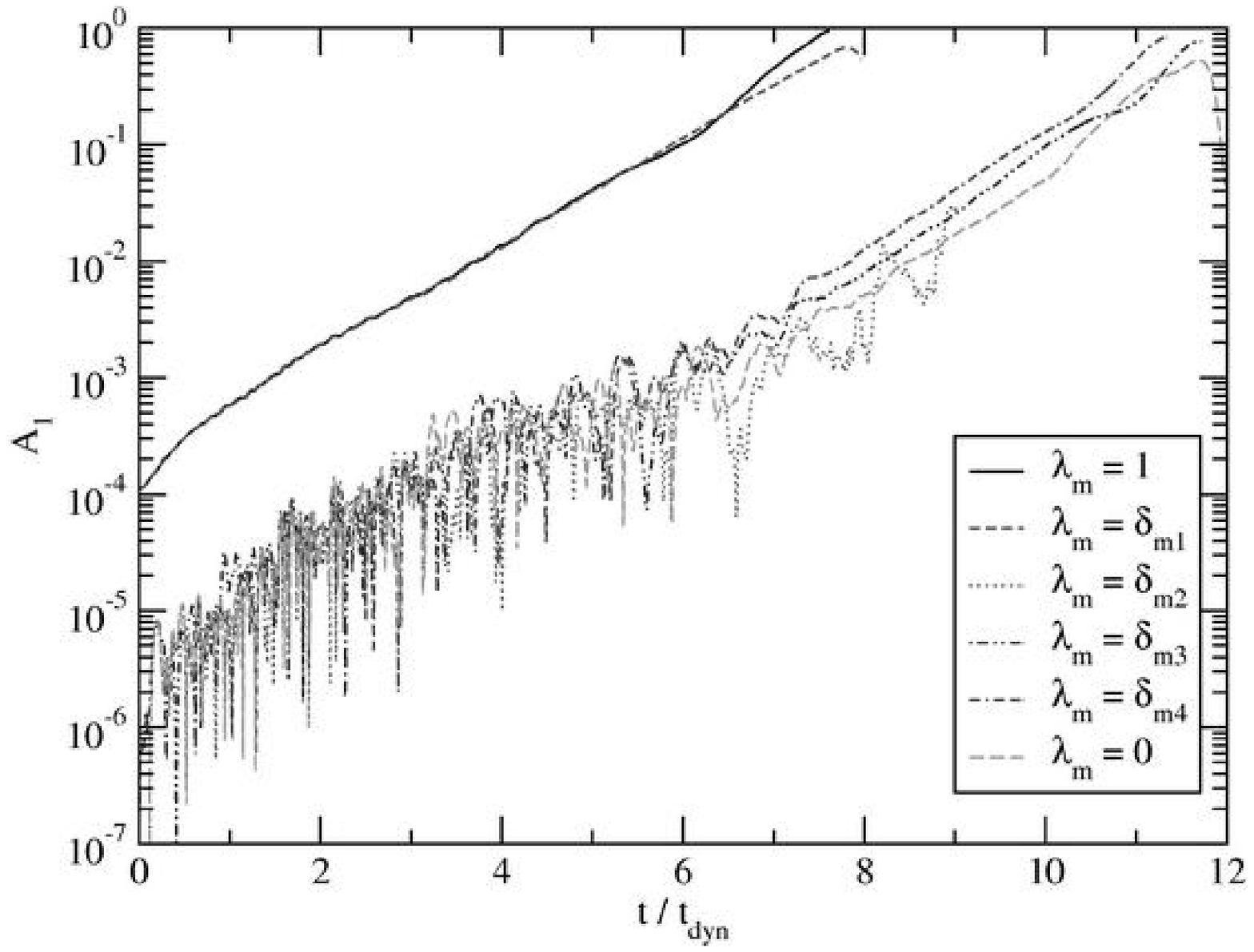}
\includegraphics[width=\columnwidth]{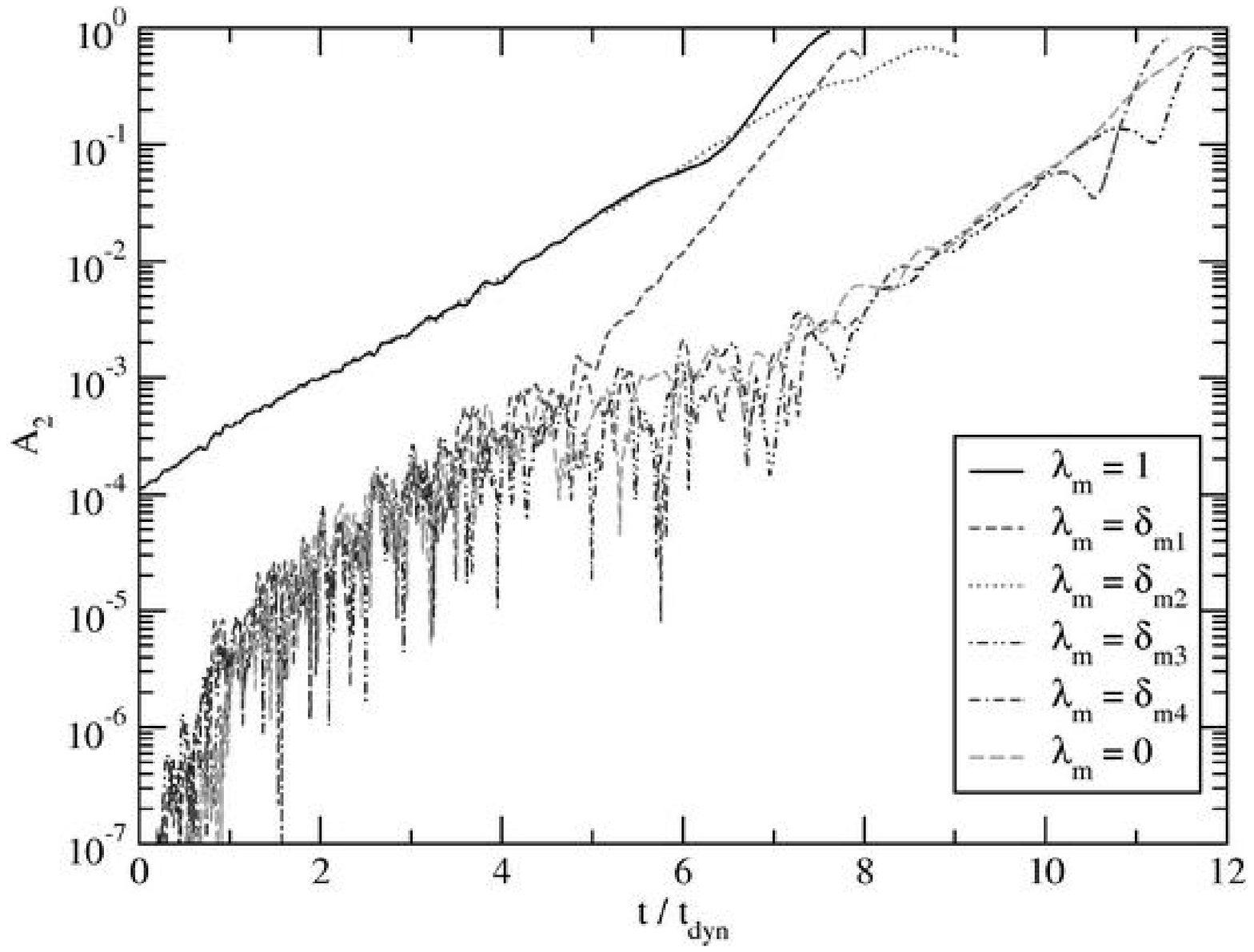}
\end{center}
\caption{Comparison of mode amplitude versus time for different initial perturbations.
The upper panel shows the mode amplitude $A_1$, and the lower one shows $A_2$.
Note that, in the case $\lambda_m = \delta_{m1}$, the $m = 2$ mode is dominated
by non-linear effects from the fragmentation.}
\label{fig:ref_64_modes_comparison}
\end{figure}

\begin{figure}
\begin{center}
\includegraphics*[width=\columnwidth]{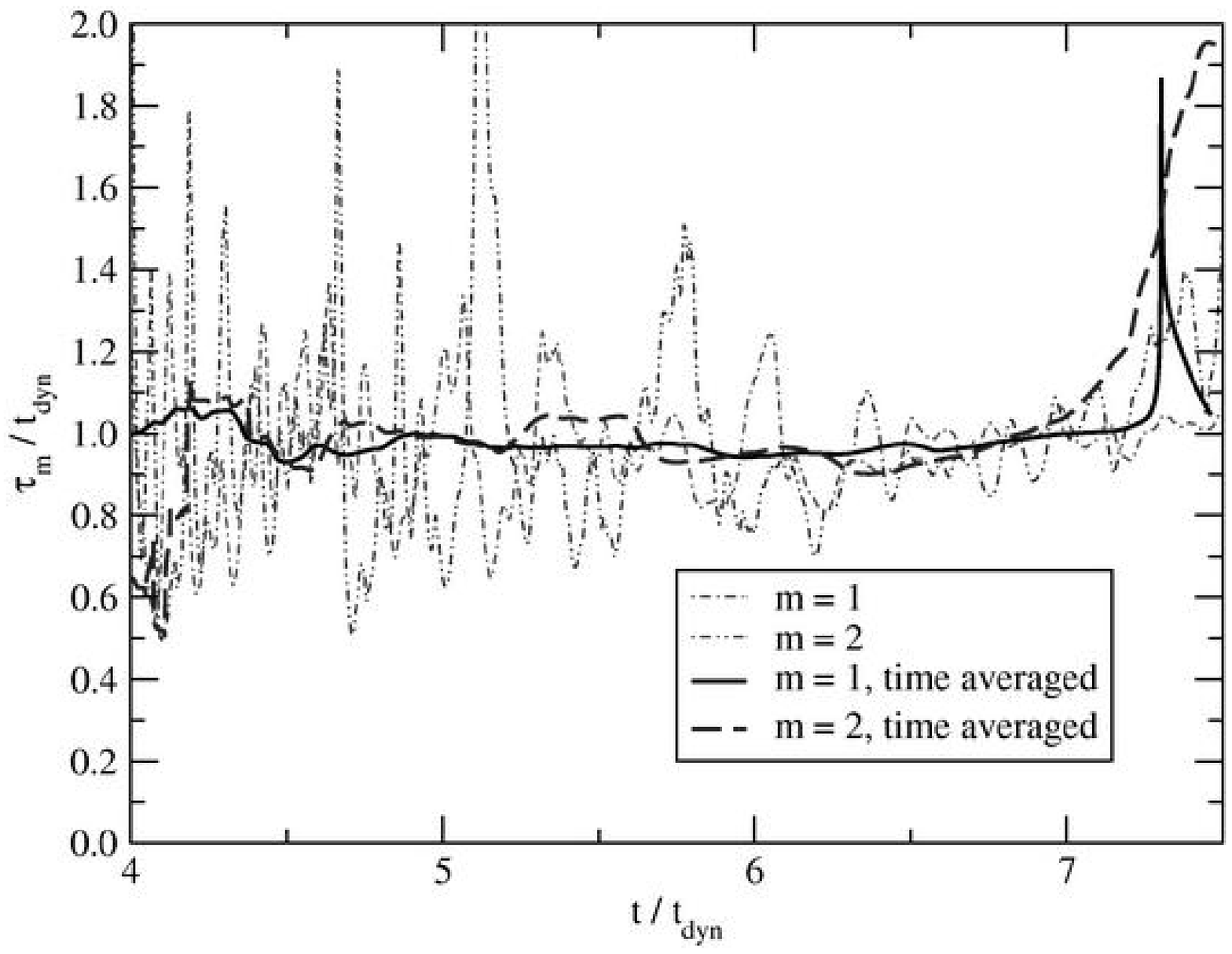}
\includegraphics*[width=\columnwidth]{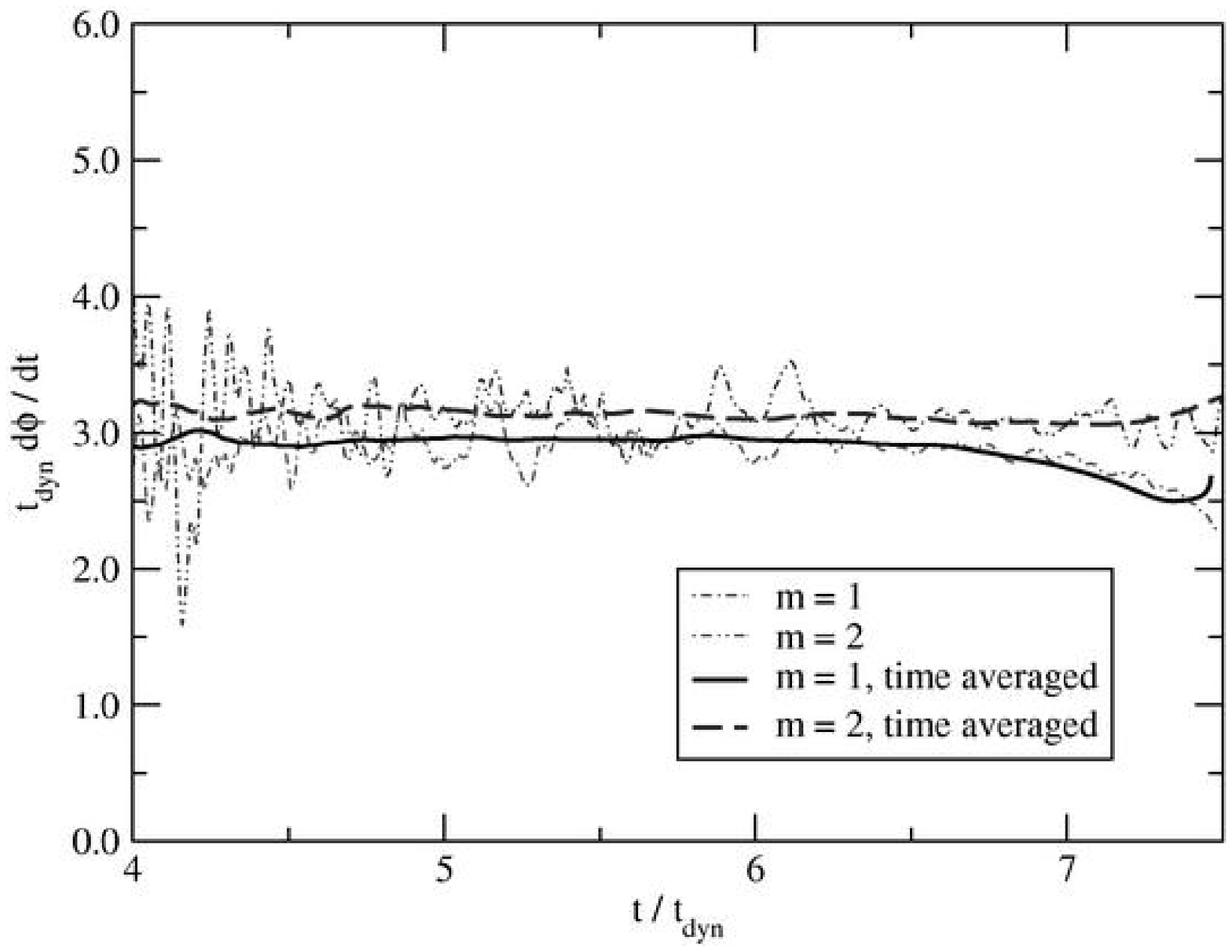}
\end{center}
\caption{Local growth times $\tau_m(t) := (d\ln A/dt)^{-1}$ (top) and
frequencies $d\phi_m/dt$ (bottom) for modes $m = 1$
and $m = 2$, both taken from simulations with the corresponding initial
perturbation. The plots denoted by ``time averaged'' are running averages 
over $t_{dyn}$.}
\label{fig:ref_64_growth_frequencies}
\end{figure}

Fig.~\ref{fig:ref_64_rho_alp} shows the development of the non-axisymmetric
instability in the equatorial plane when using a perturbation of the form given
by eqn.~\ref{eq:perturbation} and $\lambda_m = 1$ for $m = 1,\ldots,4$. 
The density perturbation is not apparent in the initial model, but, after
a few dynamical timescales, an instability has developed which entirely destroys 
the structure of the star. (Note that the artifacts outside the stellar surface
are caused by the introduction of an artifical atmosphere.)
In this case, one collapsing off-center fragment forms
in the system. Judging from Fig.~\ref{fig:ref_64_rho_alp} 
there is a ``collapse of the lapse,'' which is a well-known effect when using 
singularity-avoiding slicings, and which indicates the development of a black hole. 

To investigate the instability more closely, the Fourier mode extraction discussed in
Section~\ref{sec:methods} has been applied to the coordinate radius of highest density
in the initial model, which is at $r = 0.25 R_e$.
We concentrate on this radius for reasons
already  discussed: different extraction radii will be considered in 
Section~\ref{sec:global_nature}. The top left panel of Fig.~\ref{fig:ref_64_modes} 
displays the evolution of the amplitude
of the first four Fourier modes $A_m$, $m = 1,\ldots,4$, in this evolution. Although all
four modes have been injected with the same amplitude ($\approx 10^{-4}$), the 
$m = 4$ mode displays a significant initial growth of about an order of magnitude,
and then oscillates around this level until $t/t_{dyn} \approx 6$, where non-linear
effects become important. The high level of $m = 4$ noise is
very likely an artefact of the Cartesian grids used for the simulation.
Support for this argument can be obtained by considering that the equatorial
section of the grid has a discrete $C_4$ symmetry, and by comparing Fig.~3 and
4 in \cite{Centrella:2001xp}, where results from the development of a similar
non-axisymmetric instability, though in Newtonian gravity, were achieved on 
cylindrical and Cartesian grids. We note that the discrete model appears to be stable against
this perturbation, and also against the $m = 3$ perturbation. The remaining modes are unstable, 
and the growth times of both modes are similar. 
In this specific case the $m = 1$ structure dominates the late-time evolution, and leads
to the spiral arm structure and fragmentation visible in Fig.~\ref{fig:ref_64_rho_alp}.

The structure of the numerical noise depends on the grid geometry, resolution,
finite difference operators and discrete methods for treating hydrodynamics,
the outer boundary conditions and the artificial
atmosphere. Therefore, it is important to know to which degree the four
Fourier modes are coupled during the evolution. Since the initial
perturbation is considered to be ``small,'' which holds when compared to
the $m = 4$ numerical noise level discussed above, we expect that coupling becomes
important as soon as the amplitudes $A_m$ get close to unity. To determine
this in the reference model, a number of simulations have been performed
with perturbations of the form given by eqn.~\ref{eq:perturbation}, but
with $\lambda_m = \delta_{mj}$ for different $j \in \{1,\ldots,4\}$ to
select individual modes, and one simulation where no perturbation is
applied. The mode amplitudes in these simulations are shown in
Fig.~\ref{fig:ref_64_modes}, and the $m = 1$ and $m = 2$ modes
are compared to the perturbation with $\lambda_m = 1$ in 
Fig.~\ref{fig:ref_64_modes_comparison}. As long as the mode development is
not dominated by another mode which at a higher amplitude, e.g. as in the
case of the $m = 2$ mode in the top right panel in Fig.~\ref{fig:ref_64_modes}, 
the growth times are comparable for different perturbations. 

The high-amplitude, strongly non-linear development at late times is
sensitive to the perturbation function, as is visible from the evolutions 
shown in Fig.~\ref{fig:ref_64_m1_rho} and \ref{fig:ref_64_m2_rho}.
In the case of the $m = 1$ perturbation, a single fragment develops
and collapses in a similar manner to the case $\lambda_m = 1$. With
an $m = 2$ perturbation, however, two orbiting fragments develop,
contract, and subsequently encounter a runaway instability in the 
center (bottom right panel in Fig.~\ref{fig:ref_64_m2_rho}). Any perturbation
with different values for $\lambda_1$ and $\lambda_2$ might produce
a mixture of this spiral arm and binary-system fragmentation instability.
One might argue that a fine-grained parameter space study in
the space of $\lambda_1$ and $\lambda_2$ is necessary to obtain
a more complete understanding of the remnants. However, the reference
polytrope is already well 
inside the unstable region of the parameter space.
We will see in Section~\ref{sec:evol_plane} that, on a sequence of increasing $T/|W|$, 
the $m = 1$ mode dominates first.
As has been shown in \cite{Zink2005a}, this process
of fragmentation is black-hole forming\footnote{The case
$\lambda_m = \delta_{m2}$ exhibits a ``collapse of the lapse'' at
late times, which suggests that in this case a black hole has formed, too.
Unfortunately, it was not possible to locate an apparent horizon in this case
due to numerical difficulties.}. Unfortunately, around the time of horizon
formation the evolution fails. This
appears to be due to the growth of small scale features that cannot be
adequately resolved leading to unphysical oscillations. It seems
possible that a more robust gauge condition, artificial dissipation \cite{Baiotti06}
or excision techniques \cite{Lehner2005a, Zink2005b} could avoid these problems.

To investigate the growth time $\tau$ of the modes, which, in the context
of linear theory, is defined by the relation $A_m(t) = A_m(t=0) \exp(t/\tau_m)$,
we plot the function $\tau_m(t) := (d\ln A/dt)^{-1}$ in the top panel of
Fig.~\ref{fig:ref_64_growth_frequencies} for $m = 1$ and $m = 2$, using
different initial perturbations of the reference polytrope as before. As expected for a dynamical
instability, the growth times are of the order of the dynamical timescale. 
Finally, in the bottom panel of Fig.~\ref{fig:ref_64_growth_frequencies} the mode frequencies
$\omega_m(t) = d\phi_m/dt$ (as extracted from the Fourier decomposition of the density on
a the coordinate radius of highest initial density) are plotted versus time. These, and the connected pattern
speeds $\omega / m$, will be important in the discussion of the corotation band
in Section~\ref{sec:reference_corot}.

\subsubsection{Some results on the global nature of the instability}
\label{sec:global_nature}

\begin{figure}[p!]
\begin{tabular}{c}
\includegraphics[width=\columnwidth]{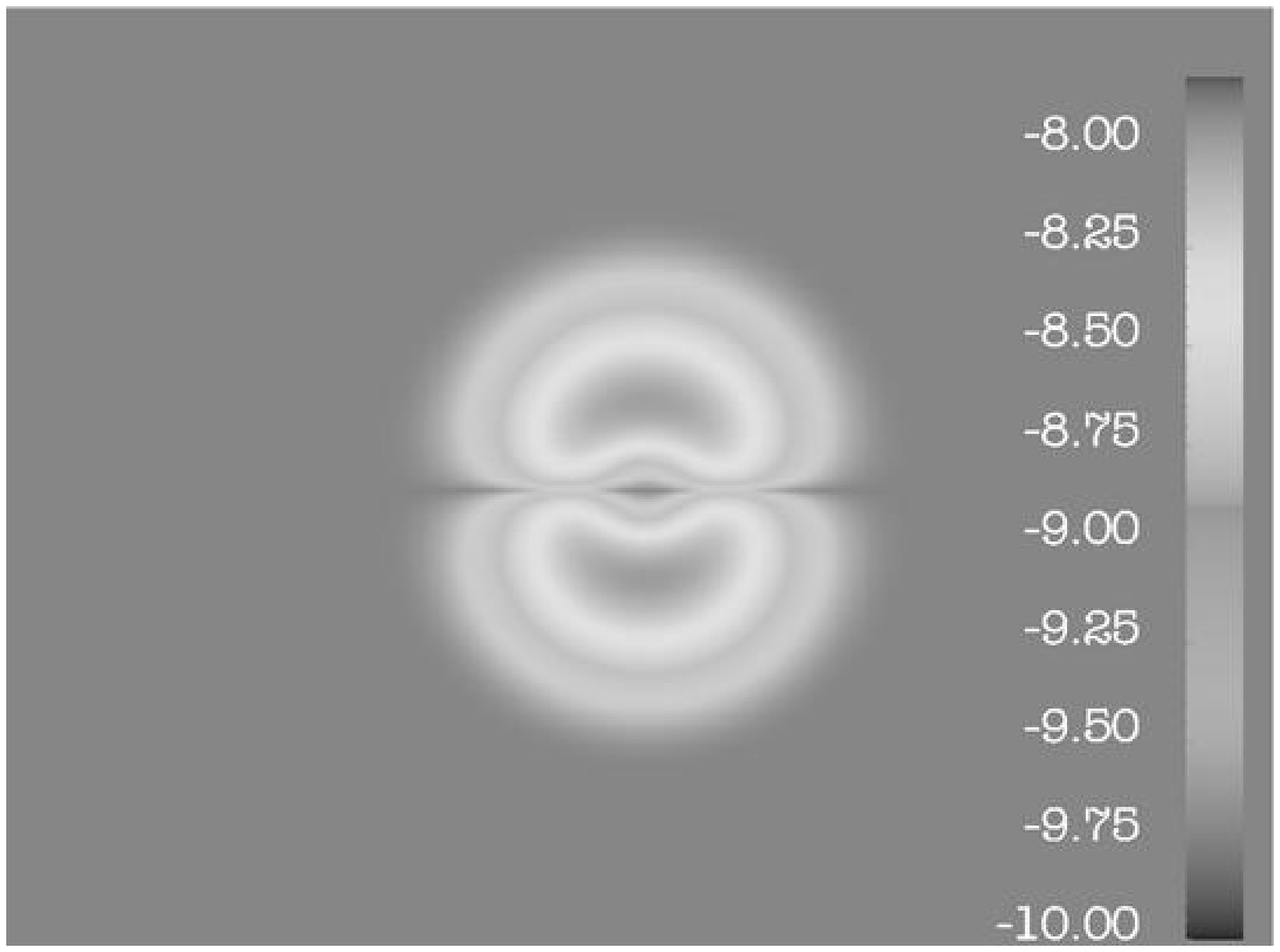} \\
\includegraphics[width=\columnwidth]{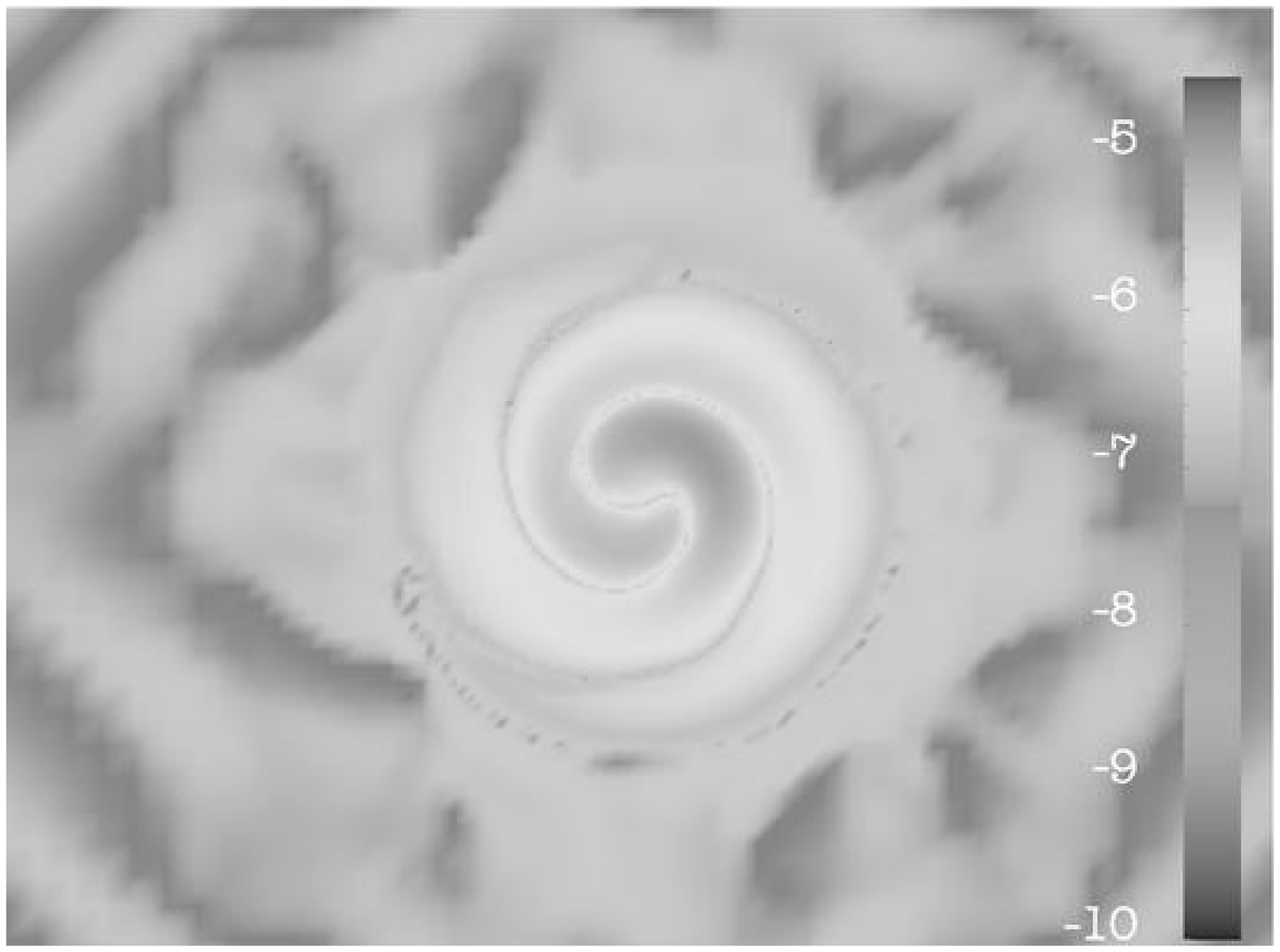} \\
\includegraphics[width=\columnwidth]{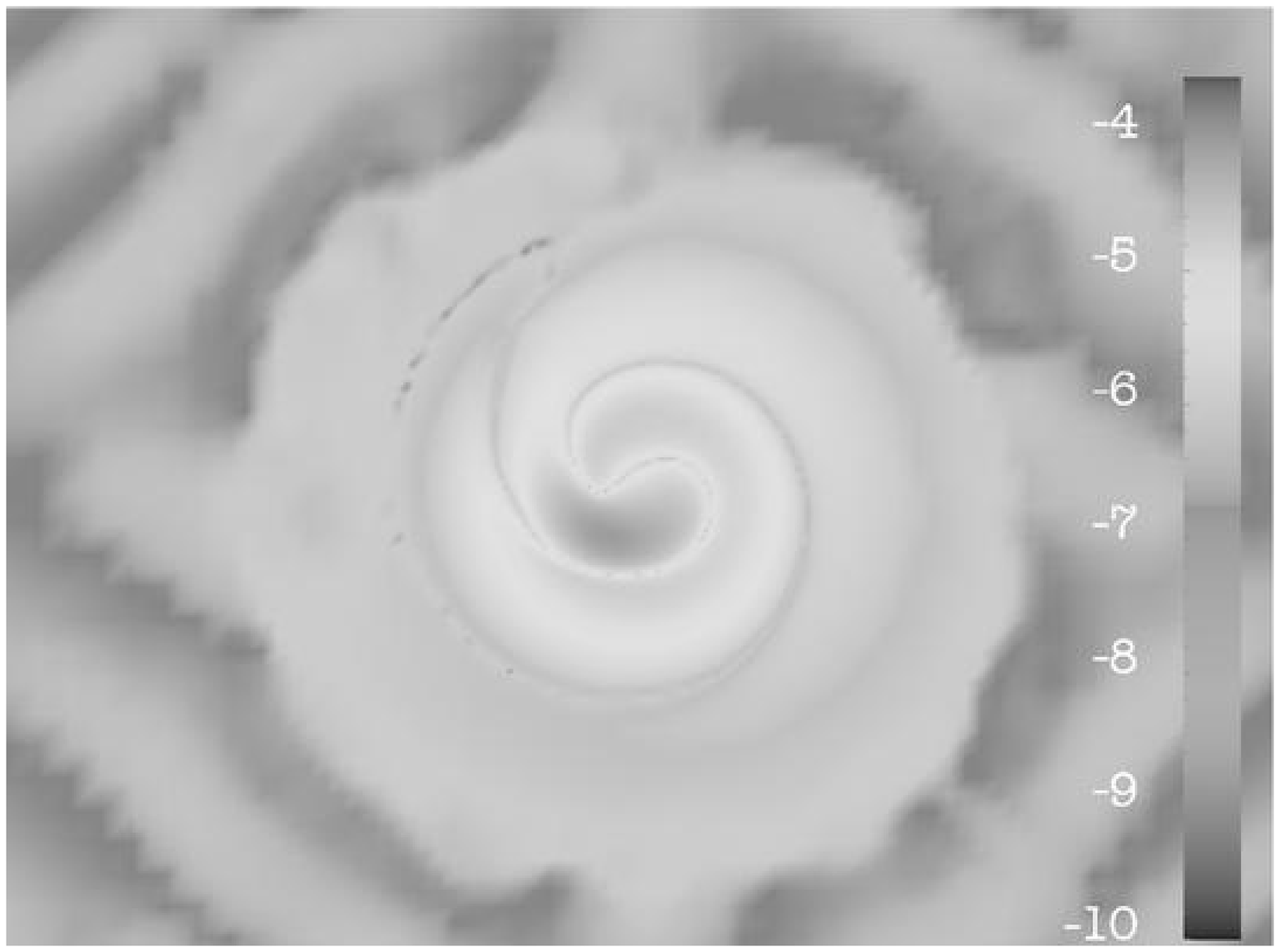} \\
\end{tabular}
\caption{Contour plot of the function $\log_{10} (|\rho(x,y,z,t) - 
\rho_0(x,y,z)| + \epsilon)$ in the equatorial plane, where $\rho$ denotes the rest-mass density
of the $m = 1$ evolution shown in Fig.~\ref{fig:ref_64_m1_rho},
$\rho_0$ denotes the the density function of the equilibrium polytrope, 
and $\epsilon$ is a small number ($\epsilon = 10^{-10}$ in this plot).}
\label{fig:ref_64_m1_rho_diff}
\end{figure}

\begin{figure}
\begin{center}
\includegraphics[width=\columnwidth]{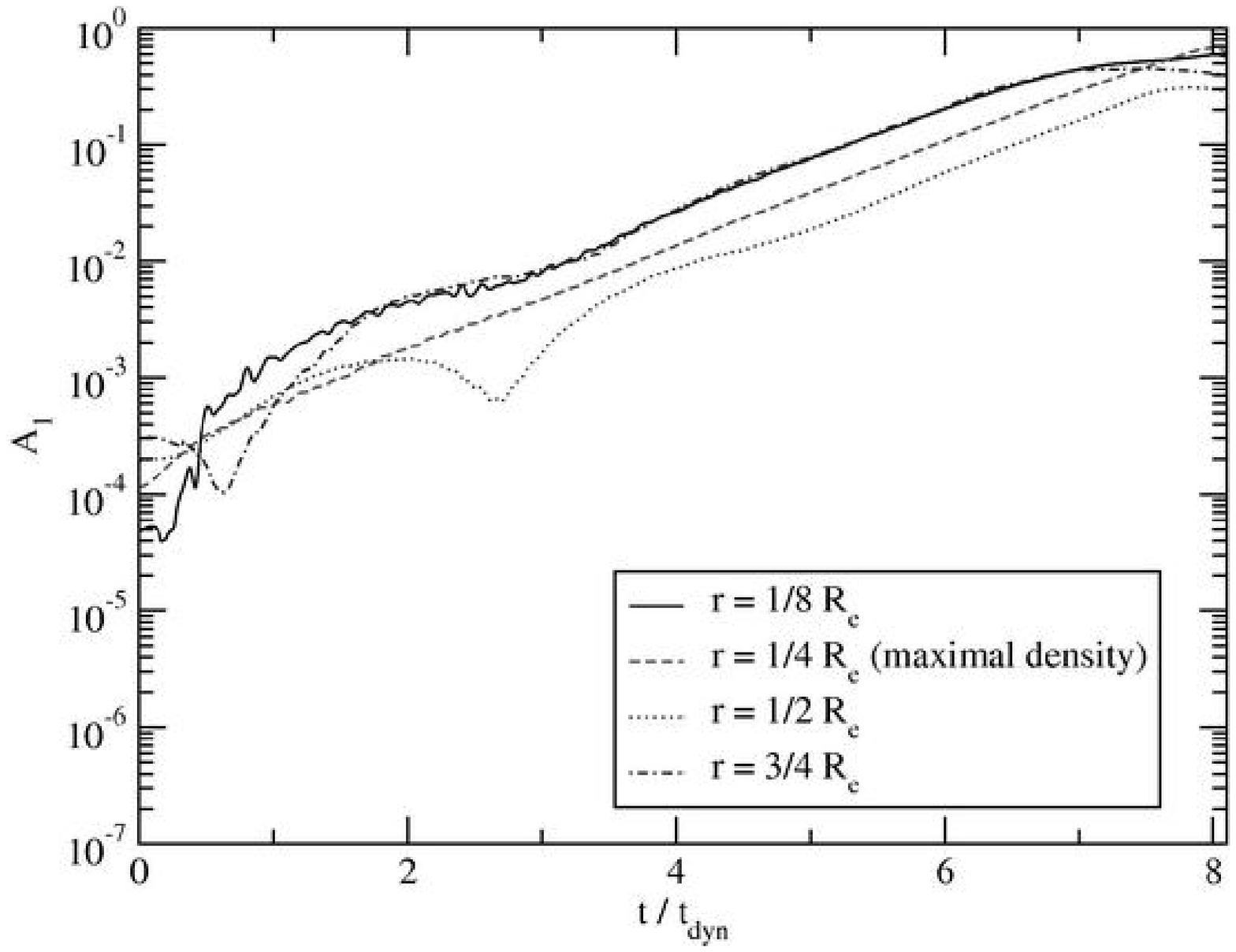}
\includegraphics[width=\columnwidth]{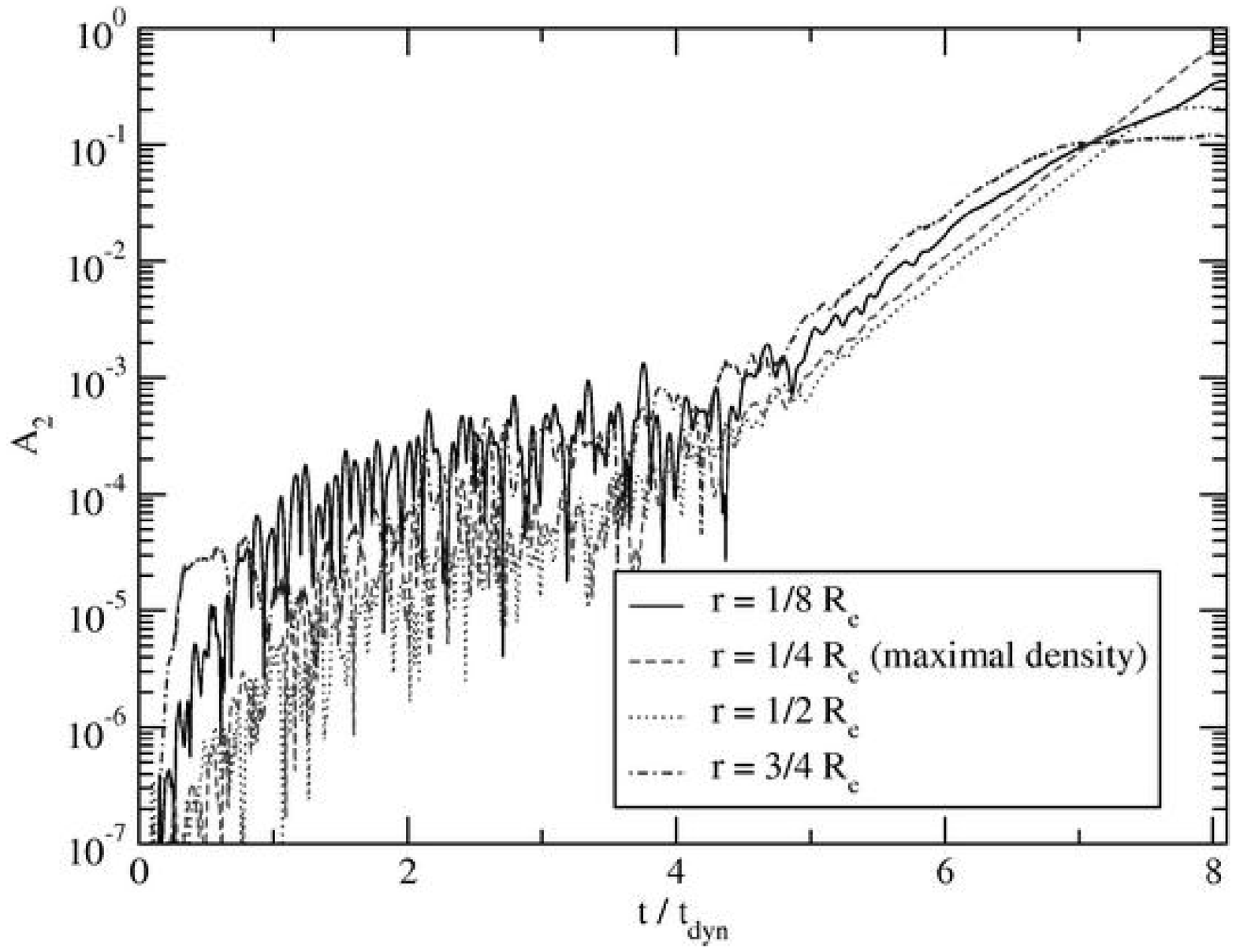}
\end{center}
\caption{Evolution of the mode amplitudes $A_1$ and $A_2$ in the reference polytrope perturbed
by $\lambda_m = \delta_{m1}$, for different mode extraction radii. The radius of highest initial 
density is indicated.}
\label{fig:m1_radii} 
\end{figure}

\begin{figure}
\begin{center}
\includegraphics[width=\columnwidth]{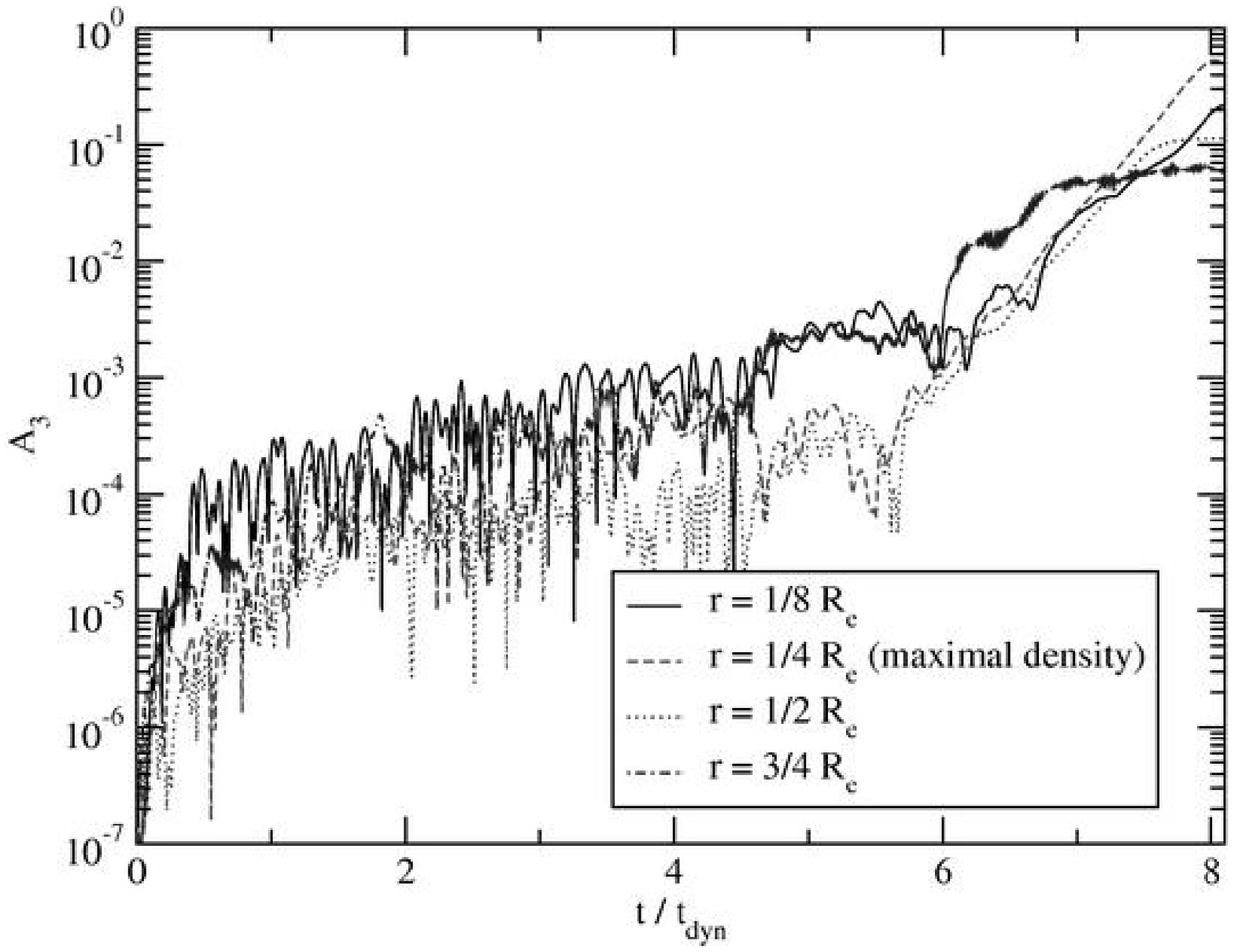}
\includegraphics[width=\columnwidth]{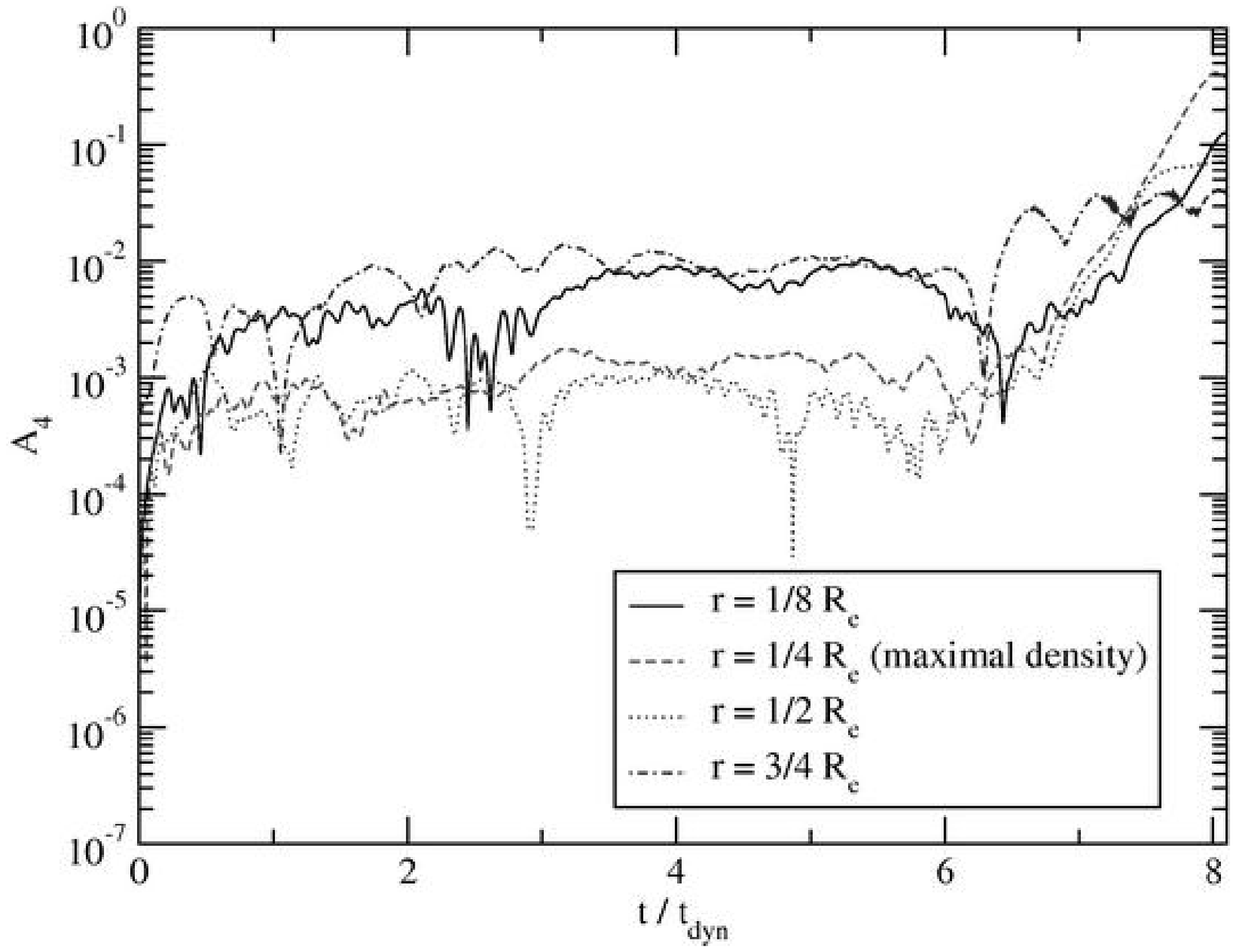}
\end{center}
\caption{Same as Fig.~\ref{fig:m1_radii}, but for the mode amplitudes $A_3$ and $A_4$.}
\label{fig:m3_radii} 
\end{figure}

\begin{figure}
\begin{center}
\includegraphics[width=\columnwidth]{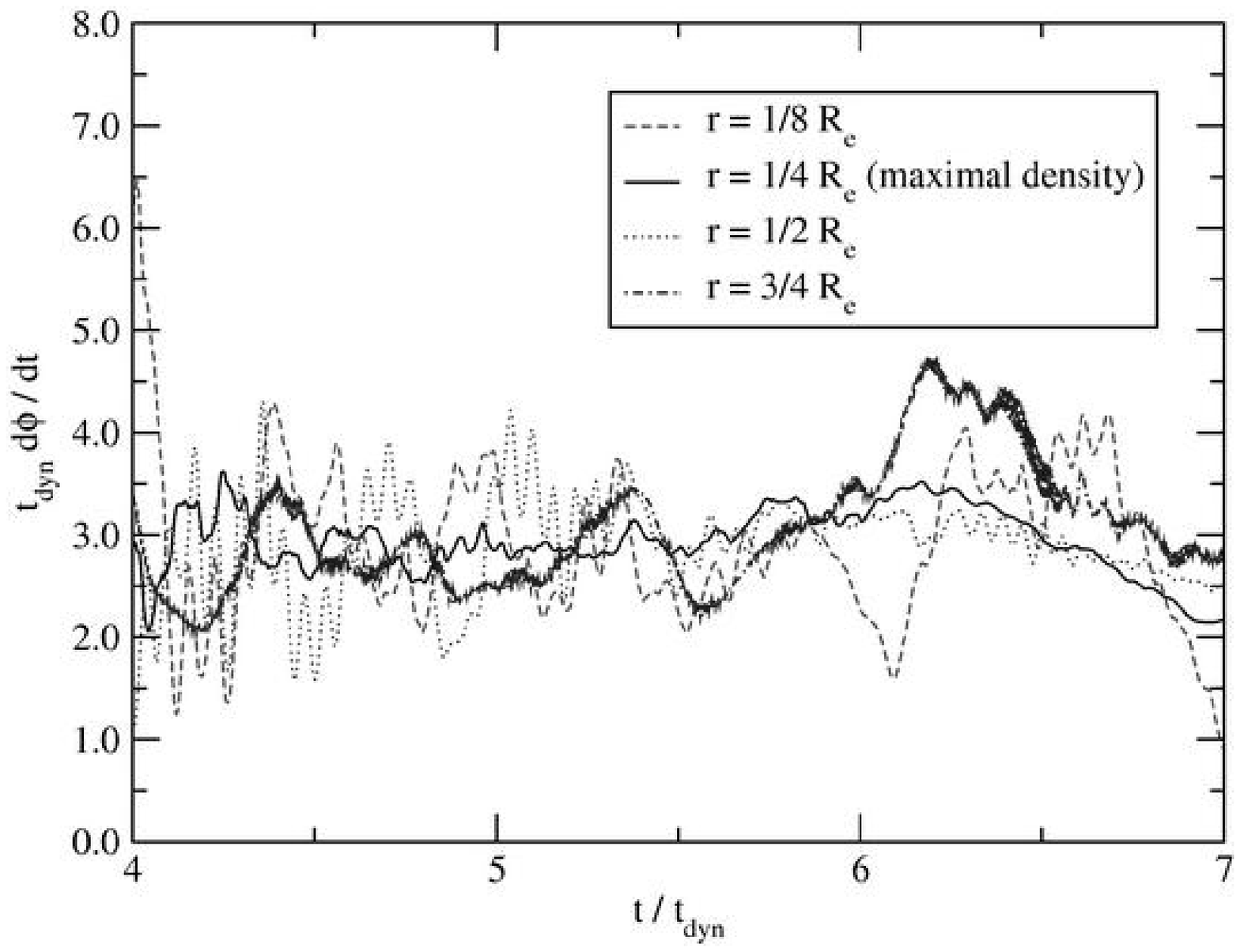}
\includegraphics[width=\columnwidth]{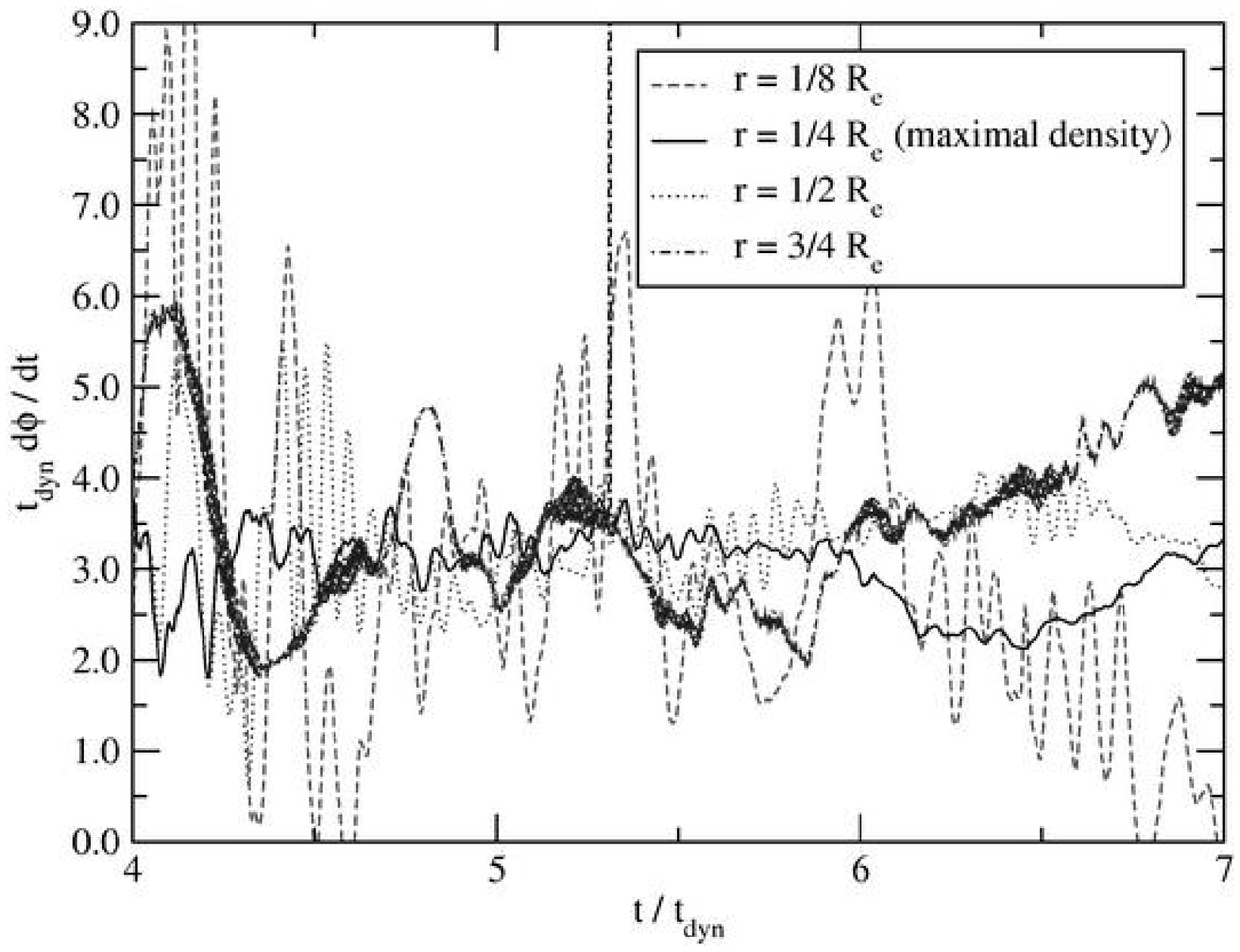}
\end{center}
\caption{Local mode frequency $t_{dyn} d\phi/dt$ for the modes $m = 1$ (top)
and $m = 2$ (bottom) in the reference polytrope, extracted at 
different radii in the equatorial plane.}
\label{fig:m1_radii_phase} 
\end{figure}

While we do not focus on the global nature of the quasi-normal modes of 
general relativistic quasi-toroidal polytropes here, this
section will give some indication about the structure of the instability. 
Consider Fig.~\ref{fig:ref_64_m1_rho_diff}
which displays the equatorial distribution of the function 
$\log_{10} (|\rho(x,y,z,t) - \rho_0(x,y,z)| + \epsilon)$, where 
$\rho_0$ is the density function of the equilibrium polytrope
and $\epsilon = 10^{-10}$. These logarithmic difference plots exhibit
the node lines of the unstable mode corresponding to $m = 1$ in the equatorial plane, and
show the spiral-arm structure of the fragmentation instability.

The mode amplitudes at different extraction radii in the equatorial
plane are shown, for a perturbation with $\lambda_m = \delta_{m1}$, 
in Fig.~\ref{fig:m1_radii} and \ref{fig:m3_radii}. The development of the
unstable modes is not very sensitive to the extraction radius, at least
as long as the amplitude of the dominant mode is not close to unity. 
Fig.~\ref{fig:m1_radii_phase} suggests that the local mode frequency does
not depend strongly on the radius, at least within the considerable uncertainties 
of the plot.


\subsubsection{The location of the unstable modes in the corotation band}
\label{sec:reference_corot}

\begin{figure}
\begin{center}
\includegraphics*[width=\columnwidth]{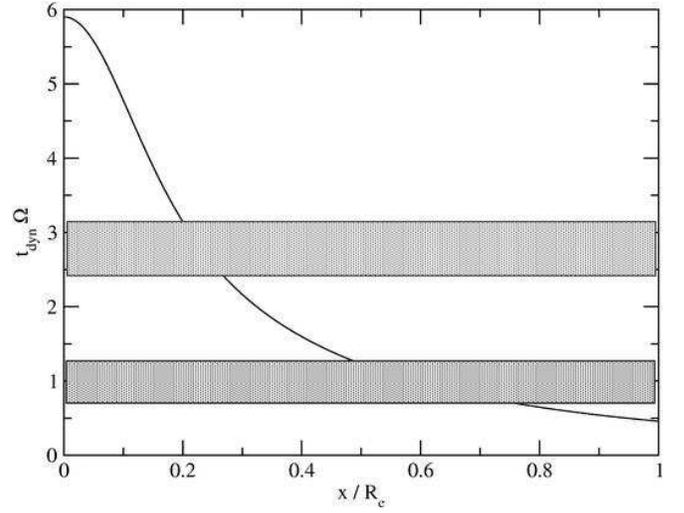}
\end{center}
\caption{Angular velocity of the reference polytrope over the $x$ axis
(black line), and approximate location (with error bar) of the pattern speed of the $m = 1$ mode 
(upper rectangle), and the $m = 2$ mode (lower rectangle). Both modes are inside the corotation
band.}
\label{fig:ref_64_corot} 
\end{figure}

To determine the location of the instability with respect to the corotation band,
we define a coordinate angular velocity of the initial model by 

\be
\Omega(\varpi) \equiv \alpha v^\phi - \beta^\phi.
\ee

This can be compared to the mode pattern speed $1/m \, d\phi / dt$,
which we will assume to be valid for the whole star (cf. Fig.~\ref{fig:m1_radii_phase}),
to determine whether a certain mode has a corotation point. In 
Fig.~\ref{fig:ref_64_corot}, the angular velocity is plotted in addition
to the numerical approximation of the location of the $m = 1$ and $m = 2$ pattern speeds.
We find that both modes have corotation points: the $m = 1$ mode near 
the radius of highest density at $0.25 \, R_e$, and the $m = 2$ mode near 
$0.5 \ldots 0.6 \, R_e$.


\subsubsection{Grid resolution and convergence}

\begin{figure}
\begin{center}
\includegraphics[width=\columnwidth]{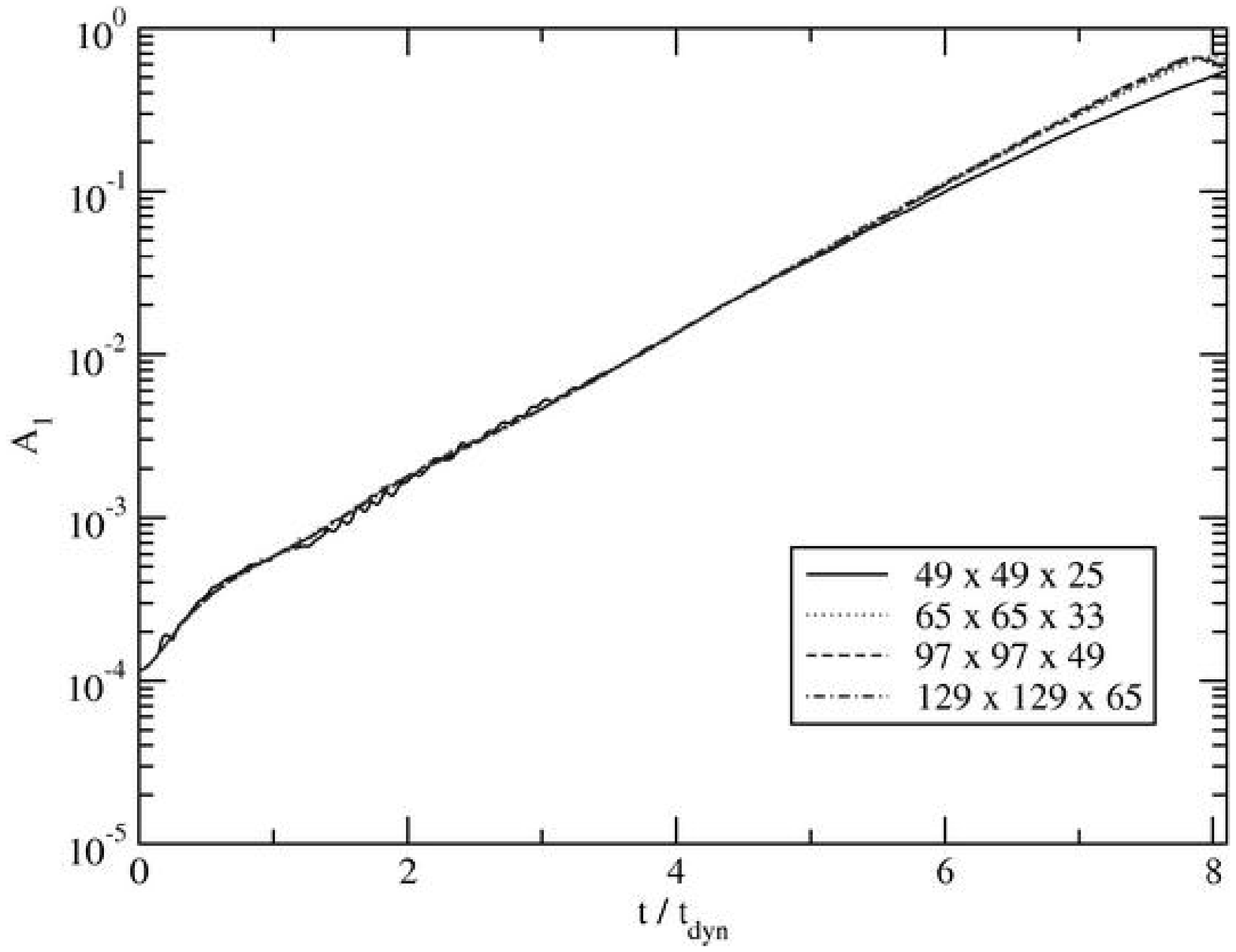}
\includegraphics[width=\columnwidth]{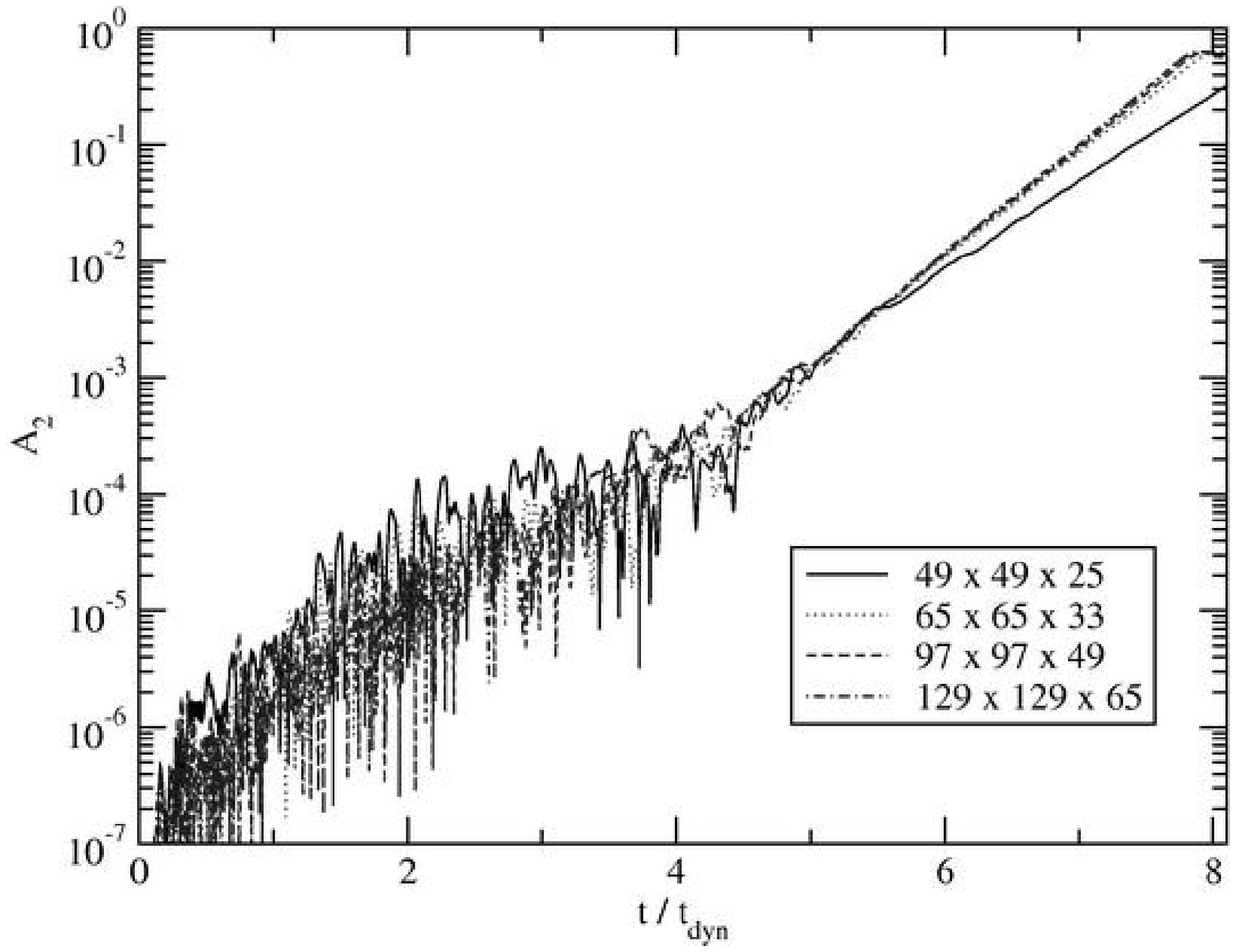}
\end{center}
\caption{Evolution of the mode amplitudes $A_1$ (top) and $A_2$ (bottom) for different grid resolutions. 
The grid sizes in the legend refer to the four outer grid patches; the innermost 
patch covering the high-density central toroidal region of the star has a resolution of 
$65  \times 65  \times 33$, 
$97  \times 97 \times 49$, $129 \times 129  \times 65$ or $193 \times 193 \times 97$, 
correspondingly.}
\label{fig:convergence_m1} 
\end{figure}

\begin{figure}
\begin{center}
\includegraphics[width=\columnwidth]{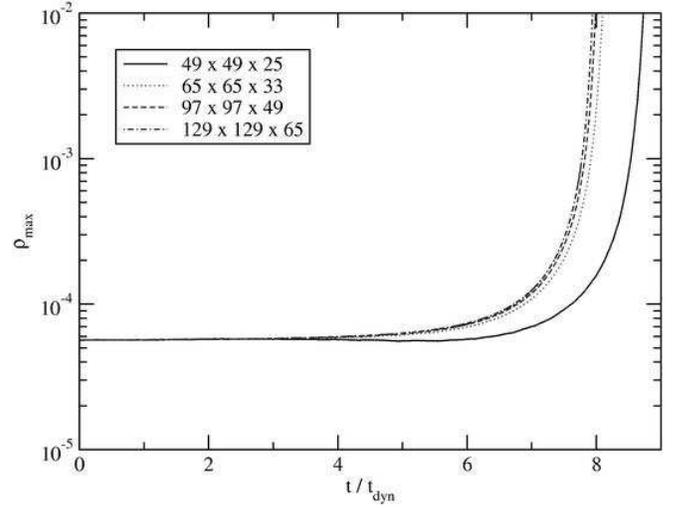}
\end{center}
\caption{Same as Fig.~\ref{fig:convergence_m1}, but for the evolution of the maximum of the rest-mass
density $\rho_{max}$.}
\label{fig:convergence_rho_max} 
\end{figure}

\begin{figure}
\begin{center}
\includegraphics[width=\columnwidth]{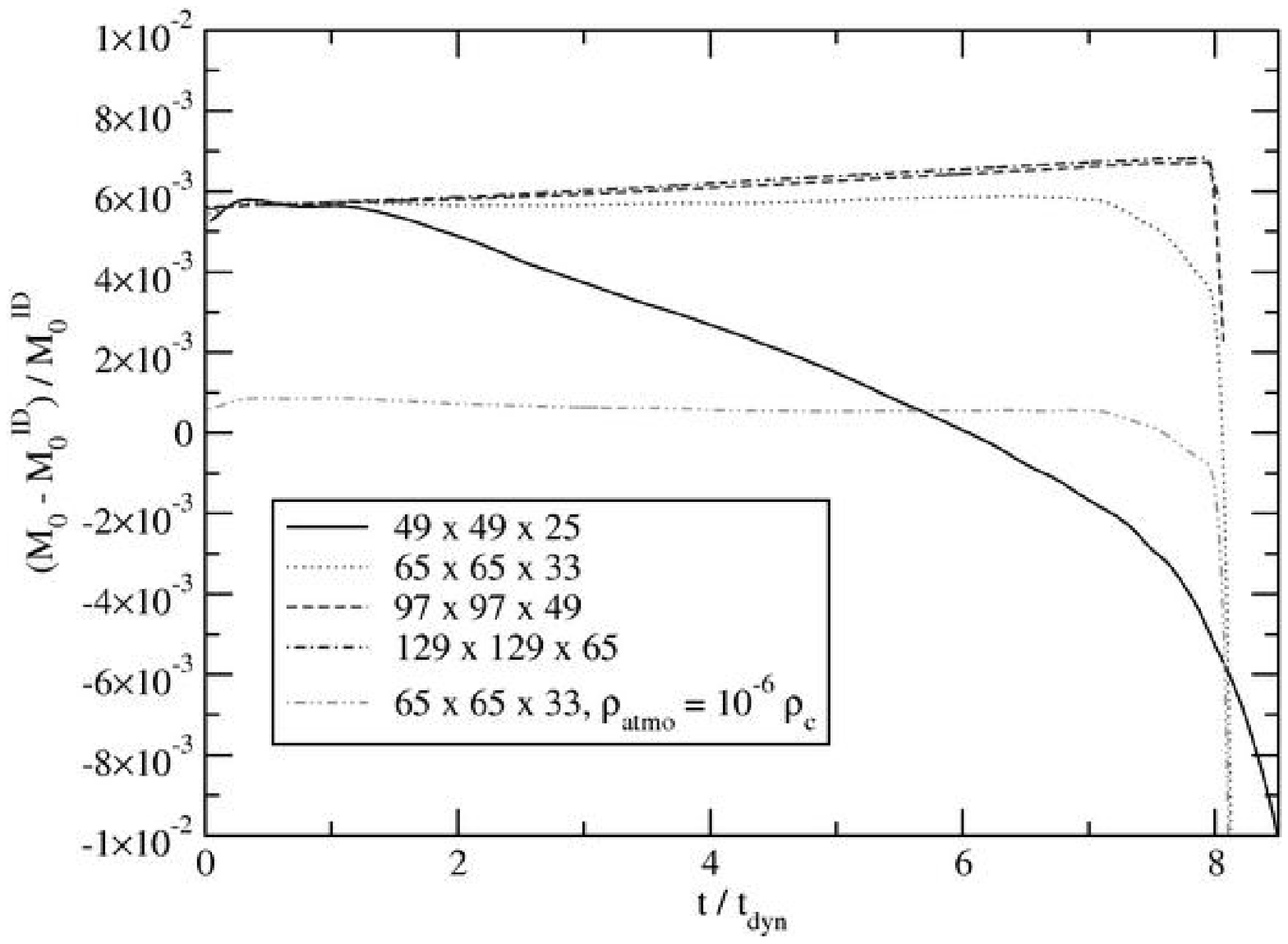}
\end{center}
\caption{Same as Fig.~\ref{fig:convergence_m1}, but for the evolution of the total rest mass
$M_0$. The values are normalized to the initial rest mass ${M_0}^{ID}$
  obtained from the initial data solver. The last graph is from a simulation with a less dense artificial atmosphere.}
\label{fig:convergence_rest_mass} 
\end{figure}

For any parameter study with numerical methods, it is important to have an understanding of the
amount of grid resolution needed to extract the physical features under consideration. A typical
way to gauge this is to evolve a system with different resolutions and to compare the results. 
For the black-hole forming fragmentation 
instability shown in Section~\ref{sec:reference_general}, it is expected that different phases
of the evolution have substantially different resolution requirements. During the nearly exponential
growth of the instability at low amplitudes (which we will call \emph{linear regime}), 
the equilibrium structure of the star as a whole needs to be covered appropriately. The instability 
is, at first, a low-frequency effect on the star, and as such it is not expected to dominate the resolution
requirements. However, if the fragment evolves into a black hole (in the
\emph{non-linear regime}), it needs to be resolved with significantly more grid points.

As explained in Section~\ref{sec:methods}, the star is covered by a grid with
fixed mesh refinement. 
Typically, five grid patches are used,
centered on each other, and with an increase of resolution by a factor 2 each. Only the central patch
with highest resolution,
which covers the region of highest density, is 0.75 times as extended as the second finest one
to reduce artefacts from inter-patch boundaries. To test convergence,
the reference model has been evolved with  $49 \times 49 \times 25$ (grid spacing $h \approx 0.29 M$), 
$65  \times 65  \times 33$ ($h \approx 0.22 M$), 
$97  \times 97 \times 49$  ($h \approx 0.15 M$) and $129 \times 129  \times 65$ ($h \approx 0.11 M$) 
zones per outer grid patch; the
innermost patch covering the high-density toroidal region has $65  \times 65  \times 33$, 
$97  \times 97 \times 49$, $129 \times 129  \times 65$ or $193 \times 193 \times 97$ grid zones. Also,
the initial data was calculated with a grid of $401 \times 201$, $601 \times 301$,
$1201 \times 601$ and $2401 \times 1201$ zones.

The results from evolving the reference model with an $m = 1$ perturbation at different resolutions
are shown in Figs.~\ref{fig:ham_convergence}, \ref{fig:convergence_m1}, \ref{fig:convergence_rho_max} 
and \ref{fig:convergence_rest_mass}. The convergence of the Hamiltonian constraint has already been
discussed in Section~\ref{sec:quasi_toroidal_polytropes}.
The time development of the mode amplitudes has a noticeably different growth
only at the lowest resolution. 
The evolution of the maximum of the rest-mass
density (Fig.~\ref{fig:convergence_rho_max}) exhibits a similar behaviour. Finally, the total
rest mass of the system is conserved from within $1.4\%$ (lowest resolution) to $0.1\%$ (higher
resolutions). The drift in the rest-mass can be explained by our use of an artificial atmosphere: The rest-mass
density of the atmosphere is $10^{-5} \rho_c = 3.38 \cdot 10^{-11}$, which corresponds to an
approximate total mass of $M_{0,atmo} \approx 3.8 \cdot 10^{-3}$ in a domain of coordinate volume $1040^3$ 
(not taking into account the volume form). This translates into a systematic shift in the total
rest mass of the system as apparent when comparing to an evolution with a lower atmospheric density
(bottom panel in Fig.~\ref{fig:convergence_rest_mass}), and a drift caused by the intrinsic atmospheric
dynamics and the interaction with the outer boundary. Note that this is to be considered
a (non-sharp) upper limit on the systematic errors induced by the atmosphere, because one could always extend
the total computational domain arbitrarily without affecting the core region significantly. 

Judging from the resolution study in Figs.~\ref{fig:ham_convergence}, \ref{fig:convergence_m1}, 
\ref{fig:convergence_rho_max} and \ref{fig:convergence_rest_mass}, we think that the lowest resolution 
considered here (which already covers the
equatorial radius of the star with about 60 grid points when comparing to uniform grids) is sufficient
to get qualitatively correct results. Quantitatively, the errors in rest mass are in a range of a few percent. 
The next highest resolution
of $65  \times 65  \times 33$ ($h \approx 0.22 M$) seems accurate to within about one percent. This resolution 
will therefore
be used for the parameter study below. This is reasonable since the structure of the quasi-toroidal models 
has similar features, and therefore similar requirements concerning resolution. Nevertheless, selected
models have been tested for convergence independently from the reference polytrope.

\subsubsection{Influence of the artificial atmosphere}

\begin{figure}
\begin{center}
\includegraphics[width=\columnwidth]{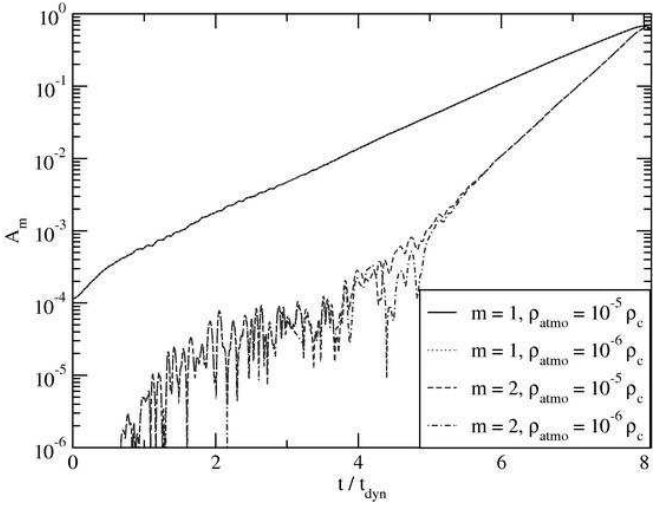}
\end{center}
\caption{Evolution of the mode amplitudes $A_1$ and $A_2$ in the reference polytrope, for
artificial atmospheres of different density $\rho_{atm} / \rho_c$.}
\label{fig:m1_atmo_comparison} 
\end{figure}

The standard artificial atmosphere we employ in our simulations has a density several orders of
magnitude lower than the average density in the star, so we expect that it does not influence
the dynamical properties of the star significantly. The atmospheric density is set in terms of the
central density of the star: we have used a ratio of $10^{-5}$ in most simulations. 
To test the influence of this parameter, we have evolved the reference model
also with a ten times lower atmospheric density (i.e. $10^{-6} \rho_c$), and with
an $m = 1$ perturbation. The results are shown in Fig.~\ref{fig:convergence_rest_mass}  
and \ref{fig:m1_atmo_comparison}. The latter shows that the dominant $m = 1$ mode 
is not influenced by the atmospheric setting, while the $m = 2$ amplitude shows dependence on the 
atmosphere setting only as long as its amplitude is on the level of the numerical noise.

\subsection{Evolution of the sequence of axis ratios}

\begin{figure}
\begin{center}
\includegraphics[width=\columnwidth]{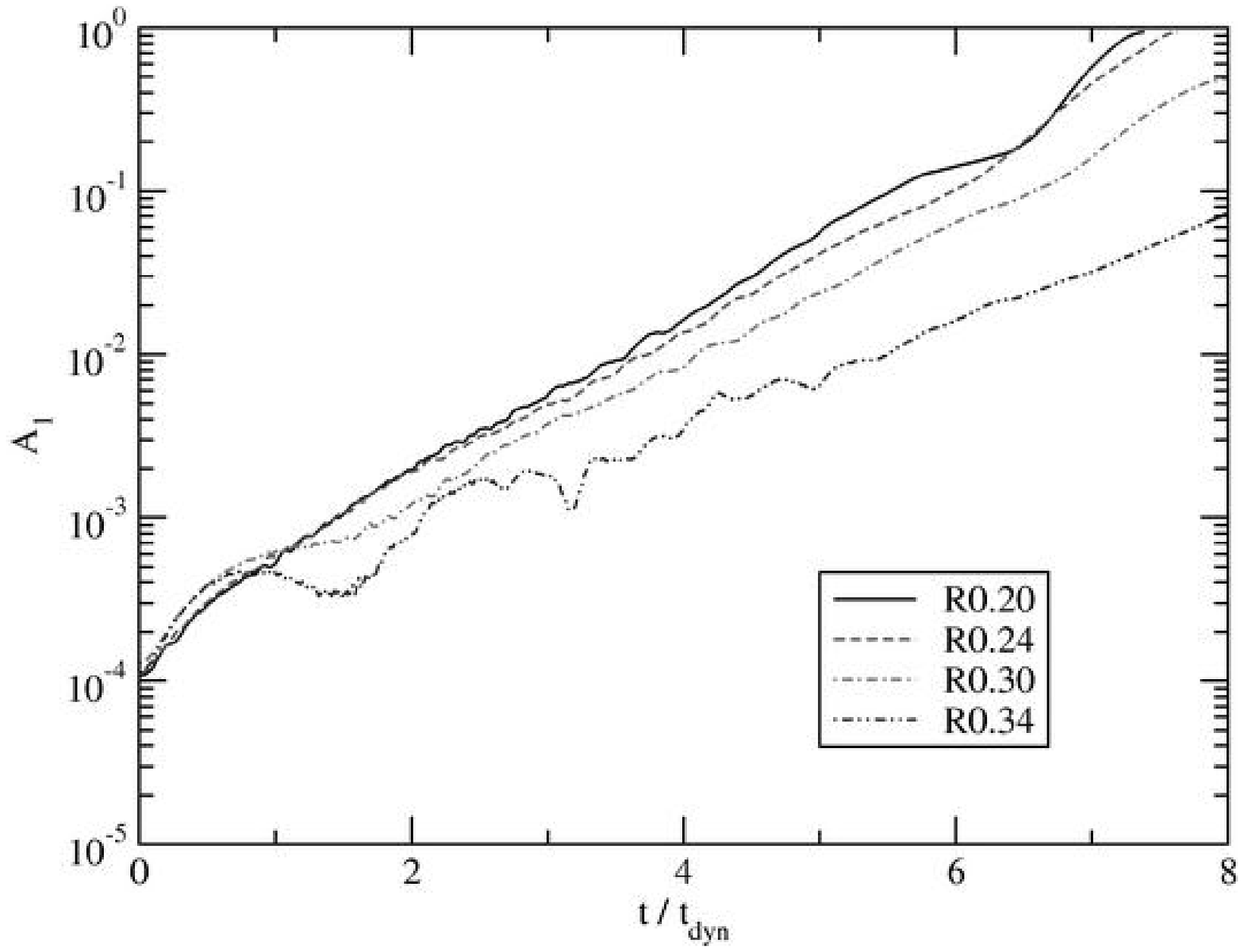}
\includegraphics[width=\columnwidth]{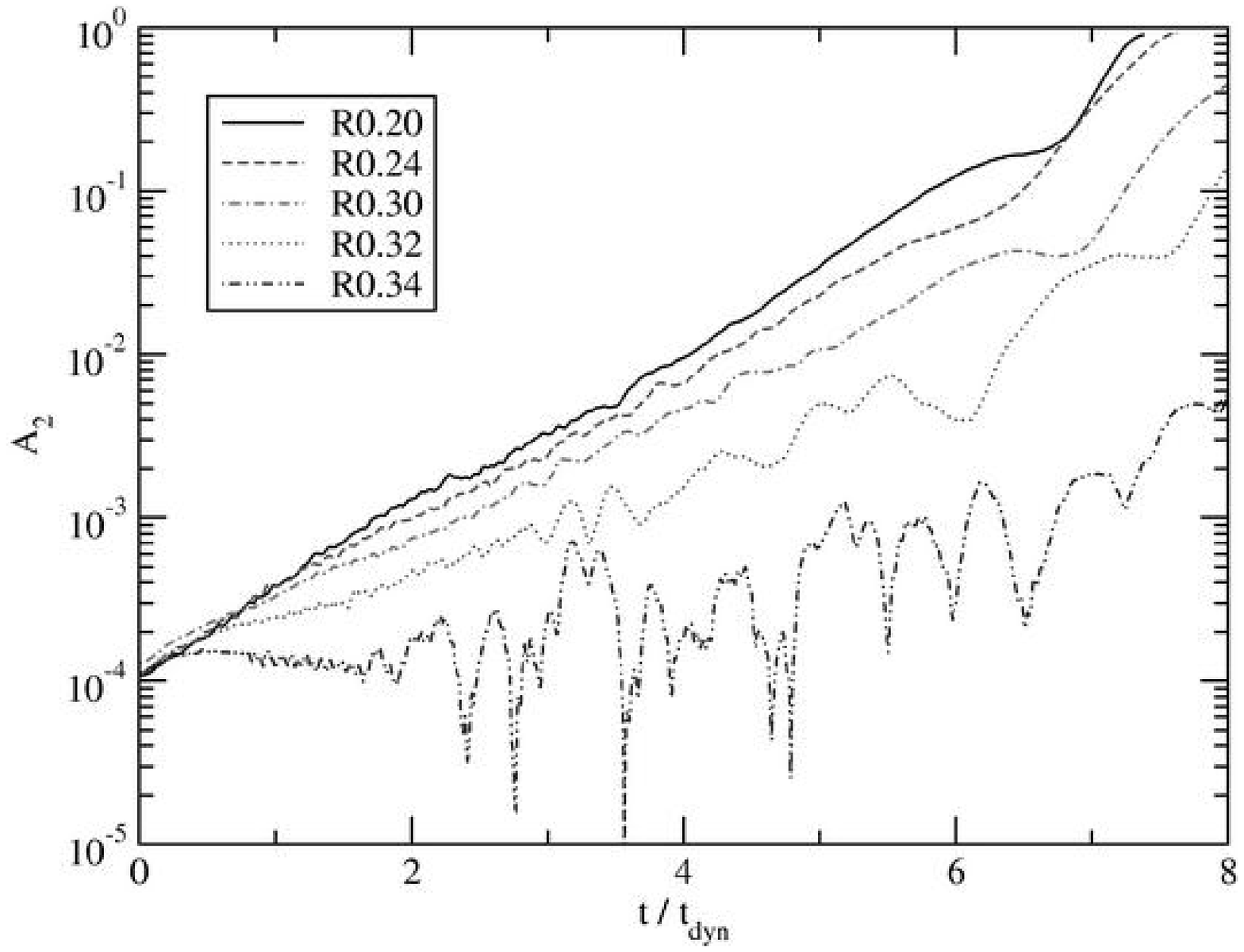}
\end{center}
\caption{Evolution of the mode amplitudes $A_1$ and $A_2$ for different members of the
\emph{R} sequence (cf. Table~\ref{tab:sequence_ratio}).}
\label{fig:ratio_m1_m2}
\end{figure}

The \emph{R} sequence has been described in Section~\ref{sec:id_ratio}. From
Table~\ref{tab:sequence_ratio}, it is apparent that higher values of $r_p/r_e$
are connected to lower $T/|W|$. In Maclaurin spheroids, this is related to a
stabilization of the initial model. Consider Fig.~\ref{fig:ratio_m1_m2}: The
growth time of the modes $m = 1$ and $m = 2$ increases with lower $r_p/r_e$, which
indeed is a sign for approaching a limit of stability. 

\subsection{Evolution of the sequence of stiffnesses}

\begin{figure}
\begin{center}
\includegraphics[width=\columnwidth]{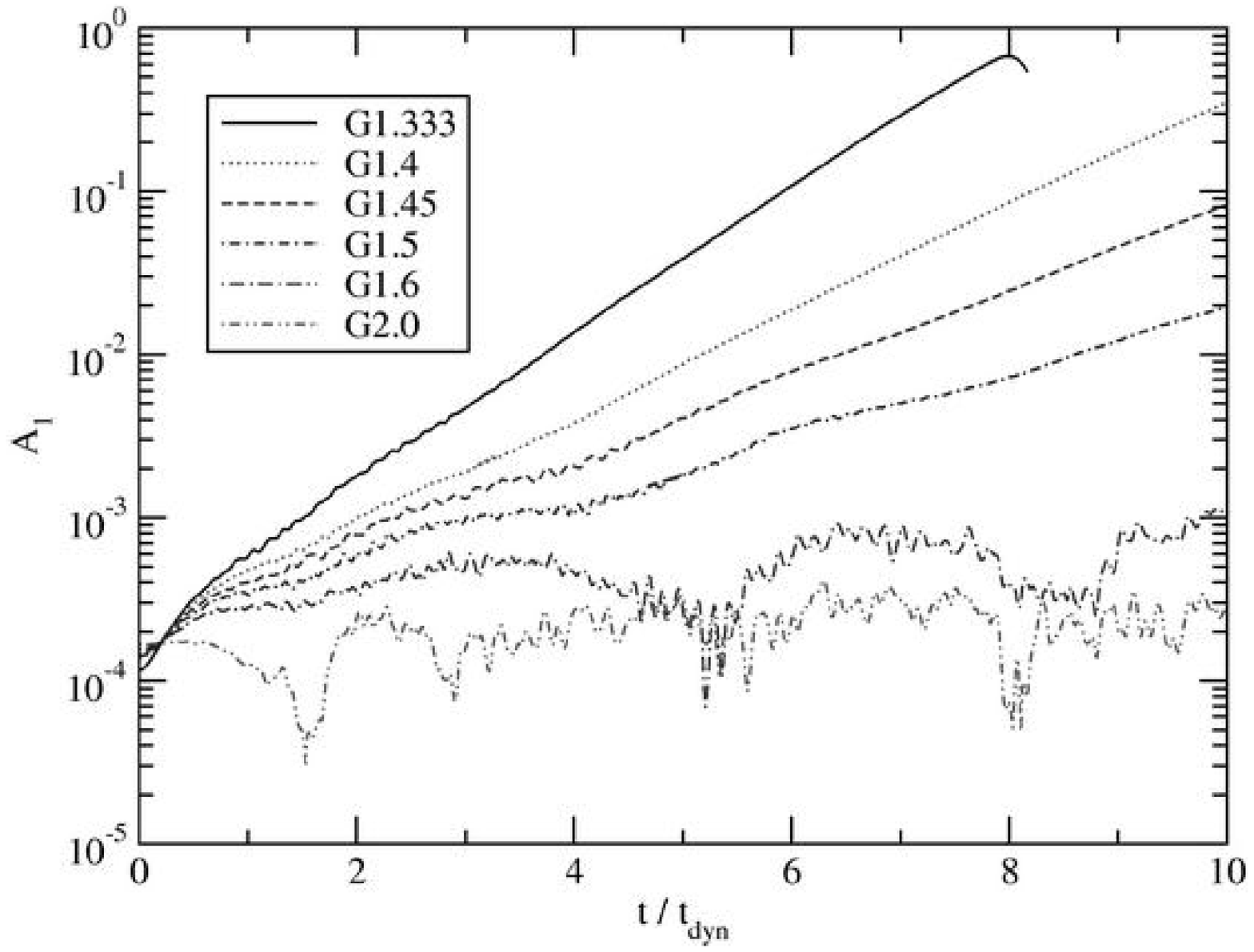}
\includegraphics[width=\columnwidth]{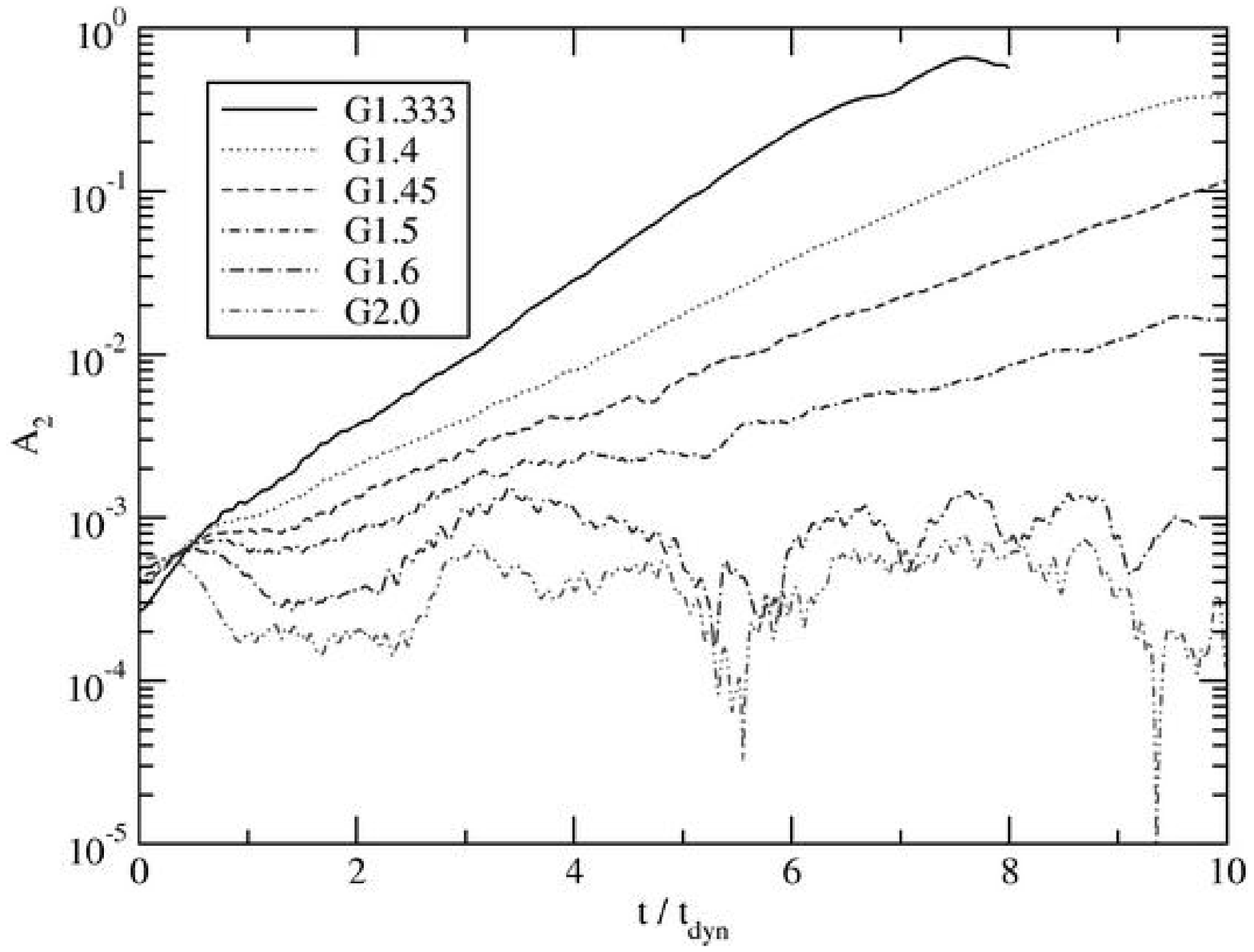}
\end{center}
\caption{Evolution of the mode amplitudes $A_1$ and $A_2$ for different members of the
\emph{G} sequence (cf. Table~\ref{tab:sequence_gamma}).}
\label{fig:gamma_m1_m2}
\end{figure}

The change in the instability along the \emph{G} sequence described in Section~\ref{sec:id_gamma}
is shown in Fig.~\ref{fig:gamma_m1_m2}. With increasing $\Gamma$, and $T/|W|$ decreasing
from $0.227$ to $0.159$ (cf. Table~\ref{tab:sequence_gamma}), both the $m = 1$ and the 
$m = 2$ modes are stabilized. The member \emph{G2.0} with $\Gamma = 2$ is of special
interest, since this choice is often used to obtain a simple polytropic equilibrium model
of neutron stars. While it is known that strong differential rotation can induce
bar-mode instabilities in neutron stars \cite{Shibata:2000jt, Saijo:2000qt, Baiotti06b}, the particular
model $G2.0$ does not appear to be $m = 2$ unstable (note, however, the limitations of our method 
to determine stability expressed in Section~\ref{sec:instability_slow}).

\subsection{Evolution of the sequence of compactnesses}
\label{sec:evol_comp}

\begin{figure}
\begin{center}
\includegraphics[width=\columnwidth]{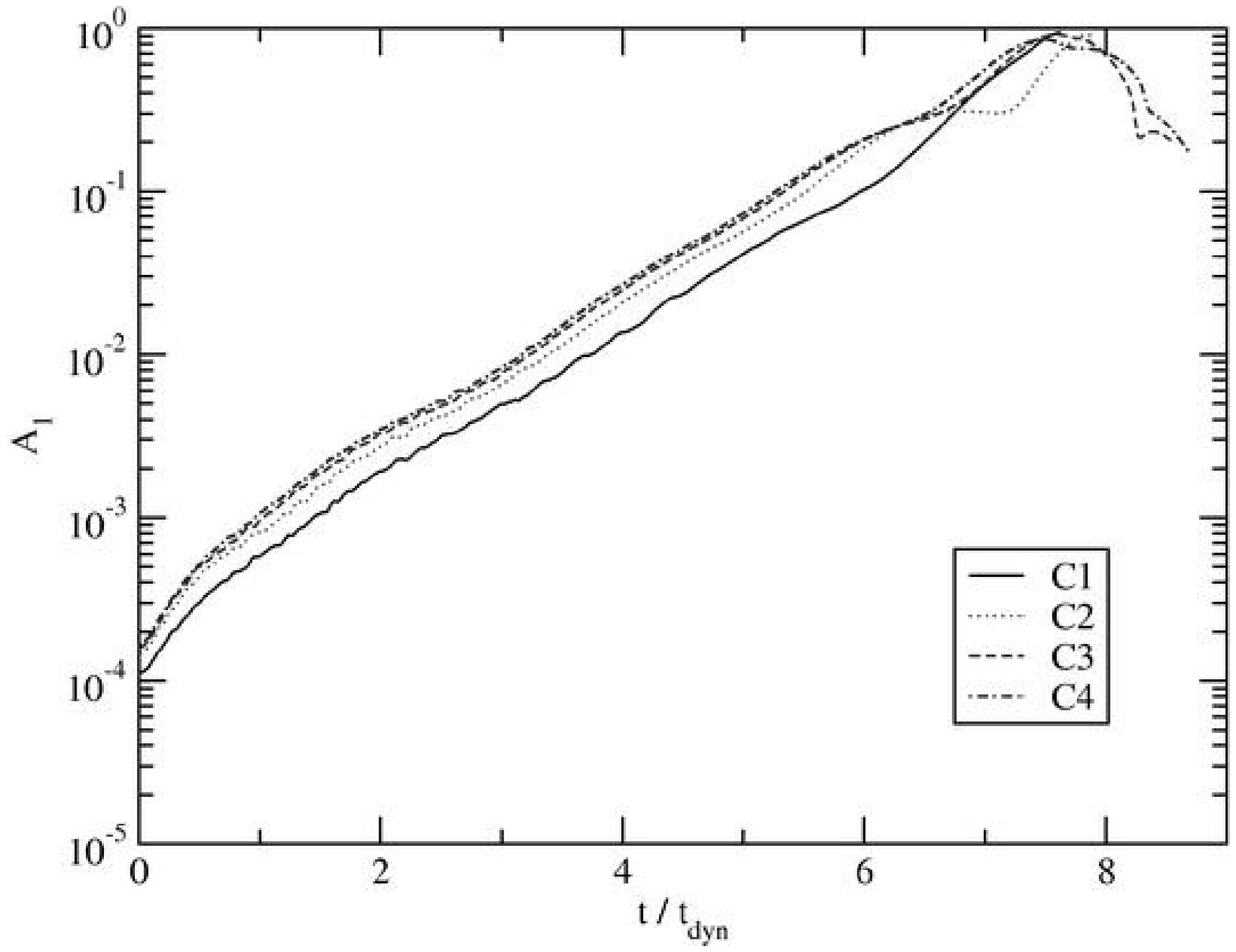}
\includegraphics[width=\columnwidth]{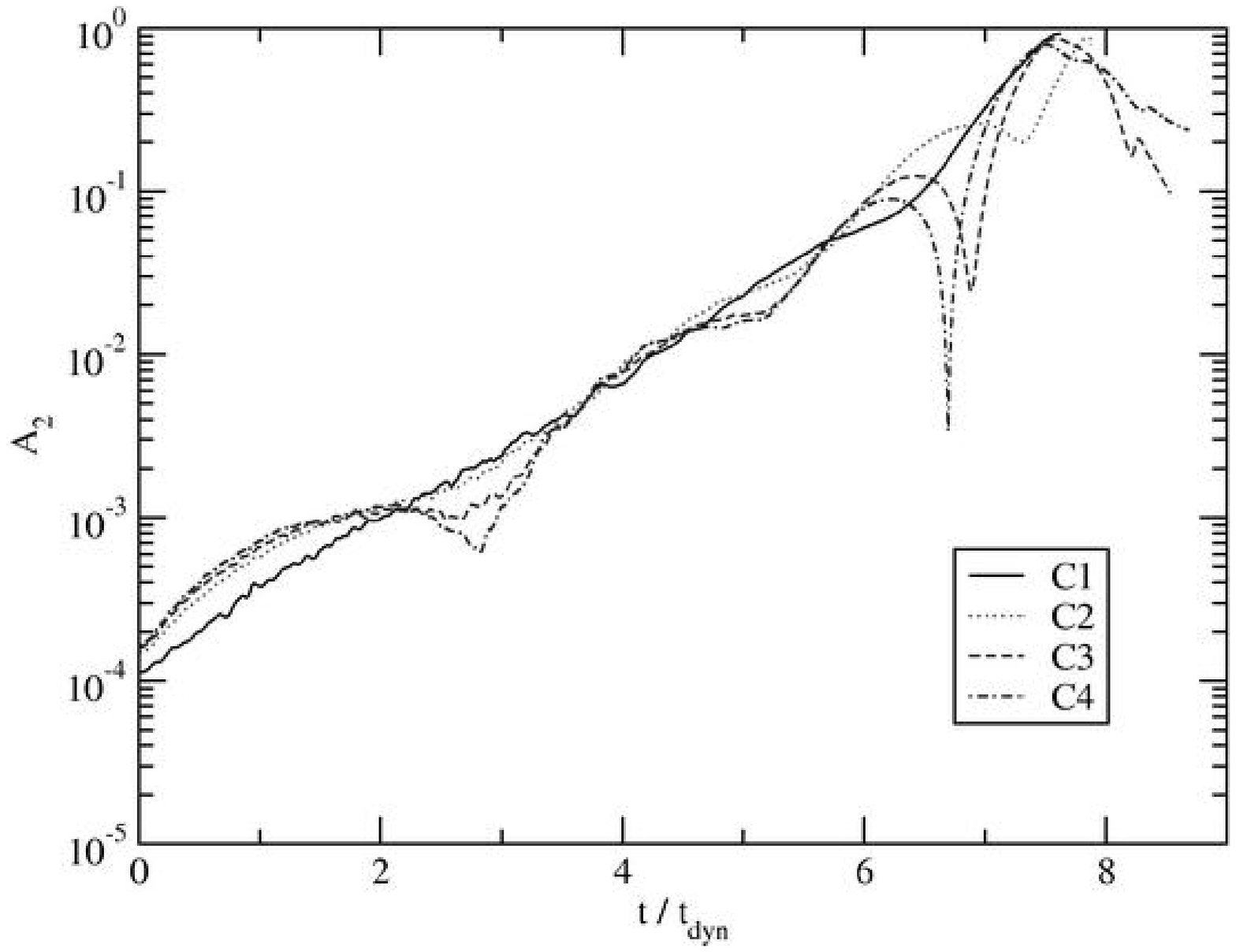}
\end{center}
\caption{Evolution of the mode amplitudes $A_1$ and $A_2$ for different members of the
\emph{C} sequence (cf. Table~\ref{tab:sequence_comp}).}
\label{fig:comp_m1_m2}
\end{figure}

\begin{figure}
\begin{center}
\includegraphics[width=\columnwidth]{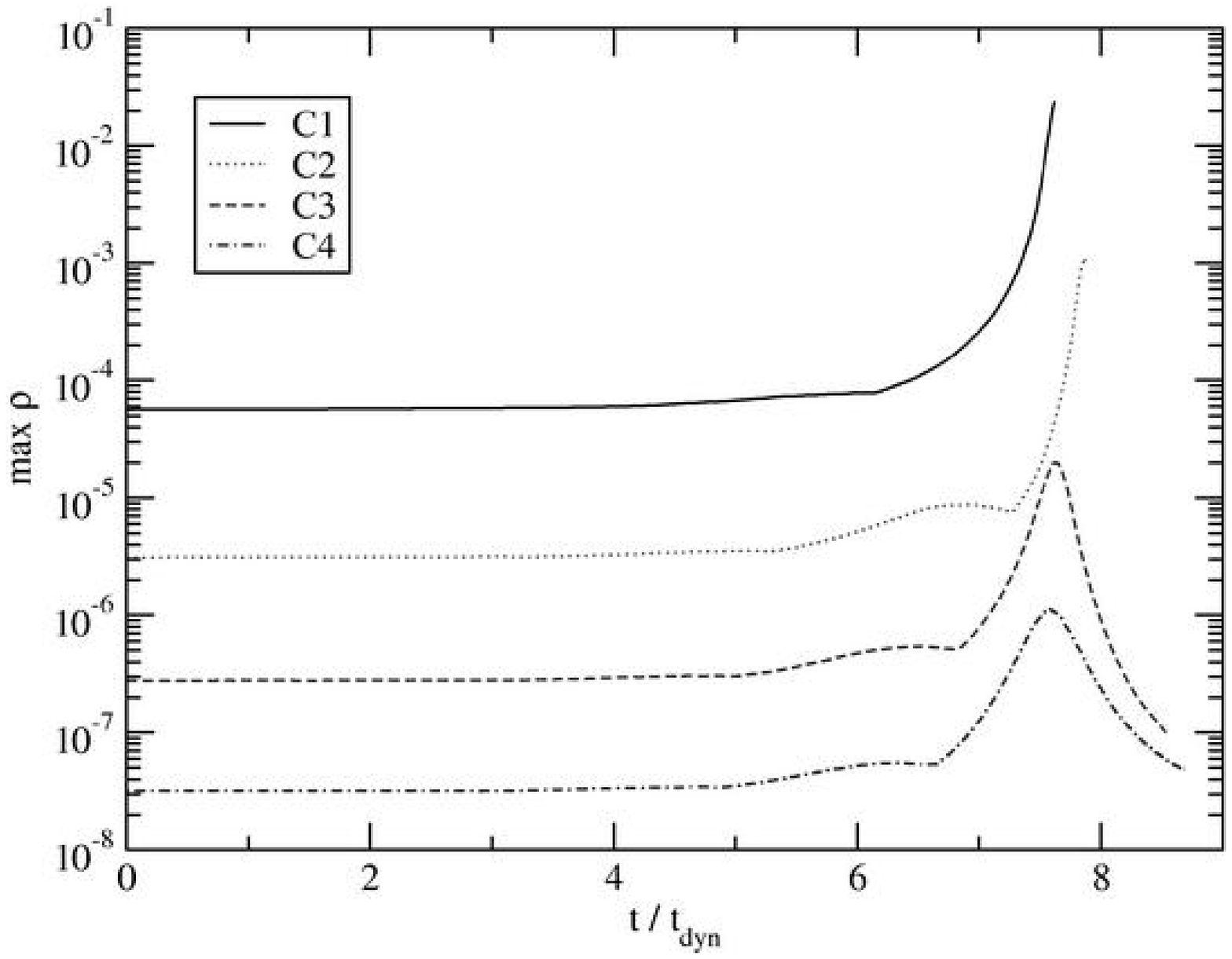}
\includegraphics[width=\columnwidth]{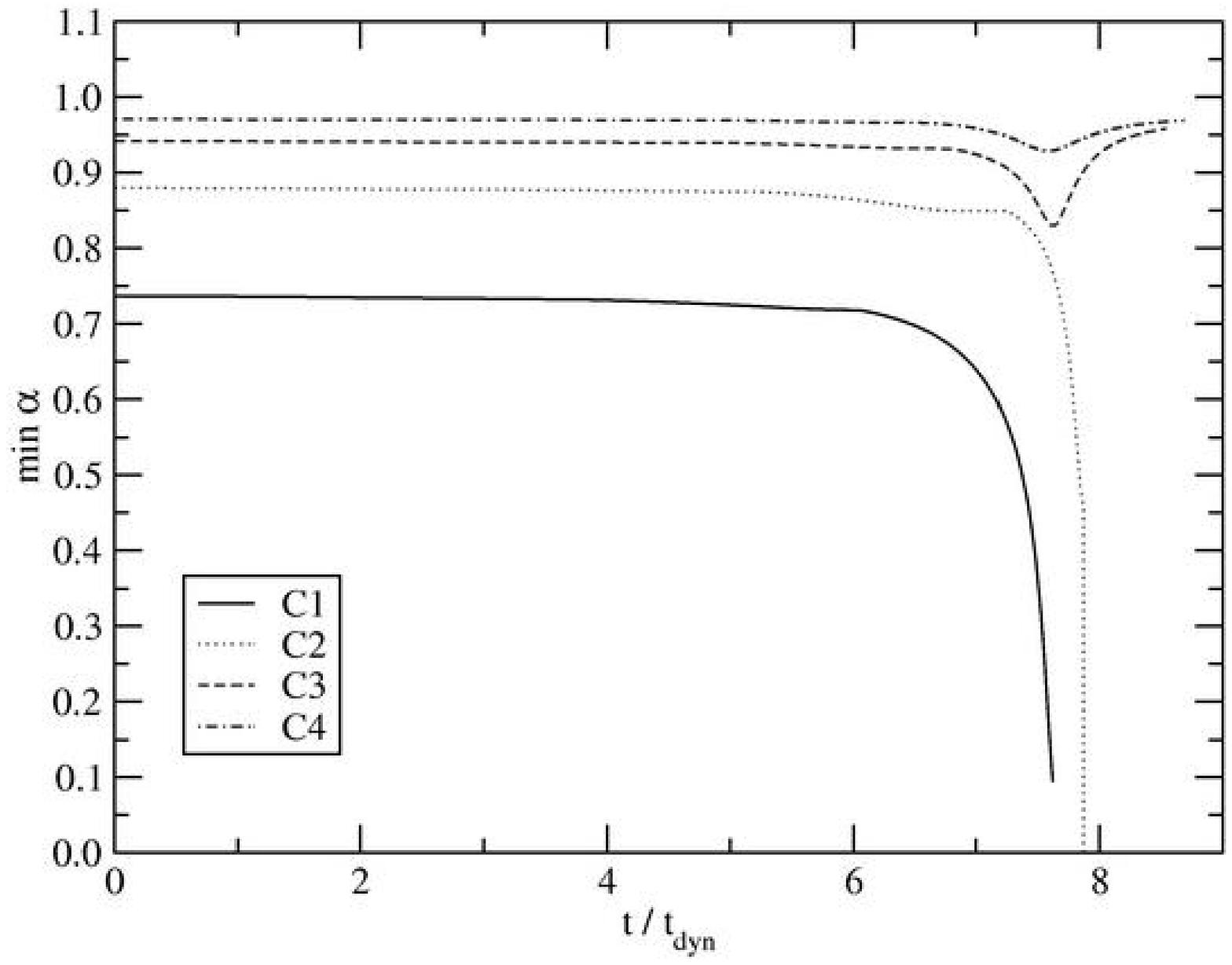}
\end{center}
\caption{Evolution of the maximum of the rest-mass density $\rho$ and the minimum
of the lapse function $\alpha$ for different members of the \emph{C} sequence.}
\label{fig:comp_rho_alp}
\end{figure}

\begin{figure}
\begin{center}
\begin{tabular}{c}
\includegraphics[width=\columnwidth]{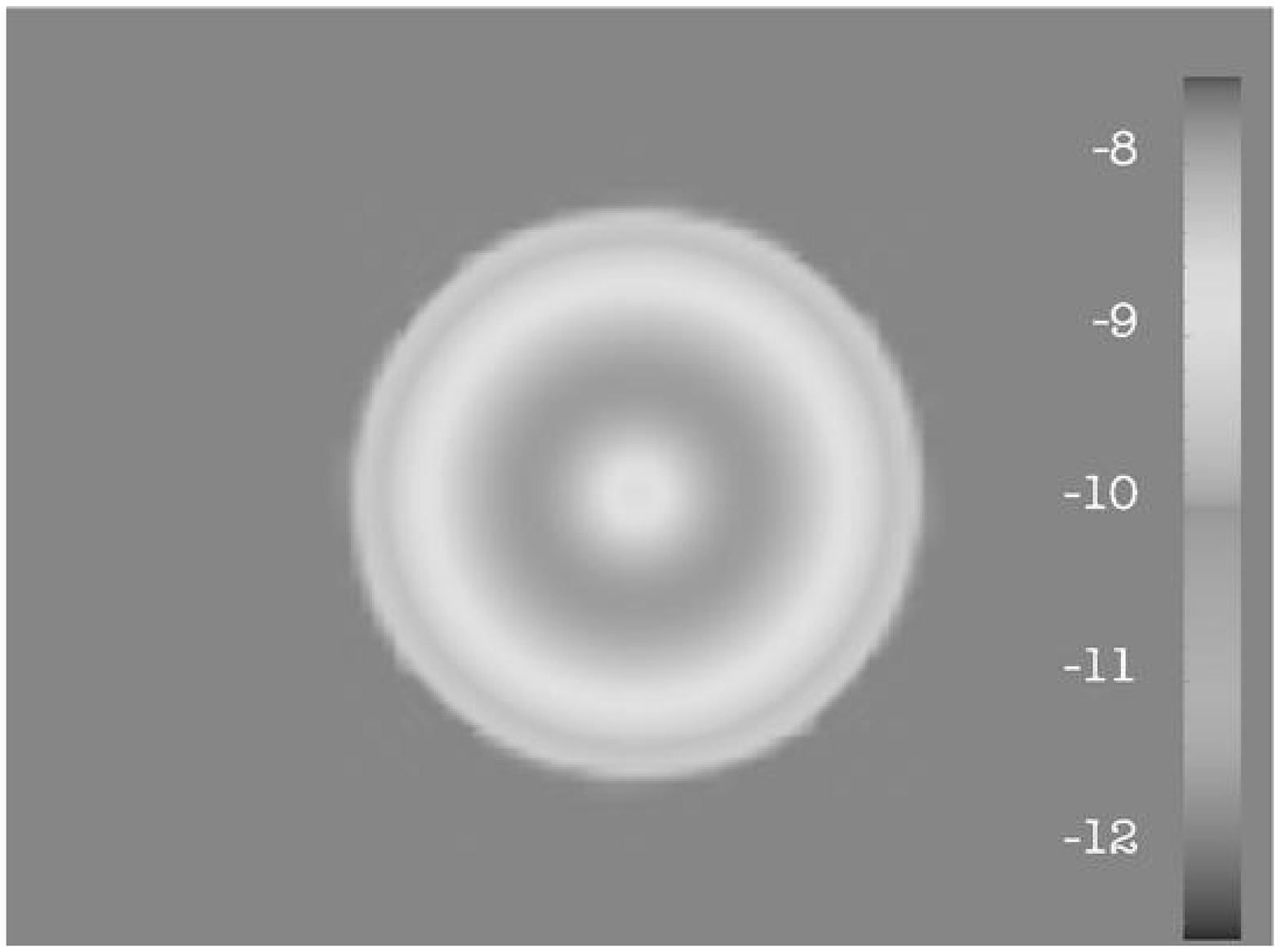} \\
\includegraphics[width=\columnwidth]{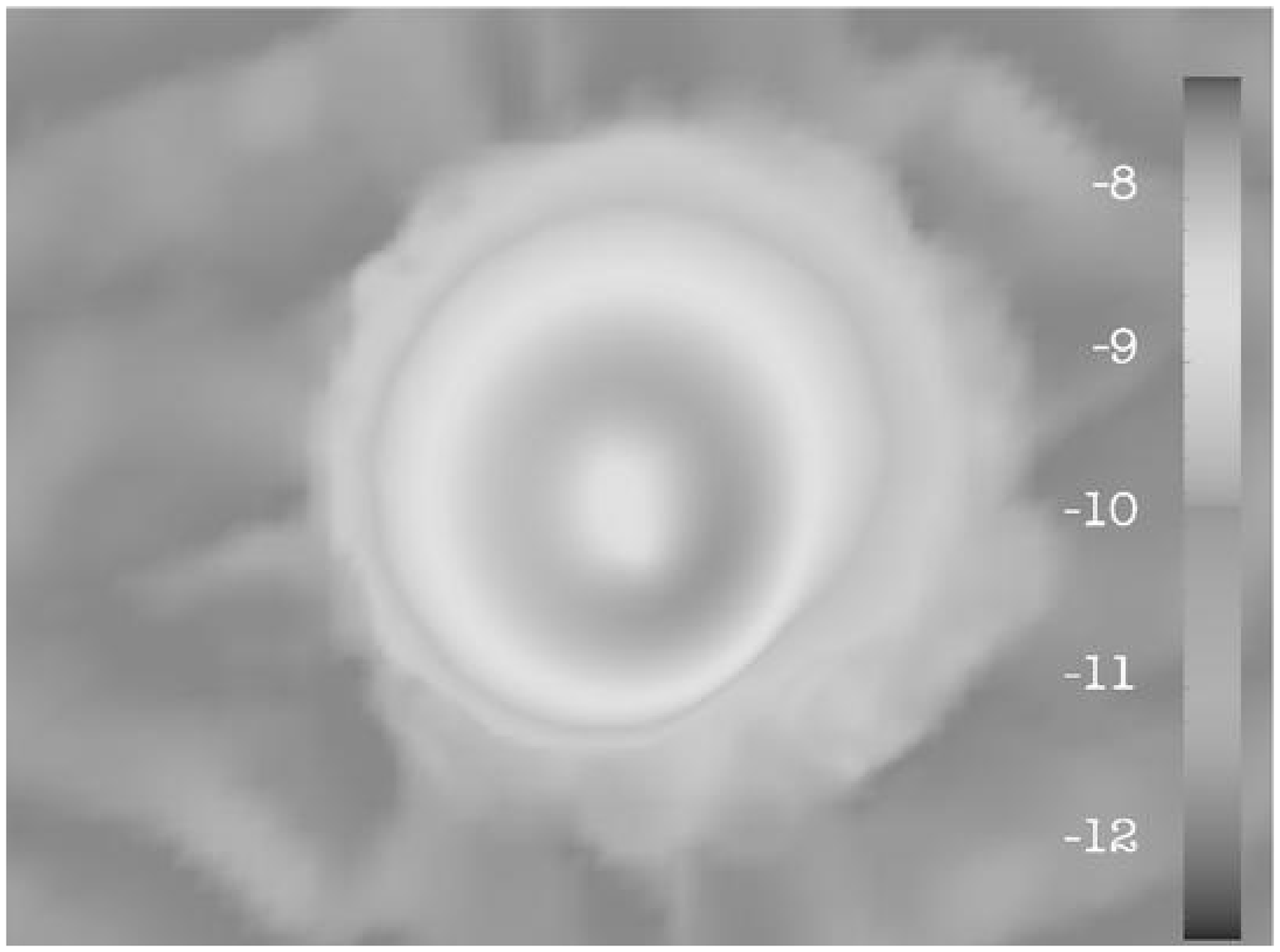} \\ 
\includegraphics[width=\columnwidth]{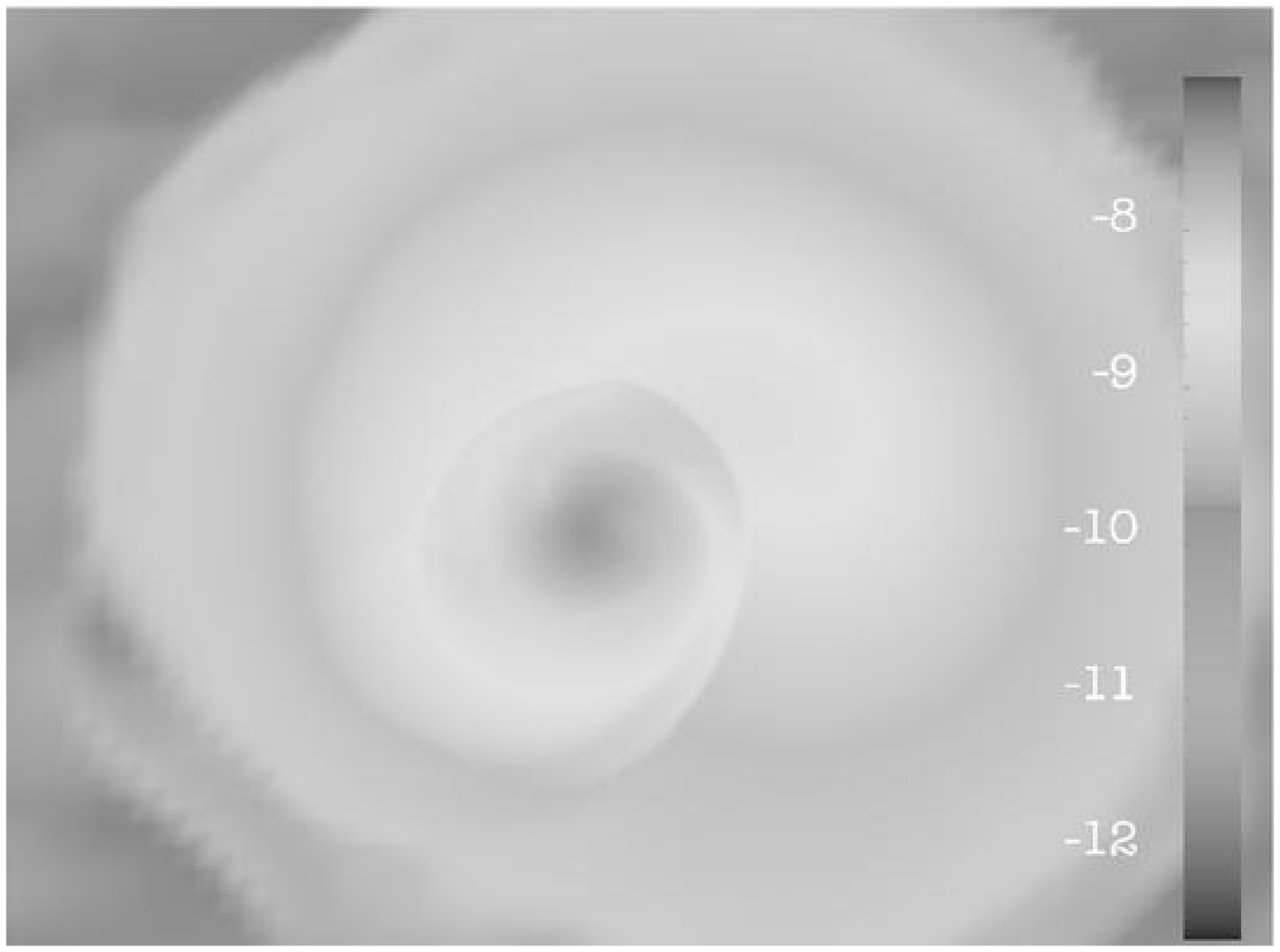} \\
\end{tabular}
\end{center}
\caption{Evolution of density in the equatorial plane of the model
\emph{C8}. Shown is the decadic logarithm of the rest-mass density. The
snapshots were taken at times $t/t_{dyn} = 0$ (top), $6.28$ (middle) 
and $8.28$ (bottom). 
In contrast to the more compact model \emph{C1}, which is the reference
polytrope investigated earlier (cf. Fig.~\ref{fig:ref_64_rho_alp}), the fragment 
re-expands after a maximal compression.}
\label{fig:comp_8_rho}
\end{figure}

The mode amplitudes $A_1$ and $A_2$ for different members of the \emph{C} sequence 
(cf. Section~\ref{sec:id_comp} and Table~\ref{tab:sequence_comp}) are shown in 
Fig.~\ref{fig:comp_m1_m2}.
The plot demonstrates that different choices of $\rho_c$ do not have a significant 
effect on the growth time of the mode, which is in contrast to the effects
of $\Gamma$ and $r_p/r_e$ discussed above. However, while the linear development 
of the mode is similar for different compactnesses,
the non-linear behaviour is not. Consider Fig.~\ref{fig:comp_rho_alp}: The reference
polytrope \emph{C1} and the model \emph{C2}, which is about half as compact, both
show an unbounded growth in the maximum of the  density and a collapse of the lapse, indicating
black hole formation. The models \emph{C4} and \emph{C8}, however, appear to avoid
black hole formation and re-expand after a state of maximum compression. 
Fig.~\ref{fig:comp_8_rho} shows these different types of evolution for the model \emph{C8} 
in more detail: In contrast to the black hole-forming case \emph{C1}, the fragment 
re-expands after the collapse in this evolution. 
This is not unexpected, since the Newtonian limit, for an equation of state
$\Gamma = 4/3$, admits a stable equilibrium state of the fragment if it has
sufficient rotation. We conclude therefore that, even when the growth time
of the instability is quite similar for stars of different compactness, the 
outcome of the fragmentation can differ drastically.\footnote{These results have been confirmed
with lower and higher resolutions.}

\subsection{Evolution of quasi-toroidal models of constant central rest-mass density}
\label{sec:evol_plane}

\begin{figure}
\begin{center}
\includegraphics[width=\columnwidth]{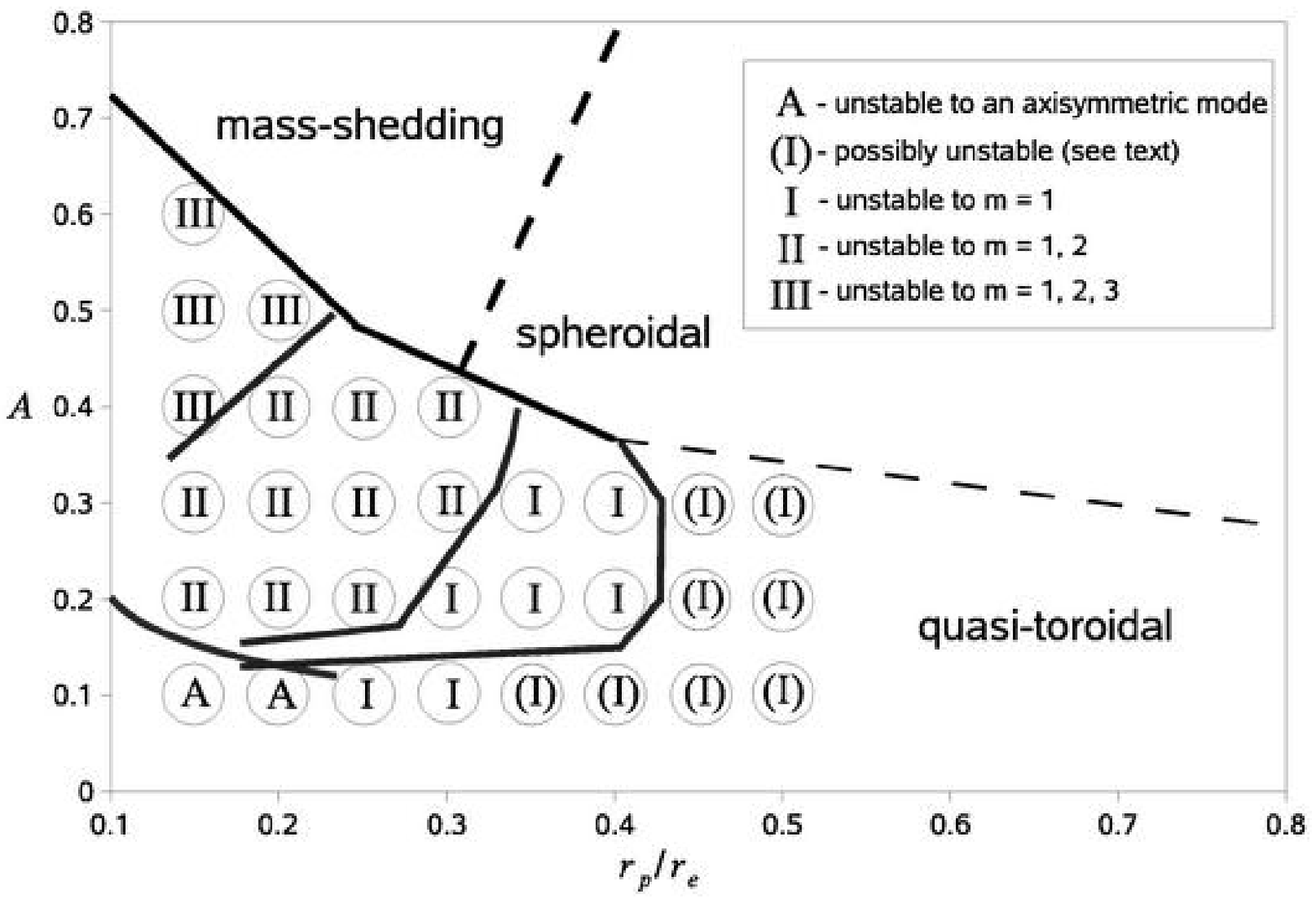}
\end{center}
\caption{Stability of quasi-toroidal models with $\rho_c = 10^{-7}$
(cf. Fig.~\ref{fig:models_plane}). A Latin number denotes
the highest azimuthal order of the unstable modes, i.e. ``I'' for $m = 1$ unstable,
``II'' for $m = 1,2$ unstable, and ``III'' for $m = 1,2,3$ unstable. Models denoted
by ``(I)'' are either long-term unstable with growth times $\tau \gg t_{dyn}$, or stable (see text),
and models denoted by ``A'' exhibit an axisymmetric instability. The red line
in the lower left is the approximate location of the sequence $J/M^2 = 1$ 
(cf. Fig.~\ref{fig:models_plane}),
and the three blue lines inside the quasi-toroidal region are the approximate 
locations of sequences with 
$T/|W| = 0.14$ (right), $T/|W| = 0.18$ (middle) and $T/|W| = 0.26$ (left).}
\label{fig:model_stability}
\end{figure}

\begin{figure}
\begin{center}
\includegraphics[width=\columnwidth]{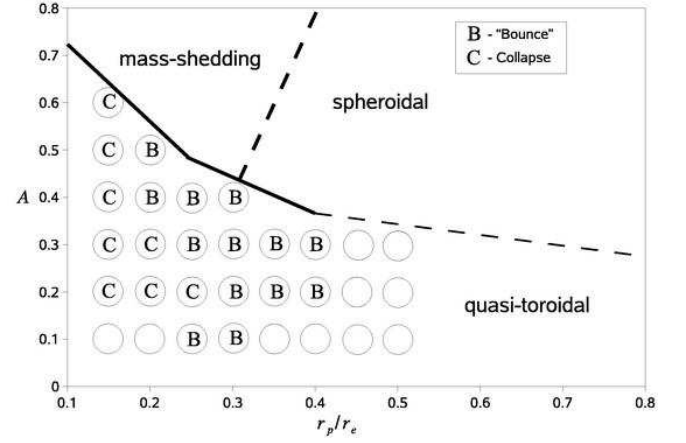}
\end{center}
\caption{Remnants of the models from Fig.~\ref{fig:model_stability}, which are unstable
with respect to non-axisymmetric modes. The non-linear behaviour has been analyzed
by observing the evolution of the function $\min \alpha$ (see also 
Fig.~\ref{fig:comp_rho_alp}). Models which show a minimum in this function
are marked by ``B'' for ``bounce,'' while models exhibiting an exponential
collapse of the lapse are marked by ``C'' for ``collapse.''}
\label{fig:model_remnant}
\end{figure}

\begin{figure}
\begin{center}
\includegraphics[width=\columnwidth]{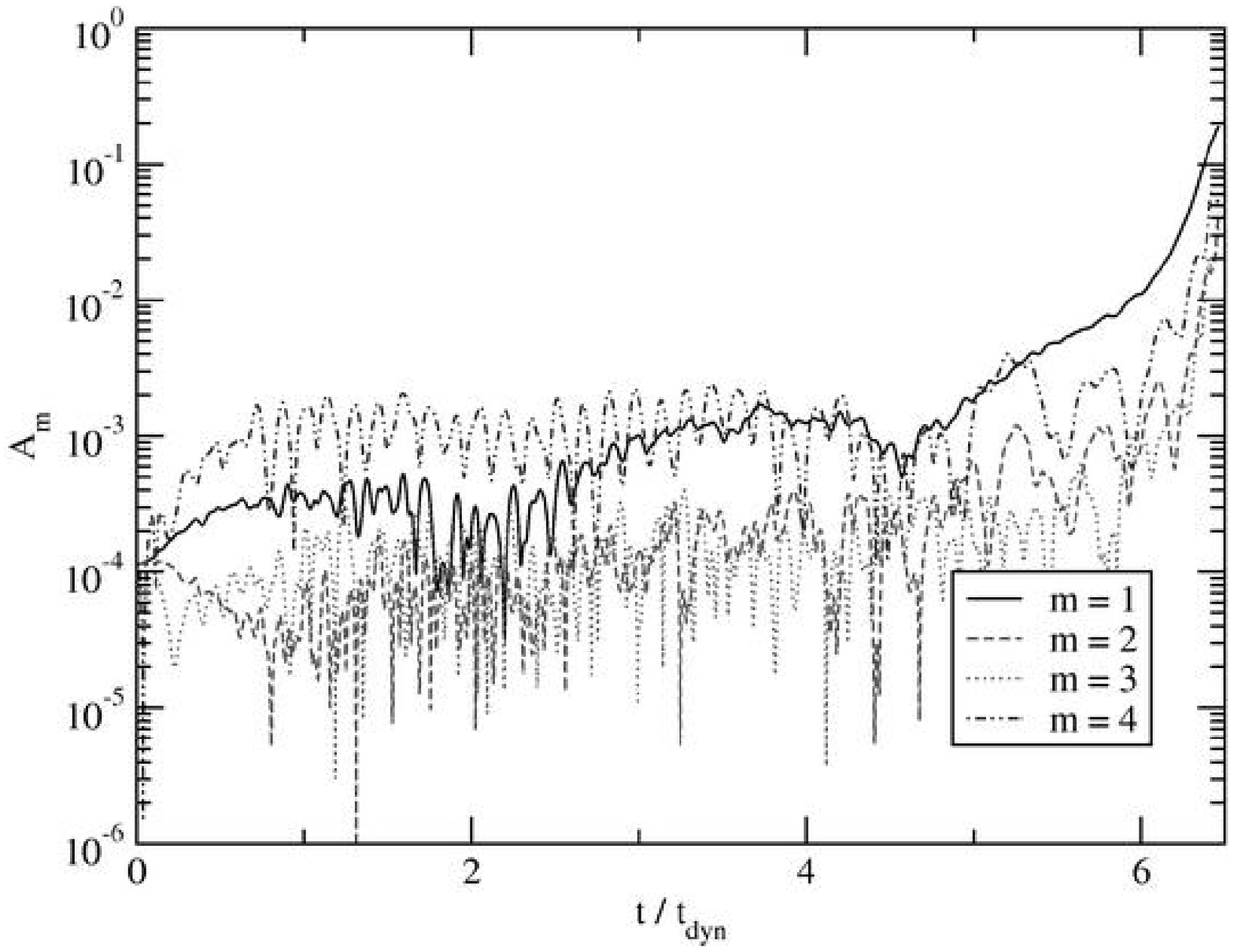}
\end{center}
\caption{Mode amplitudes in the model \emph{A0.1R0.15} (cf. Table~\ref{tab:selected_models}), 
extracted at the radius of highest initial rest-mass density.}
\label{fig:a_0.1_ratio_0.15_64_modes}
\end{figure}

\begin{figure}
\begin{center}
\includegraphics[width=\columnwidth]{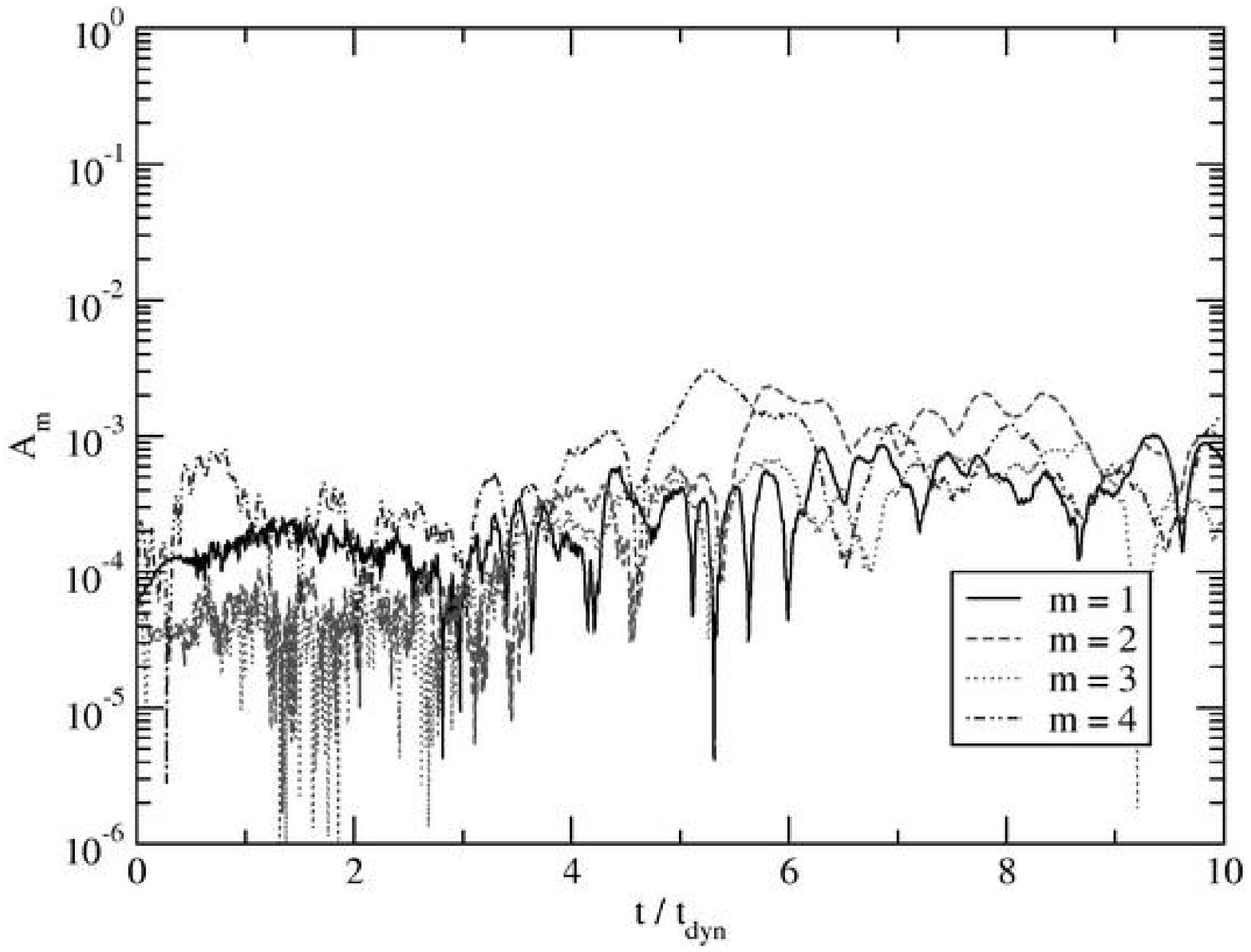}
\end{center}
\caption{Mode amplitudes in the model \emph{A0.1R0.50} (cf. Table~\ref{tab:selected_models}), 
extracted at the radius of highest initial rest-mass density.}
\label{fig:a_0.1_ratio_0.50_64_modes}
\end{figure}

\begin{figure}
\begin{center}
\includegraphics[width=\columnwidth]{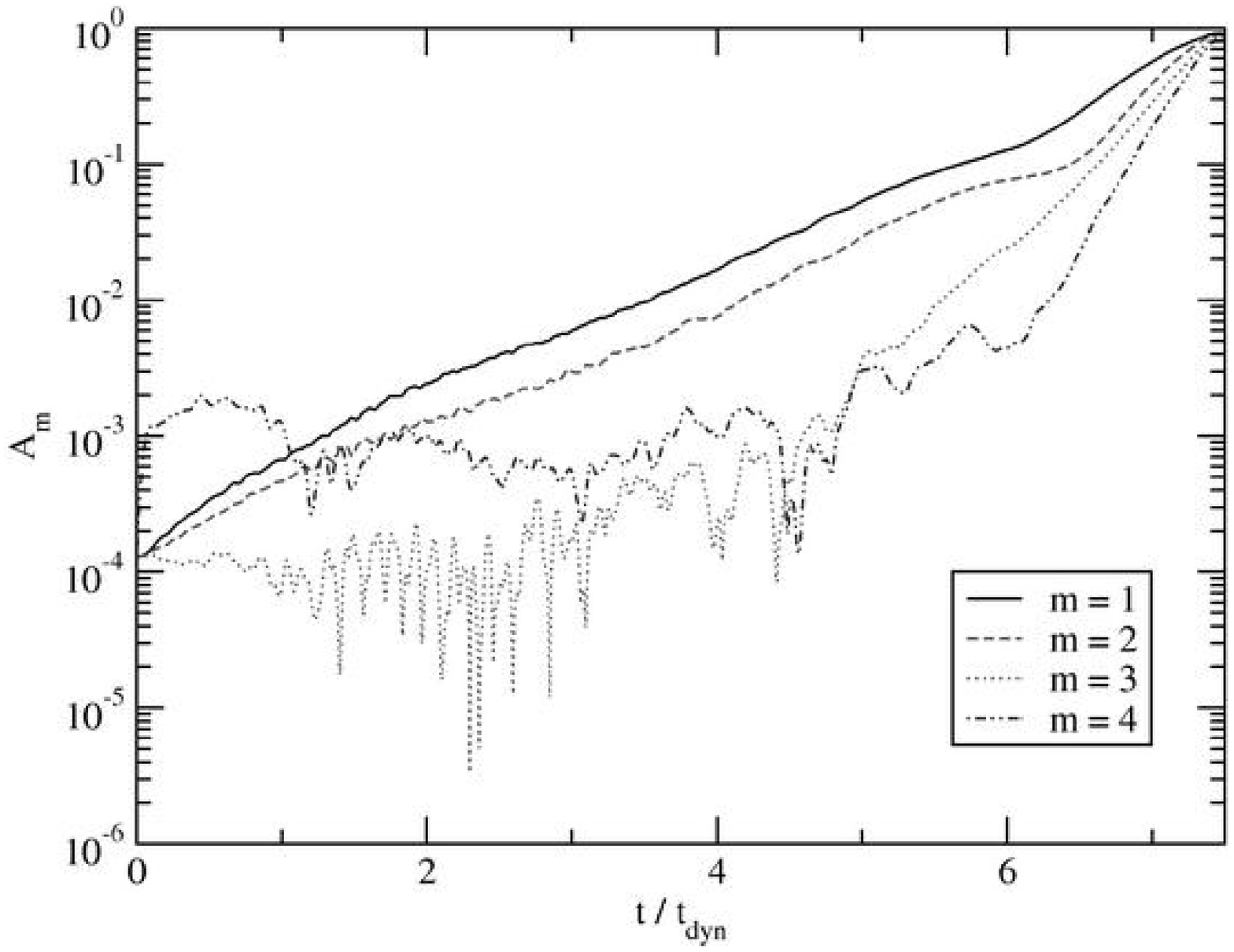}
\end{center}
\caption{Mode amplitudes in the model \emph{A0.3R0.15} (cf. Table~\ref{tab:selected_models}), 
extracted at the radius of highest initial rest-mass density.}
\label{fig:a_0.3_ratio_0.15_64_modes}
\end{figure}

\begin{figure}
\begin{center}
\includegraphics[width=\columnwidth]{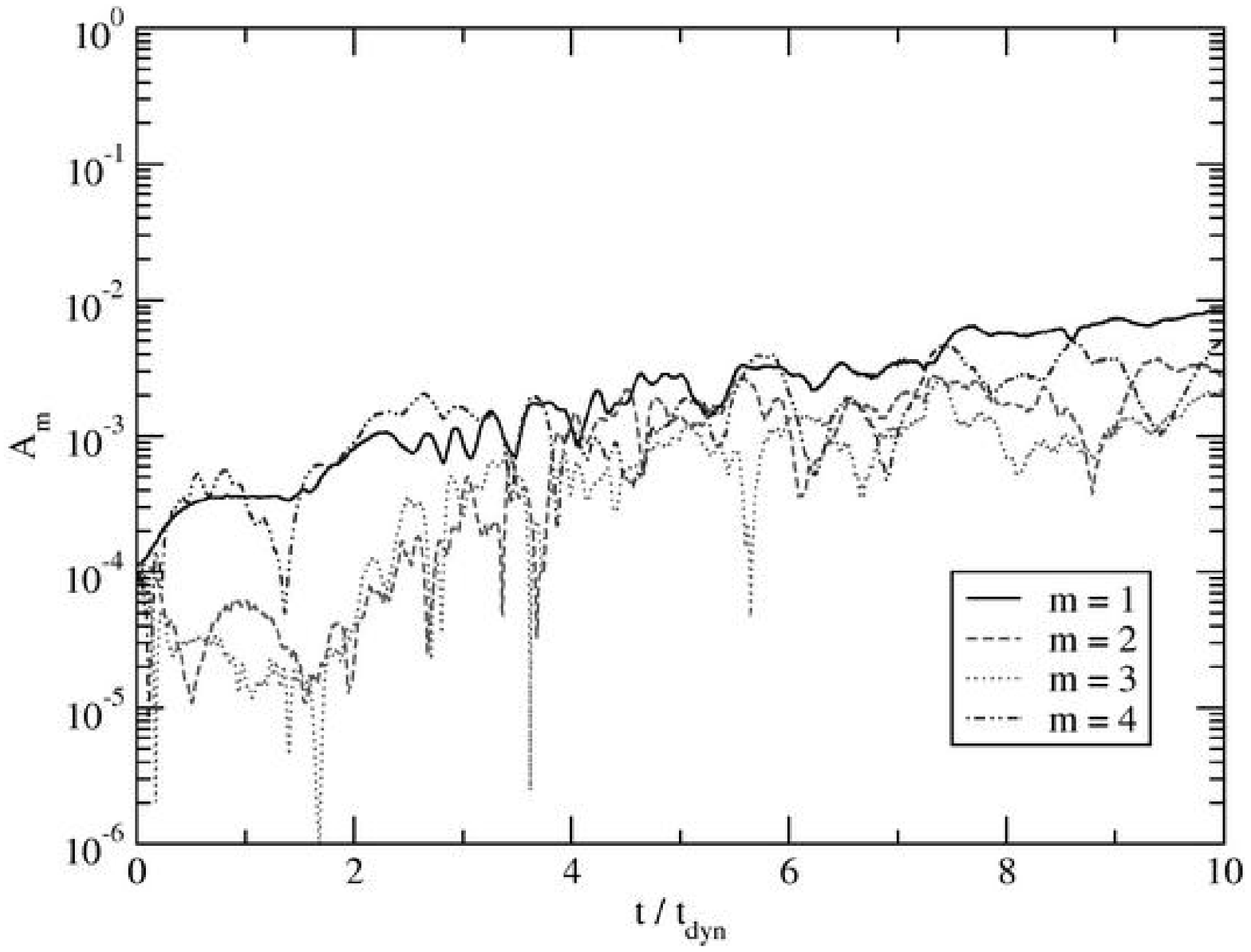}
\end{center}
\caption{Mode amplitudes in the model \emph{A0.3R0.50} (cf. Table~\ref{tab:selected_models}), 
extracted at the radius of highest initial rest-mass density.}
\label{fig:a_0.3_ratio_0.50_64_modes}
\end{figure}

\begin{figure}
\begin{center}
\includegraphics[width=\columnwidth]{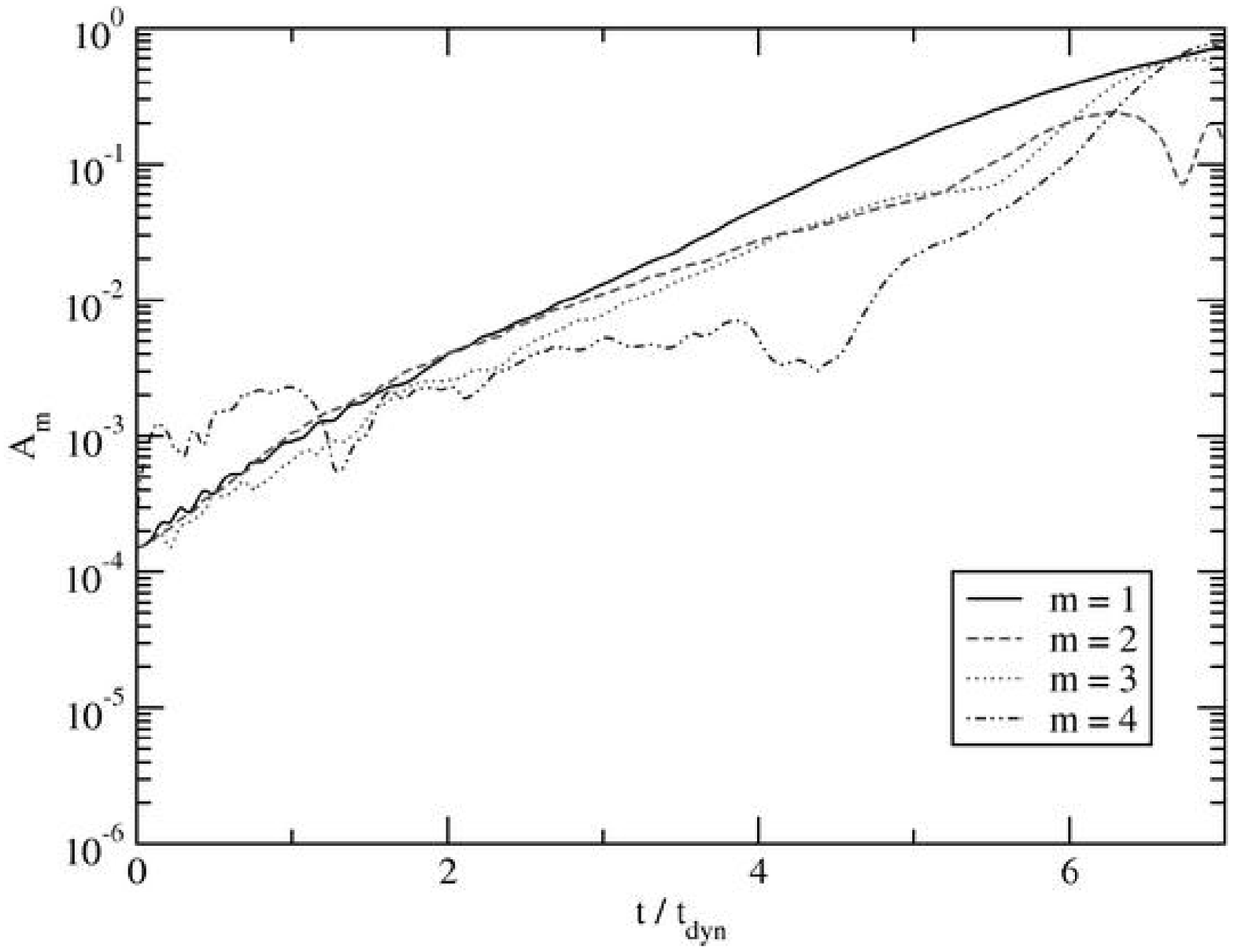}
\end{center}
\caption{Mode amplitudes in the model \emph{A0.6R0.15} (cf. Table~\ref{tab:selected_models}), 
extracted at the radius of highest initial rest-mass density.}
\label{fig:a_0.6_ratio_0.15_64_modes}
\end{figure}

\begin{figure}
\begin{center}
\begin{tabular}{c}
\includegraphics[width=\columnwidth]{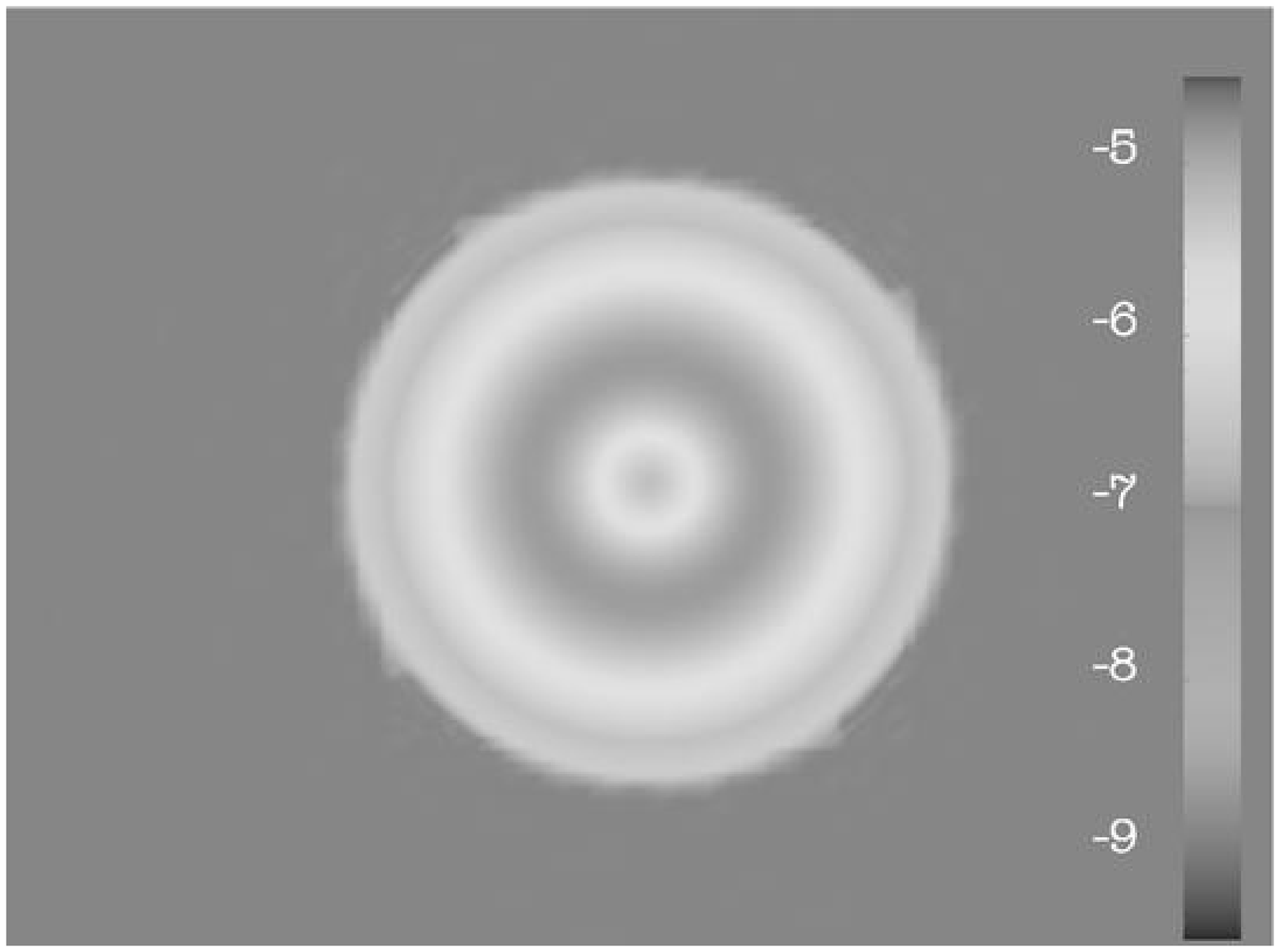} \\
\includegraphics[width=\columnwidth]{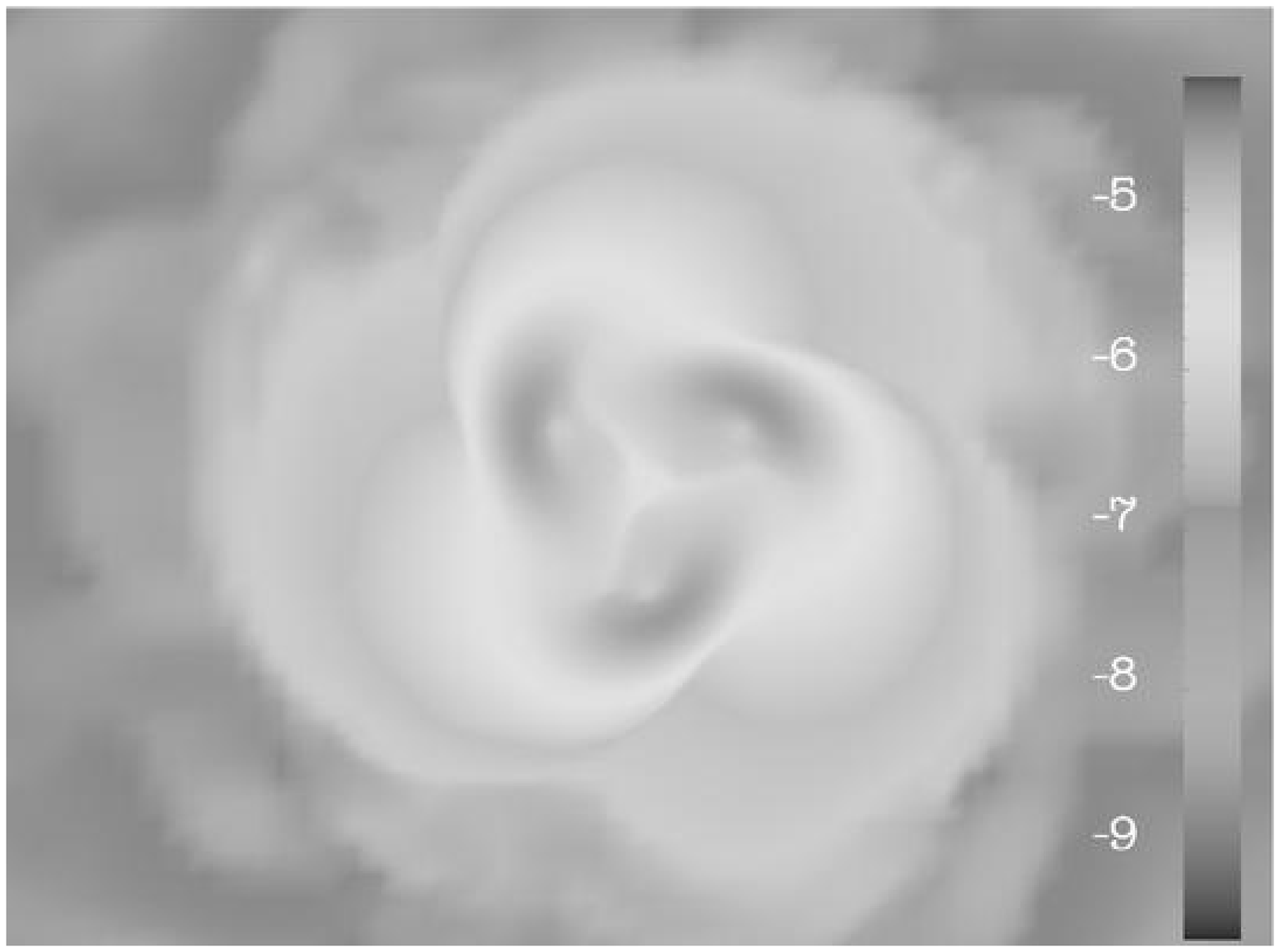} \\
\includegraphics[width=\columnwidth]{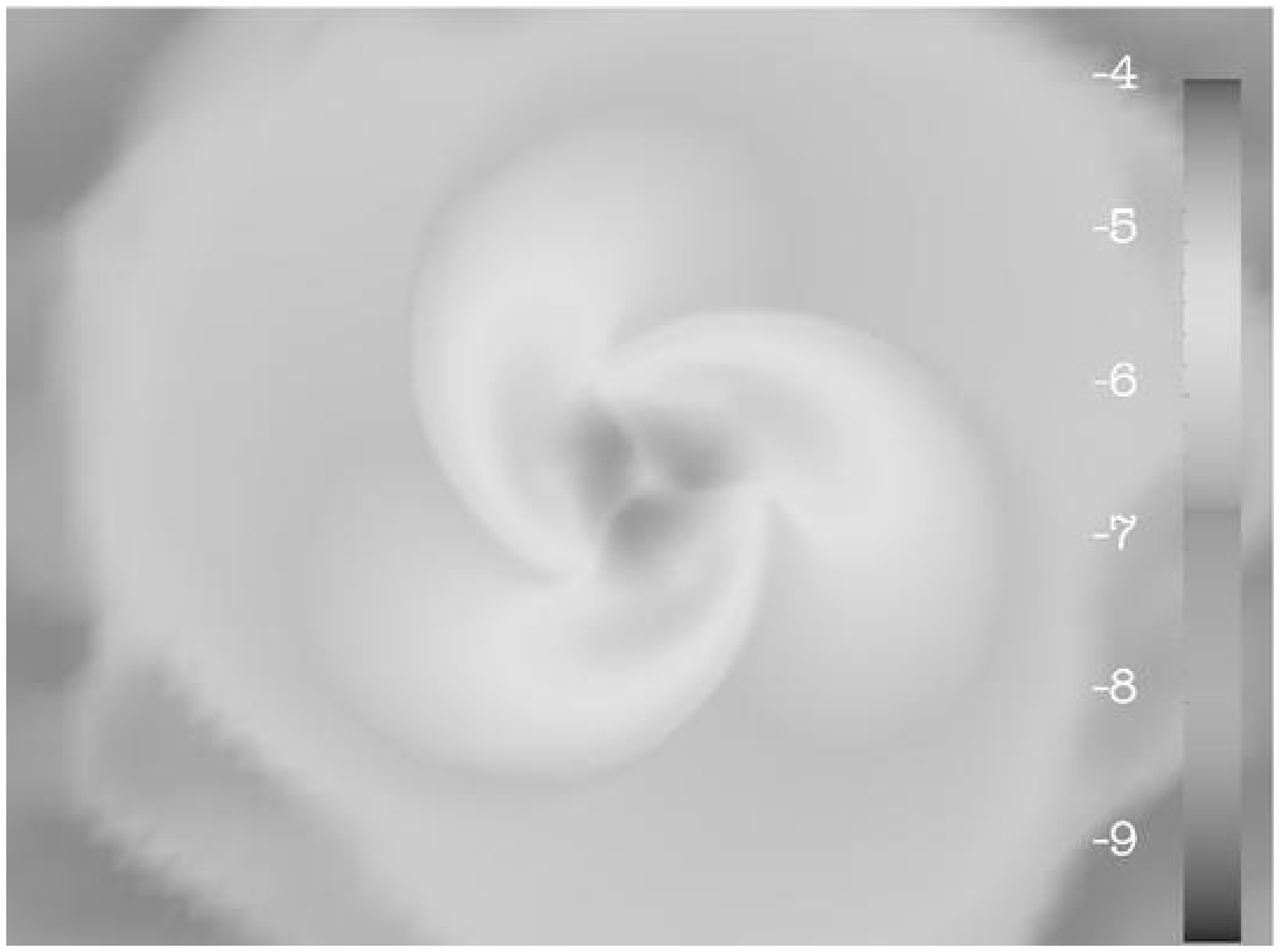} \\
\end{tabular}
\end{center}
\caption{Equatorial density evolution of the model \emph{A0.6R0.15}, 
with an $m = 3$ perturbation. (Note that, in Fig.~\ref{fig:a_0.6_ratio_0.15_64_modes},
a perturbation with $\lambda_m = 1$ has been used instead).
Shown is the decadic logarithm of the rest-mass density.
The snapshots were takes at times $t/t_{dyn} = 0$ (top), $6.28$ (middle) and 
$7.60$ (bottom). Three fragments develop and subsequently encounter
collapse similar to the two-fragment case (cf. Fig.~\ref{fig:ref_64_m2_rho}). 
The evolution of the model 
perturbed with $m = 1$ and $m = 2$ is similar to the corresponding one in the 
reference polytrope.}
\label{fig:a_0.6_ratio_0.15_m3_rho}
\end{figure}

\begin{figure}
\begin{center}
\includegraphics[width=\columnwidth]{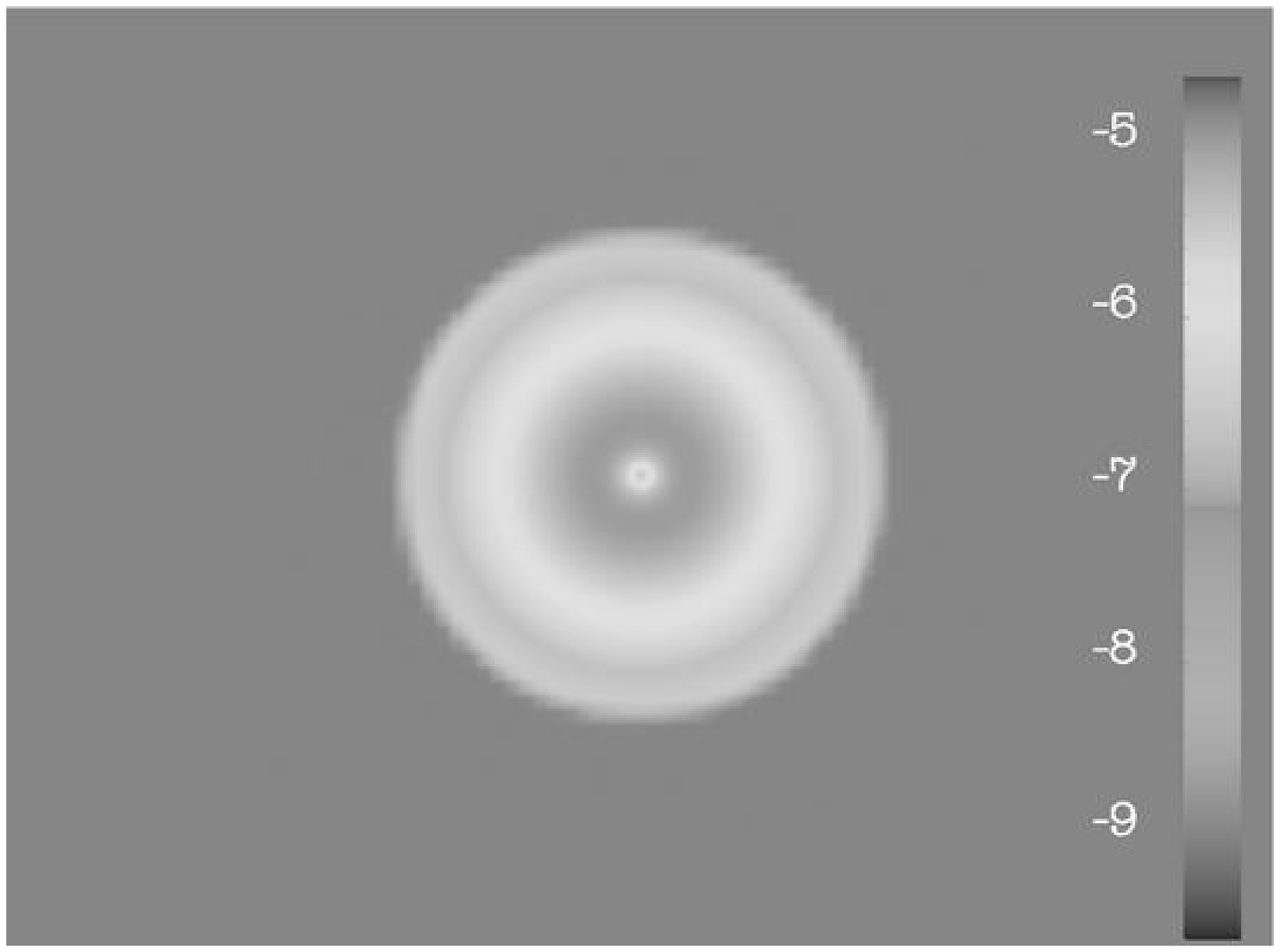}
\includegraphics[width=\columnwidth]{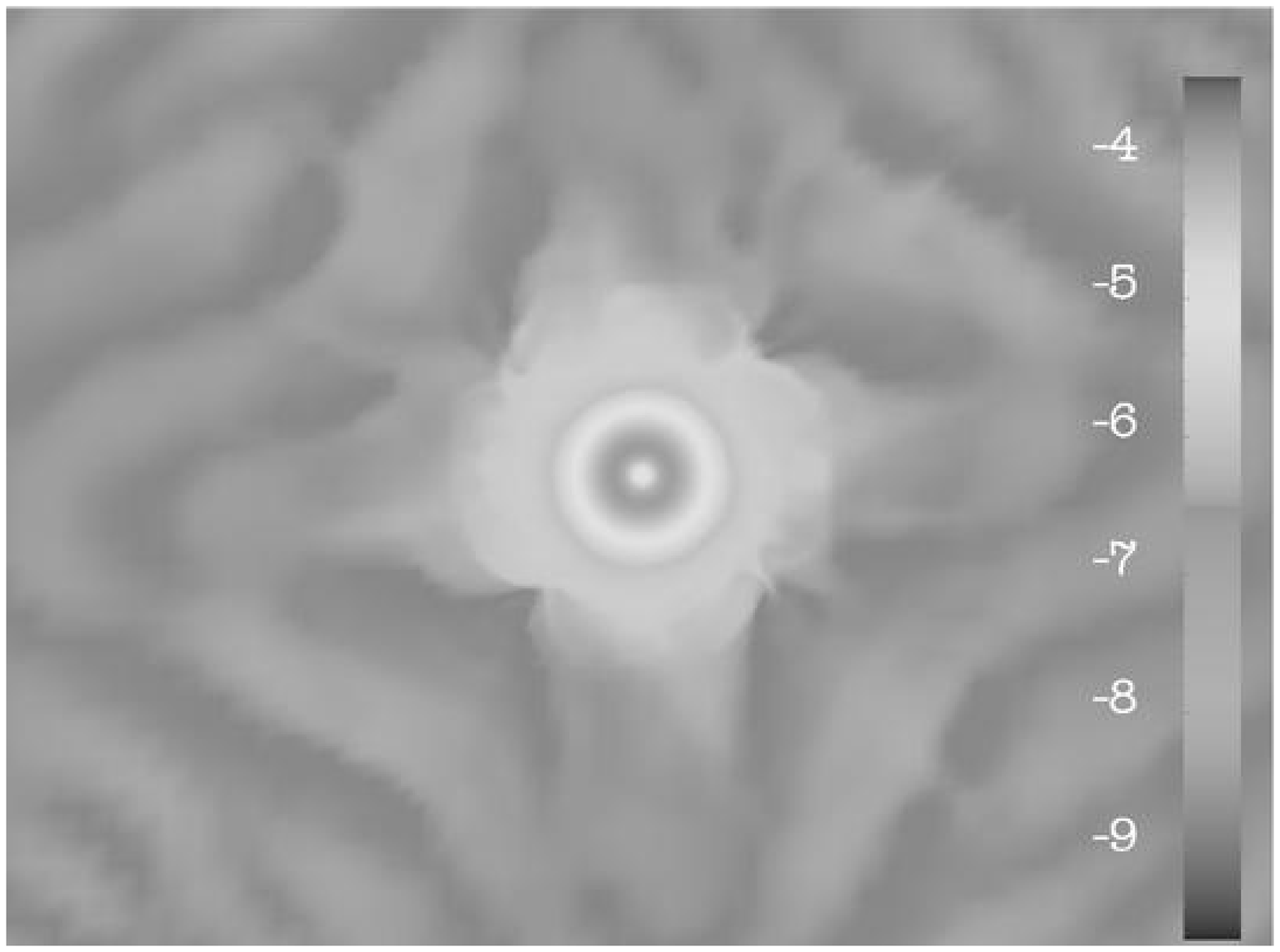}
\end{center}
\caption{Equatorial density evolution of the model \emph{A0.1R0.15}.
Shown is the decadic logarithm of the rest-mass density.
The snapshots were takes at times $t/t_{dyn} = 0$ (top) and 
$6.28$ (bottom). The model exhibits an axisymmetric instability.}
\label{fig:a_0.1_ratio_0.15_rho}
\end{figure}

The structure of the parameter space plane $\Gamma = 4/3$ and $\rho_c = 10^{-7}$ has
been discussed in Section~\ref{sec:id_plane}. As already noted, the necessity of investigating
only one plane is determined primarily by the computational cost of three-dimensional general
relativistic hydrodynamical simulations. Also, the choice of the central
density does not seem to affect the almost exponential development of a non-axisymmetric unstable mode 
in the linear regime considerably, even for very compact quasi-toroidal polytropes (cf. Section~\ref{sec:evol_comp}), which 
is in contrast to axisymmetric modes. Finally, we note that the models
with $\rho_c  = 10^{-7}$ are already quite compact, with $R_e/M \approx 10 \ldots 100$, and
$r_p/M \approx 2 \ldots 70$.

To investigate the stability of these models, the initial data indicated by circles in the bottom
right panel of Fig.~\ref{fig:models_plane} have been evolved, imposing a perturbation of the form given by 
eqn.~\ref{eq:perturbation} with $\lambda_m = 1$, and with a resolution of 
$65 \times 65 \times 33$ grid points in the outer patches, and $97 \times 97 \times 49$ in the innermost 
patch. Selected models have been tested against individual $m = j$ perturbations with 
$\lambda_m = \delta_{mj}$, with different resolutions, and different densities of the artificial 
atmosphere, to test consistency and convergence. Also, central rest-mass densities different from
$10^{-7}$ were investigated in a few models.

Fig.~\ref{fig:model_stability} gives an overview of the stability properties of the 
selected models. The Latin numbers ``I'' to ``III'' refer to the highest $m$ with an 
unstable mode, i.e. in addition to the reference polytrope, which belongs to the class
``II,'' we find models which are unstable to an $m = 3$ perturbation, and models
which appear to be stable against $m = 2$ (within the restrictions illustrated in 
Section~\ref{sec:instability_slow}). The models denoted with an ``A'' have been found to 
be unstable to an axisymmetric mode, and collapse before any non-axisymmetric instability 
develops. Finally, the models marked with ``(I)'' are either stable or long-term unstable with a growth
time $\tau \gg t_{dyn}$. Each model has been evolved for up to $10 \, t_{dyn}$ to determine its 
stability. This limit is arbitrary, but imposed by the significant resource requirements of 
these simulations. If no mode amplitude exceeds the level of the $m = 4$ noise during this
time, the model is marked with a ``(I).'' This does not imply that the model is actually
stable, and we will investigate a specific model denoted by ``(I)'' later. We will find it 
to be unstable to an $m = 1$ mode with slow growth (Section~\ref{sec:instability_slow}).

The additional lines in Fig.~\ref{fig:model_stability} are approximate
isolines of the functions $T/|W|$ for the values $0.14$, $0.18$ and $0.26$ 
and of the function $J/M^2$ for the value
$1$. As long as the models do not rotate too differentially, $T/|W|$ still
seems to be a reasonable indicator of the non-axisymmetric stability of the polytropes, even
though they are quasi-toroidal and relativistic.

The nature of the non-linear behaviour of models exhibiting a non-axisymmetric instability
is indicated in Fig.~\ref{fig:model_remnant}. We use the evolution of the minimum
of the lapse function to classify the models, see also Fig.~\ref{fig:comp_rho_alp}. Models
denoted by ``B'' have a global minimum in the lapse, while models denoted by ``C'' do not. Given that the
compactness of the models increases with smaller axis ratios in this plot (cf. 
Fig.~\ref{fig:models_plane}), we expect that a black hole forms for each member of 
the ``C'' class. To determine this uniquely, each of these models should be tested using the
adaptive mesh-refinement technique presented in \cite{Zink2005a}, which is, however,
beyond the scope of this study.\footnote{We also note that models denoted by ``B'' might
actually form a black hole by delayed collapse.}

In Fig.~\ref{fig:a_0.1_ratio_0.15_64_modes} to \ref{fig:a_0.6_ratio_0.15_64_modes}, we
have plotted the mode amplitudes $A_m$ for selected models
(cf. Table~\ref{tab:selected_models}). The evolution of the model \emph{A0.3R0.15}
(Fig.~\ref{fig:a_0.3_ratio_0.15_64_modes}) is quite similar to that of the reference polytrope.
Model \emph{A0.6R0.15} is further inside the unstable region, and exhibits also an
$m = 3$ instability: the density evolution of this mode is plotted in 
Fig.~\ref{fig:a_0.6_ratio_0.15_m3_rho}.  The models \emph{A0.1R0.50} and 
\emph{A0.3R0.50} are stable within the numerical restrictions mentioned above. 

The model \emph{A0.1R0.15} seems to have an unusual evolution
of the mode amplitudes; this also applies to the model \emph{A0.1R0.20}, which is not
shown here. 
The density evolution of
\emph{A0.1R0.15} (Fig.~\ref{fig:a_0.1_ratio_0.15_rho}) shows that the
model has encountered an axisymmetric instability before any non-axisymmetric
modes can grow to appreciable amplitudes. We note that both models \emph{A0.1R0.15}
and \emph{A0.1R0.20} have $J/M^2 < 1$, in contrast to most other models in the
parameter space plane considered here; only the model \emph{A0.1R0.25} has $J/M^2 = 0.961$. In
Fig.~\ref{fig:model_stability}, the isoline $J/M^2 = 1$ is marked. It approximately
separates the region of axisymmetric from that of non-axisymmetric instability.

\subsection{The location of the instability in the corotation band}
\label{sec:corotation}

\begin{figure}
\begin{center}
\begin{tabular}{c}
\includegraphics*[width=\columnwidth]{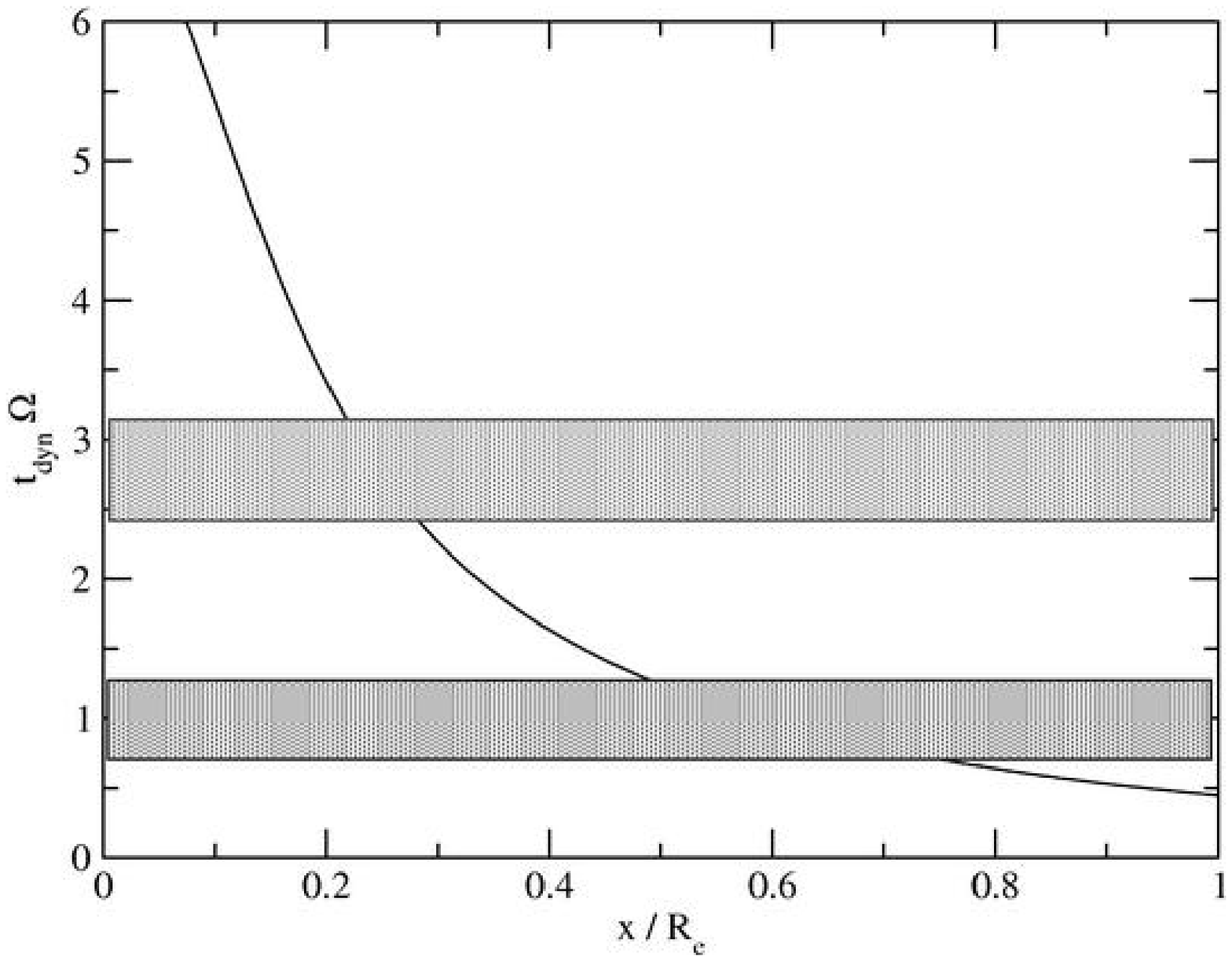} \\
\includegraphics*[width=\columnwidth]{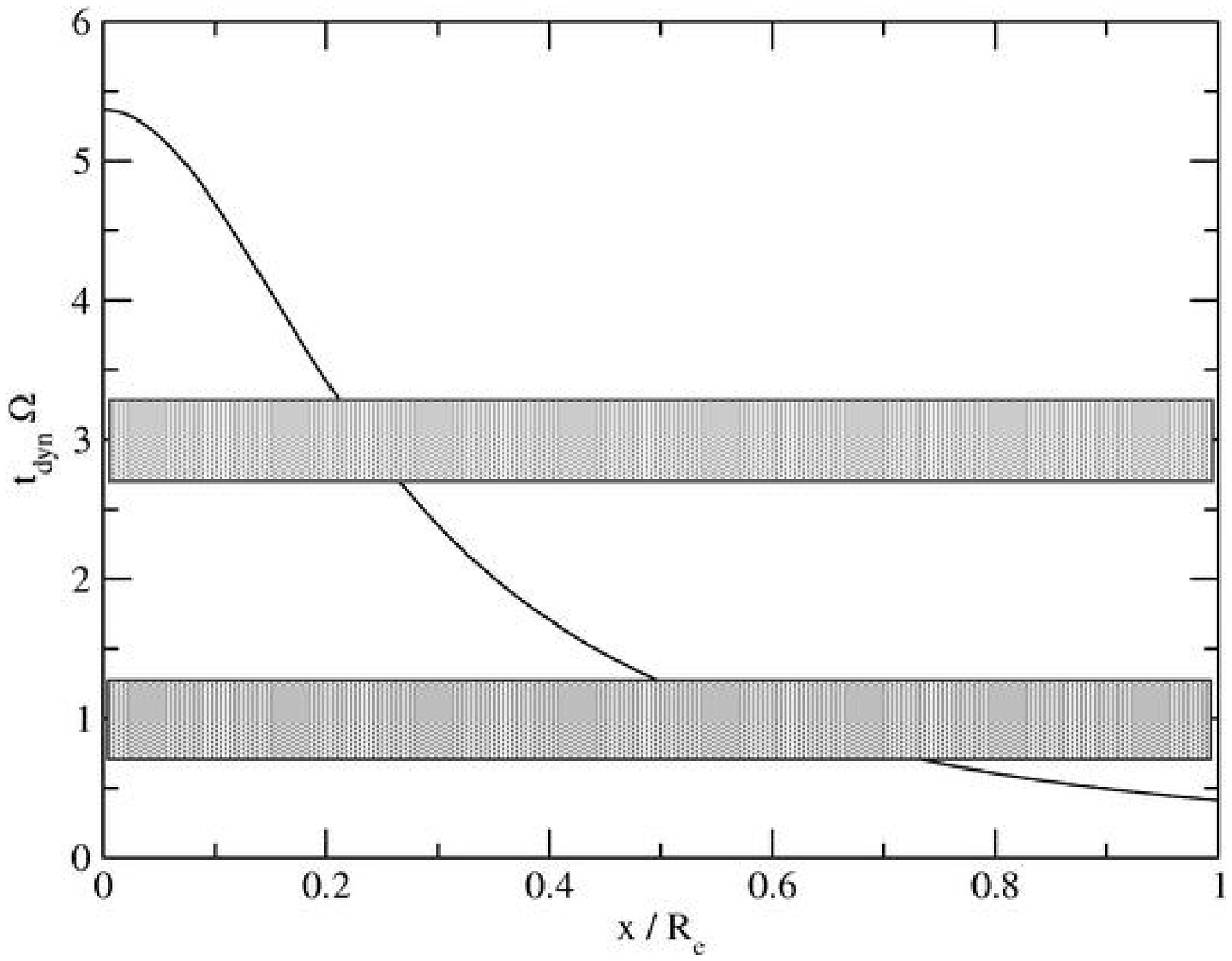} \\
\includegraphics*[width=\columnwidth]{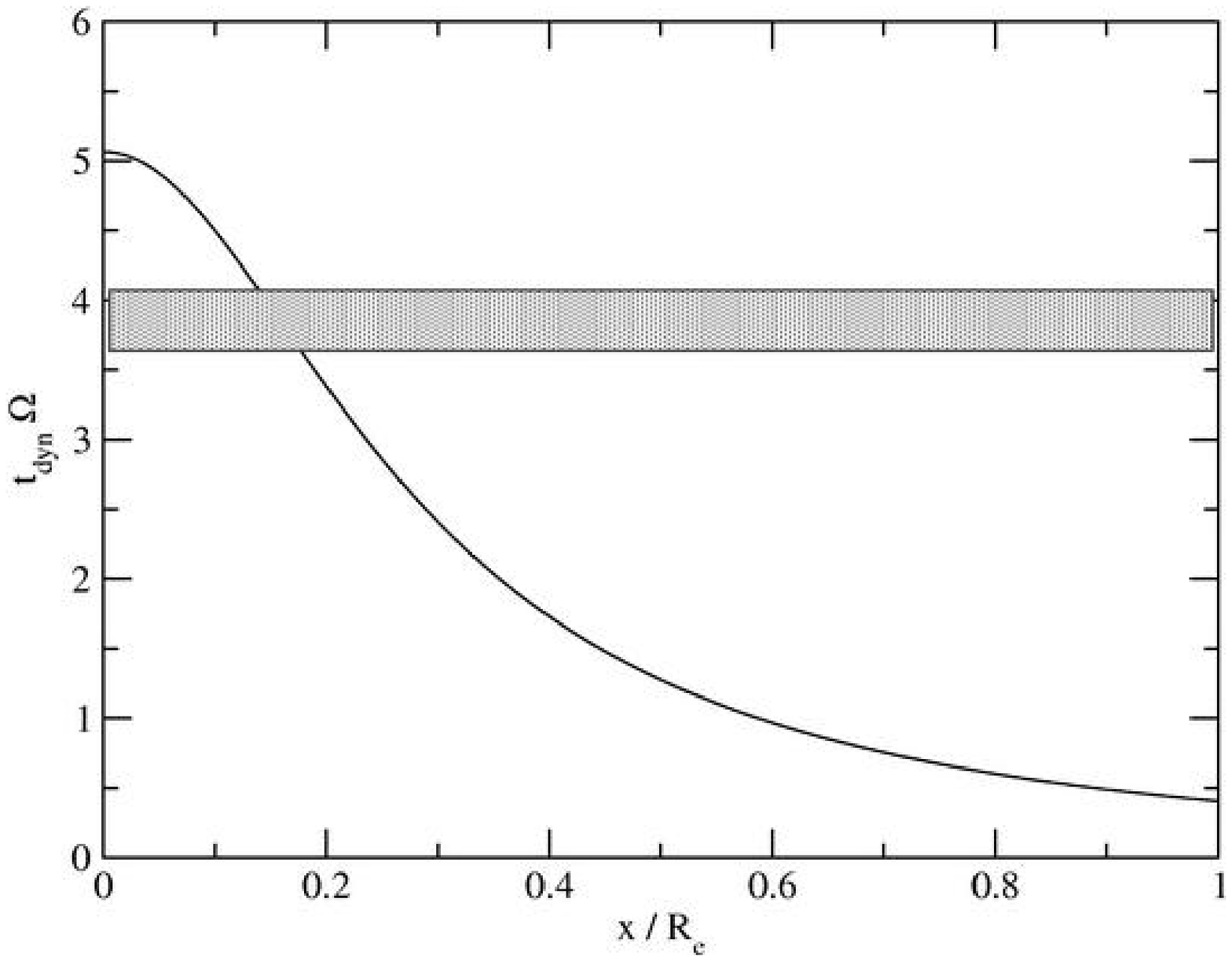} \\
\end{tabular}
\end{center}
\caption{Angular velocity of the polytropes \emph{A0.3R0.15} (top), 
\emph{A0.3R0.30} (middle) and \emph{A0.3R0.40} (bottom) over the $x$ axis 
(black line), and approximate 
location of the pattern speed of the $m = 1$ mode 
(upper rectangle) and the $m = 2$ mode (lower rectangle), cf. also Fig.~\ref{fig:ref_64_corot}.}
\label{fig:corotation}
\end{figure}

Fig.~\ref{fig:corotation} shows the location of the unstable modes in the
corotation band, for three different models on the sequence $\Gamma = 4/3$,
$\rho_c = 10^{-7}$, and $A = 0.3$. All modes are in corotation, but there is
evidence that, for decreasing $T/|W|$, the corotation point moves towards the
axis of rotation. This gives support to the arguments presented in \cite{Watts:2003nn},
where the existence of low-$T/|W|$ and spiral-arm instabilities in differentially
rotating polytropes are connected to corotation\footnote{There is evidence that,
on a sequence of increasing rotation parameter, some modes in the discrete spectrum
become unstable when entering the corotation band (which has a continuous spectrum), 
or might merge with other modes inside the band and become unstable. These
are mechanisms not present in uniformly rotating polytropes.}. Limitations of resources did
not permit us to investigate the boundary of the corotation region, where 
growth times of many $t_{dyn}$ are expected. However, with the results of
Sections~\ref{sec:evol_comp} and \ref{sec:limit_sequence}, 
a purely Newtonian investigation could be sufficient to reproduce the linear regime of the 
instability.

\subsection{Evolution of a model with a slow growth of the $m = 1$ instability}
\label{sec:instability_slow}

\begin{figure}
\begin{center}
\includegraphics[width=\columnwidth]{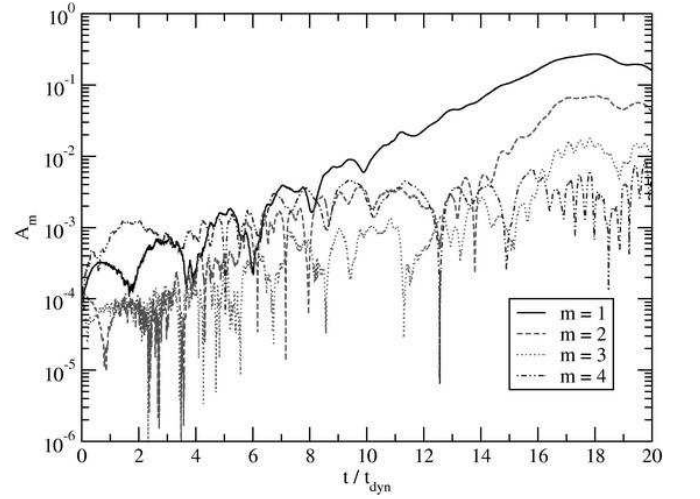}
\end{center}
\caption{Long-term evolution of the mode amplitudes for the model \emph{A0.2R0.45},
which is unstable to an $m = 1$ perturbation. The mode, however, grows
rather slowly over a time of $20 \, t_{dyn}$.}
\label{fig:a_0.2_ratio_0.45_64_modes}
\end{figure}

As already discussed, the nature of the 
class ``(I)'' models in Fig.~\ref{fig:model_stability} could not be investigated in detail due to the high
computational cost when evolving general relativistic, three-dimensional
models. However, to illustrate the behaviour in one specific case, 
a long-term simulation of the model \emph{A0.2R0.45} has been performed
(Fig.~\ref{fig:a_0.2_ratio_0.45_64_modes}). A slowly growing $m = 1$ instability
is apparent in the evolution, which saturates at high amplitudes only after
$20 \, t_{dyn}$. While the $m = 1$ mode is clearly dominant, the $m = 2$ might
be unstable as well. A detailed investigation of these sequences should be
attempted in the limit of vanishing compactness, with a Newtonian model
and preferably, with a cylindrical grid (see also the discussion in
Section~\ref{sec:conclusion}).

\subsection{Evolution of a sequence of models with different compactness 
starting from the boundary between the regions ``I'' and ``(I)''}
\label{sec:limit_sequence}

\begin{figure}
\begin{center}
\includegraphics[width=\columnwidth]{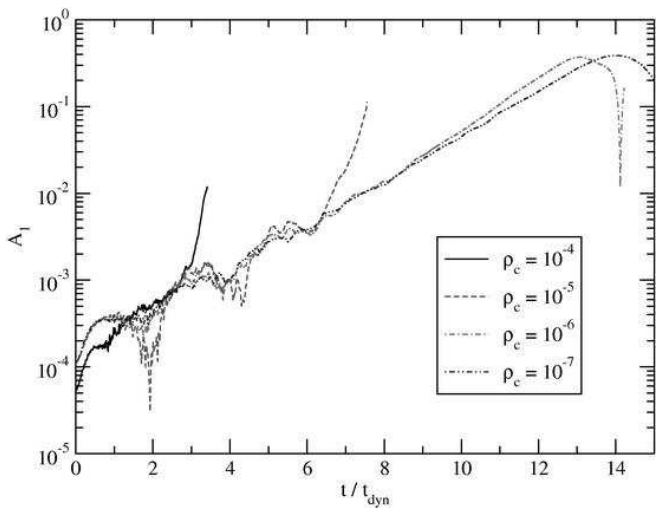}
\end{center}
\caption{Amplitude of the $m = 1$ mode for different models from a sequence
with constant $T/|W|$ limiting in the model \emph{A0.2R0.40},
which has $\rho_c = 10^{-7}$. Cf. also Table~\ref{tab:sequence_marginal}.}
\label{fig:marginal_m1}
\end{figure}

In Section~\ref{sec:evol_comp}, we have already studied the influence of the
compactness on the development of the instability in the reference polytrope.
According to the results of Section~\ref{sec:evol_plane}, the reference
 model is located inside region ``II'' of the parameter space
plane for $\rho_c = 3.38 \cdot 10^{-6}$. Thus, it is instructive
to investigate the effect of compactness on a model's evolution which is located close to
the boundary between regions ``I'' and ``(I)'' in Fig.~\ref{fig:model_stability}
(although this boundary is not sharply defined). 
A selected model, \emph{A0.2R0.40}, has been extended to the \emph{L} sequence 
of constant $\Gamma$, $A$, and $T/|W|$ (cf. Table~\ref{tab:sequence_marginal}).
The influence of compactness on the $m = 1$ mode is illustrated in 
Fig.~\ref{fig:marginal_m1}: The most compact models \emph{L1} and \emph{L2}
show a growth of the non-axisymmetric mode already early on, but collapse due to an axisymmetric
instability (both models have $J/M^2 < 1$). The growth rate of the
non-axisymmetric instability is not very sensitive to the compactness, which confirms
our findings for models of the \emph{C} sequence. One might therefore be reasonably optimistic that
the \emph{non-axisymmetric} stability properties of quasi-toroidal $N = 3$ polytropes 
are well-represented by Fig.~\ref{fig:model_stability}, even for a different choice of 
central rest-mass density. The \emph{axisymmetric} stability, and the question whether 
the collapse of the fragment will be halted or not, is sensitive to $\rho_c$.

\section{Discussion}
\label{sec:conclusion}

In this paper we have presented an extension of our earlier work on the
fragmentation of general relativistic quasi-toroidal polytropes and the production of
black holes in this scenario \cite{Zink2005a}.
The central focus was to gain an understanding of how various parameters determining 
the structure of the equilibrium polytrope affect the development of the instability
observed in \cite{Zink2005a}, and the nature of its remnant. In addition, we have
investigated the location of the unstable modes in the corotation band of the
differentially rotating models.

All investigations have been performed using three-dimensional numerical simulations
in general relativity, and assuming the stars to be self-gravitating 
perfect fluids with an adiabatic coefficient equal to the polytropic constant
$\Gamma$. The equations of general relativistic hydrodynamics have been evolved
using high-resolution shock-capturing methods, and the NOK-BSSN formalism has been
used for the metric evolution. All grids use fixed mesh refinement, and 
impose an equatorial plane symmetry. The development of unstable modes has been
followed by the use of a discrete Fourier transform of the rest-mass density computed at certain
coordinate radii in the equatorial plane, with a preference on the radius of
initial highest density.

The central results are represented in Fig.~\ref{fig:model_stability},
\ref{fig:model_remnant}, and \ref{fig:corotation}. 
For a plane of constant
rest-mass density $\rho_c = 10^{-7}$ and $\Gamma = 4/3$, we have determined the region 
where quasi-toroidal
models become dynamically unstable to non-axisymmetric fragmentation. 
From the structure of the space of initial models presented
in Fig.~\ref{fig:models_plane}, it appears that
there is a rough relation between $T/|W|$ and the highest order of unstable
modes, at least as long as the degree of differential rotation is not too
high. Since the numerical method is not well-suited to follow the development
of instabilities with growth times much longer than a dynamical timescale,
we could not determine the fate of models from class ``(I)'' with certainty.
However, we have shown in one specific case that the model is actually unstable. In the same
manner, a model from class ``I'' could be subject to a slowly growing
mode with $m > 1$; however, in this case the $m = 1$ mode would clearly be dominant.

From the investigation of a sequence emanating from the model used
in the publication \cite{Zink2005a}, we have found that the central rest-mass
density $\rho_c$, which controls the compactness of the polytrope, does not affect the
almost exponential development of the non-axisymmetric instability significantly. This is related to
the fact that $T/|W|$ is insensitive to $\rho_c$. However, $\rho_c$ determines the nature
of the final remnant: While the model in \cite{Zink2005a} forms a black hole, two models having
one fourth and one eighth as much compactness show a re-expansion of the fragment after
maximal contraction.

The regions of models in the plane $\rho_c = 10^{-7}$ and $\Gamma = 4/3$ where such a 
re-expansion was observed  is indicated in Fig.~\ref{fig:model_remnant}
by a ``B.'' If one assumes that the models \emph{not} exhibiting this behaviour,
marked with ``C'' in Fig.~\ref{fig:model_remnant}, are forming black holes
in the same manner as shown in \cite{Zink2005a}, then we can draw the following
tentative picture of black hole formation by fragmentation of single stars.

The nature of the final system, either an almost unperturbed axisymmetric star, 
a single central black hole, single or multiple non-central black holes with a disk, 
or one or several expanding remnants without trapped surfaces, depends on the symmetry 
properties of the perturbation, and the location of the equilibrium star with respect 
to three types of surfaces in the space of parameters:
\begin{enumerate}
\item \emph{Surfaces indicating the onset of the instability of a mode of a certain order
$m$.} These surfaces might be close to isosurfaces of $T/|W|$ as indicated
in Fig.~\ref{fig:model_stability}, but the resource requirements of performing
three-dimensional general relativistic simulations limit our ability
to identify slowly growing modes. However, if only modes growing on a dynamical
timescale are considered, then $T/|W|$ might yield a reasonable indicator
of the location of the limit surfaces. 
The compactness of the initial model seems to have no significant effect on the 
location of these surfaces, at least for $\Gamma = 4/3$. 
\item \emph{Surfaces indicating the onset of an axisymmetric instability.} In our
samples, all models with $J/M^2 < 1$ were unstable to quickly growing axisymmetric
modes, and hence will likely evolve to central black holes. The surface $J/M^2 = 1$ can
therefore be used as an approximate separator between axisymmetric collapse and 
stability (cf. \cite{Sekiguchi:2004ba} for a more detailed discussion of this
point).
\item \emph{Surfaces separating prompt black-hole formation from re-expansion.} In 
Fig.~\ref{fig:model_remnant}, an approximate determination of such a surface
has been attempted. In a first approach, and assuming that results for
the stability of slowly and uniformly rotating relativistic polytropes 
\cite{Durney67, Chandrasekhar68, Hartle72} can be applied to the fragments, 
we expect a fragment with a higher compactness and lower rotation rate to be destabilized, 
and, given that 
the geometric development of the fragmentation process is similar for different
choices of compactness of the equilibrium polytrope, that there is a close
connection to isosurfaces of $R_e/M$ and $J/M^2$. However, when comparing 
the structure of $R_e/M$ and $J/M^2$ (cf. Fig.~\ref{fig:models_plane}) and \ref{fig:model_remnant}, the situation
appears more complicated, and deserves further attention.
\end{enumerate}

With respect to the question of whether multiple black holes may form or not,
two comments are in order: First, in the unstable systems of class ``II''
and ``III,'' all growth times of modes with different $m$ are of comparable
magnitude, i.e. the nature of the actual time development in a specific star
will depend on the nature of the perturbation, as already mentioned. On a sequence of
increasing $T/|W|$, we have found the $m = 1$ perturbation to be dominant before higher 
order modes become unstable, which suggest the (off-center) formation of a single black 
hole with a massive accretion disk. However, as discussed in the main text, this conclusion 
applies to instabilities with growth times in the order of a dynamical timescale. More 
specifically, our simulations do not exclude, on a sequence of increasing $T/|W|$, 
a slowly growing higher order instability to be effective before the $m = 1$ mode 
is dynamically unstable. Second, when two fragments are forming and collapsing, in all 
cases considered here a runaway instability develops and leads to a central collapse 
(Fig.~\ref{fig:ref_64_m2_rho}). In this case, the gravitational wave signal
is expected to resemble the ring-down phase of a highly deformed black hole. For the 
$m=1$ collapse, we have published an approximate prediction for a partial waveform in \cite{Zink2005a},
which suggests an amplitude and frequency well inside the LISA sensitivity if the
source is a collapsing supermassive star.

Concerning the nature of the non-axisymmetric mode, we have collected evidence
that, along a sequence with decreasing $T/|W|$, the corotation point moves towards
the axis of rotation of the polytrope. This gives support to the arguments by Watts et al.
\cite{Watts:2003nn}. To also investigate the cases
with large growth times, however, a Newtonian model, preferably on a cylindrical
grid, would be of advantage to obtain more detailed results \cite{Saijo:2005gb}.

A final comment is in order concerning the black hole remnant. Since the
normalized angular momentum $J/M^2$ of the initial model is greater
than unity, the resulting black hole, unless it is ejected from its shell, may
very well be \emph{rapidly rotating}, spun up by accretion of the material
remaining outside the initial location of the trapped surface.

Possible future work on this problem can be roughly divided into four approaches.
First, the nature of the low-$T/|W|$ and $m = 1$ instabilities in
quasi-toroidal polytropes could be investigated in Newtonian gravity,
or perhaps using some perturbative approximation of general relativity, to determine the
location of the corresponding surfaces in parameter space, and to suggest regions
where the quantity $T/|W|$ is still a good indicator for the degree of instability. Since the
Newtonian polytropes can be considered to be limit points of relativistic sequences
with vanishing compactness, the systematic effects of general relativity on their
stability properties can be determined separately.

Second, the location of the surfaces separating black hole formation from 
``bounce'' behaviour, and its relation to the initial compactness $R_e/M$ and
the normalized angular momentum $J/M^2$, needs 
to be determined with more detail, specifically also for different equations of
state and rotation laws. Could a newly formed, rapidly and differentially rotating neutron star 
fragment in this way?  We have found no example of this kind here, but such a question 
deserves further attention. A more detailed investigation of the ``bouncing'' models
might also be interesting, since they could have a rich phenomenology, 
including a possible delayed collapse to black holes and complex gravitational wave 
signatures.

Third, a study of the evolution of the black hole and its shell would shed light on a 
number of interesting aspects: (i) The angular momentum of the remnant black hole, which 
is related to the structure of the accretion disk and its ability to power jets, 
(ii) the kick velocity of the black hole, and the question whether it may be ejected from
its host galaxy in some cases, and (iii) the gravitational wave signal from the formation 
process at $\mathscr{J}^+$. While we were not able to evolve the black hole for
an extended time after its formation, recent advances in discrete techniques may be able
to solve this problem, either by adding a sufficient amount of artificial dissipation
\cite{Baiotti06} or by employing multi-block grids with summation-by-parts operators
\cite{Lehner2005a, Zink2005b}.

Fourth, to connect more closely to certain astrophysical systems, a detailed model
of the micro-physical processes, particle transport, and magnetic fields is necessary
in many cases to obtain specific answers. The most important bulk property appears
to be a change in $\Gamma$ with density, since this would modify the non-linear
evolution of the fragmentation significantly. In the specific case of core
collapse, results in this context have been obtained already
\cite{Shibata:2004kb, Ott:2005gj}.

\begin{acknowledgments}

We would like to thank L. Rezzolla, J. Font, A. Watts, N. Andersson, and 
A. Nagar for stimulating discussions.
Our numerical calculations used the Cactus framework \cite{Goodale02a, url:cactus} with
a number of locally developed thorns. This work was partially supported by the
SFB-TR7 ``Gravitational Wave Astronomy'' of the DFG, by the IKYDA 2006/7 grant between
IKY (Greece) and DAAD (Germany), and by the National Center
for Supercomputing Applications under grant
MCA02N014, utilizing the machines \texttt{cobalt} and \texttt{tungsten}. 
The \texttt{peyote} cluster of the Albert Einstein Institute and the
AMD Opteron machines at the Max Planck Institute for Astrophysics were also used.
This research employed the resources of the Center for Computation and Technology at Louisiana State
University, which is supported by funding from the Louisiana legislature's Information Technology Initiative.

\end{acknowledgments}




\end{document}